\documentclass[a4paper,fleqn,usenatbib]{mnras}
\usepackage{newtxtext,newtxmath}
\usepackage[T1]{fontenc}
\usepackage{ae,aecompl}

\usepackage{graphicx}
\usepackage{amsmath}
\usepackage{amssymb}
\usepackage{float}
\usepackage{enumerate}  
\usepackage{lipsum}
\usepackage{gensymb}
\usepackage{graphicx}
\usepackage{longtable}
\usepackage{rotate}
\usepackage{rotating}
\usepackage{mathtools}
\usepackage{pdflscape}
\usepackage{scalerel}
\usepackage{times}
\usepackage{verbatim}
\usepackage{subfig}
\usepackage{lscape}
\newif\ifAMStwofonts

\usepackage{appendix}

\def\gs{\mathrel{\hbox{\rlap{\hbox{\lower4pt\hbox{$\sim$}}}\hbox{$>$}}}}
\def\ls{\mathrel{\hbox{\rlap{\hbox{\lower4pt\hbox{$\sim$}}}\hbox{$<$}}}}

\def\suzaku{{\it Suzaku}}
\def\hst{{\it HST}}

\def\swift{{\it Swift}}

\def\xmm{{\it XMM-Newton}}

\def\nustar{{\it NuSTAR}}
\def\spitzer{{\it Spitzer}}
\def\wise{{\it WISE}}
\def\galex{{\it GALEX}}

\def\mrk335{{Mrk~335}}
\def\izw1{{I~Zw~1}}

\def\A{{\rm\thinspace \AA}}

\title[Empirical analysis of \mrk335]
      {
Tracking the year-to-year variation in the spectral energy distribution of the Narrow-line Seyfert~1 Galaxy \mrk335\ }

\author[]
       {S. Tripathi,$^1$ 
        K. M. McGrath,$^1$
	L. C. Gallo,$^1$
        D. Grupe,$^2$
       	S. Komossa,$^3$ 
	M. Berton,$^{4,5}$
	G. Kriss,$^6$
\newauthor
        A. L. Longinotti$^7$ %
        \\ 
$^{1}$ Department of Astronomy and Physics, Saint Mary's University, 923 Robie Street, Halifax, NS, B3H 3C3, Canada \\
$^{2}$ Space Science Center, Morehead State University, 235 Martindale Drive, Morehead, KY 40351, USA\\
$^{3}$ Max-Planck-Institut f{\"u}r Radioastronomie, Auf dem H\"ugel 69, 53121 Bonn, Germany \\
$^{4}$ Finnish Centre for Astronomy with ESO (FINCA), University of Turku, Quantum, Vesilinnantie 5, FI-20014, University of Turku, Finland\\
$^{5}$ Aalto University Mets{\"a}hovi Radio Observatory, Mets{\"a}hovintie 114, FI-02540 Kylm{\"a}l{\"a}, Finland\\
$^{6}$Space Telescope Science Institute, 3700 S. Martin Drive, Baltimore, MD 21218, USA\\
$^{7}$ CONACyT$-$Instituto Nacional de Astrof\'isica, \'Optica y Electr\'onica, Luis E. Erro 1, Tonantzintla, Puebla, M\'exico 	
}
\date{Accepted 2020 September 11. Received 2020 August 28; in original form 2020 April 27}
\pagerange{\pageref{firstpage}--\pageref{lastpage}}
\pubyear{2020}
\begin{document}
\maketitle
\label{firstpage}

\begin{abstract}
Multi-wavelength monitoring of \mrk335 with \swift\ between 2007-2019 are used to construct annual spectral energy distributions and track year-to-year changes. Non-contemporaneous archival data prior to 2007 are used to build a bright state SED. In this work, the changes are examined and quantified to build the foundation for future SED modelling. The yearly SEDs trace a downward trend on the average, with the X-ray portion varying significantly and acquiring further lower values in the past two years when compared to the optical/UV portion of SED. The bolometric Eddington ratios derived using optical/UV to X-ray SEDs and the calculated X-ray luminosities show a gradual decrease over the monitoring period. Changes in the parameters over time are examined. Principal component analysis suggests that the primary variability is in the X-ray properties of \mrk335. When looking at the broader picture of \mrk335\ and its behaviour, the X-rays, accounting most of the variability in the 13-year data, are possibly driven by physical processes related to the corona or absorption whereas the modest optical-UV variations suggest their origin within the accretion disc. These results are consistent with the previous interpretation of \mrk335\ using the timing analyses on the monitoring data and spectral modelling of deep observations.  

\end{abstract}

\begin{keywords}
galaxies: active -- 
galaxies: nuclei -- 
galaxies: individual: \mrk335\  -- 
X-ray: galaxies 
\end{keywords}


\section{Introduction}
\label{sect:intro}

The narrow-line Seyfert~1 galaxy (NLS1) (e.g. \citealt{Osterbrock1985,Komossa2008,Gallo2018}) \mrk335\ ($z=0.025$) has revealed variations in the flux and spectral shape during recent years that makes it as one of the remarkable sources for the investigation of the nature of corona and the accretion disc. \mrk335\ displayed an intriguing behaviour when it switched from, being a typical bright `normal' AGN over decades, to an abruptly low flux system in 2007 \citep{Grupe+2007}. The historic flux drop which was about one-thirtieth of its typical bright flux state, immediately gathered attention to carefully investigate the source, resulting in the \swift\ monitoring as well as several deep follow-up observations with \xmm, \suzaku, \nustar\ and \hst\ since 2007 (e.g. \citealt{Grupe+2008, Gallo+2013, Gallo+2015, Gallo+2018, Gallo+2019, Longinotti+2008, Longinotti+2013, Longinotti+2019, Wilkins+2015, Wilkins++2015, Parker+2014, Parker+2019}). Various follow-up studies indicate that the source has never reverted to its former bright state. It is usually at one-tenth its previous brightness with occasional episodes of intense flaring by factors of ten \citep{Grupe+2014} and deep flux drops \citep{Grupe+2015}. The source exhibits high variability in the X-rays roughly by a factor of 50, while only a factor of 2-3 variations are seen in the simultaneous UV lightcurves. \mrk335\ is currently in an X-ray weak state \citep{Grupe++2018, Grupe+2018} compared to its UV brightness and continues to be monitored by \swift\ so to understand the cause of the prolonged dim state and the intermittent high amplitude flaring activity. In June 2020, \mrk335\ exhibited rapid brightening after nearly 2-years in relative quiescence \citep{Grupe+2020}. These data are presented in Komossa et al. (2020, submitted) and are not included in this work. 
 
\indent \citet{Gallo+2018} performed the structure function analysis on the elapsed eleven years' \swift\ optical/UV and X-ray data of \mrk335. The objective was to incorporate the structure function technique to characterize the optical/UV emission on the simultaneous but unevenly sampled multi-wavelength data. The study showed that the long dimmed X-ray state and the corresponding optical/UV emission favourably respond to different processes. The X-ray low flux state could be attributed to the physical changes in the corona or absorption, whereas the variability in the optical/UV band is more consistent with the thermal and dynamic timescales associated with the accretion disc.

\indent Also, the extended dim X-ray phase is not accompanied by the corresponding similar substantial variability or dimness in the optical/UV band. The optical/UV emission, on average, remain consistent with the measurements obtained prior to 2007 i.e. bright state (e.g. \citealt{Peterson+1998, Grupe+2012, Komossa+2014}); varying by relatively small amplitude by about a factor of two. This results in the X-ray weak state of \mrk335\ compared to its UV luminosity (e.g., \citealt{Gallo2006}). Several studies have attempted to explain the cause of the X-ray dimming in \mrk335. One explanation could be due to the collapse of the corona to form a possible collimated structure (e.g. \citealt{Gallo+2013, Gallo+2015, Wilkins++2015, Wilkins+2015}) or due to absorption processes and partial obscuration of the corona and the disc (e.g. \citealt{Grupe+2008, Longinotti+2013, Longinotti+2019}). Blurred reflection scenario favour some flux states of \mrk335\ where the reverberation lags are found \citep{Kara+2013}. Partial covering scenario, too, can explain the dimness, but is unable to sef-consistently describe both the low and high-flux broadband emission from the source \citep{Gallo+2015}. Absorption does play a role in \mrk335\ (e.g. \citealt{Longinotti+2013, Longinotti+2019, Parker+2019}) on some level.

\indent The first five and the eleven years of the monitoring campaign are presented by \citet{Grupe+2012} and \citet{Gallo+2018}, respectively (see also \citet{Buisson+2017}). \citet{Gallo+2018} have examined the power spectra and characteristic timescales for the eleven-year period of simultaneous multi-wavelength optical/UV and X-ray data. However, these multi-wavelength data have not been examined for the spectral energy distribution (SED) and the spectral variability analyses. 

\indent In this work we look at the available multi-wavelength data post-2007 of \mrk335\  with the goal of characterising the changes in the SED in the NLS1. The annual SED measurements can give physical insight to the changes in the accretion disc leading to changes in the simultaneous UV emission. In addition, this allows to investigate the physical parameters that drive the variations in the X-rays. Through this approach, we aim to search for correlations amongst its key SED parameters and the multi-wavelength properties over more than a decade, test for comparison of these parameters to that of a representative high state of \mrk335, and thereby, to understand the rudimentary physical processes responsible for the low flux characteristics of \mrk335.

\indent The following section features all the available multi-wavelength observations and the data reduction procedure. Section~\ref{sect:vary} describes the variations in the optical and UV data. X-ray variability of \mrk335\ is discussed in the Section~\ref{sect:xray}. Section~\ref{sect:sed} is split into the time-resolved and flux-resolved methods of SED measurements, respectively. Correlations between all the measured multi-wavelength parameters of \mrk335\ are investigated and the statistical analysis using Principal Component Analysis (PCA) is performed to describe the behaviour of \mrk335\ over the extended 13-year period in Section ~\ref{sect:corsed}. Discussion and conclusions follow in Section~\ref{sect:disc} and ~\ref{sect:conc}, respectively.


\section{Observations and data reduction}
\label{sect:data}
\subsection{X-ray Data Analysis}
In this work, we study all the \swift\ observations starting from 2007 to 2019 year-wise in Table~\ref{swift_log0}. The new Swift observations between 2018 May and 2019 October are listed in Table~\ref{swift_log}. The X-ray and UV measurements for these observations are shown in Table~\ref{swift_uvlog}. Previous observations have been discussed in \citet{Grupe+2012} and \citet{Gallo+2018}. \mrk335\ has been monitored on a regular basis since 2007 when it was discovered by \swift\ to be in an extreme X-ray low state \citep{Grupe+2007}. Due to its sun constraint, \mrk335\ can not be observed by \swift\ between February and May. 

All X-ray observations were obtained by the X-ray Telescope (XRT, \citealt{Burrows+2005}) on board \swift\ and were performed in the photon counting mode (pc, \citealt{Hill+2004}). Data corresponding to each epoch in Table~\ref{swift_log0} were used to generate X-ray spectrum, background spectrum, response and arf files. The processing followed standard methods: 
\begin{enumerate}
\item Data reduction with {\it xrtpipeline} which is a part of the HEASOFT software package.
\item Extraction of source and background spectra and event files in circular regions with radii of 94$^{''}$ and 295$^{''}$, respectively, using XSELECT.
\item Creating auxiliary response files (ARF) for each spectrum using the FTOOL {\it xrtmkarf}.
\item Due to the low number of counts we use a binning of 1 within the  FTOOL {\it grppha}. We also linked the most recent response file  {\it swxpc0to12s6\_20130101v014.rmf} to the spectrum. 
\item All spectra were analyzed by the {\it XSPEC} version 12.9.1 (Arnaud, 1996) using  Cash Statistics \citep{Cash1979}.
\item Count rates were determined using the online \swift\ XRT product tool at the \swift\ data center in Leicester \footnote{https://www.swift.ac.uk/user\_objects/}.
\item Hardness ratios were defined as $HR = \frac{hard - soft}{hard + soft}$ where {\it soft} and {\it hard} are the counts in the 0.3-1.0 keV and 1.0-10 keV bands, respectively, and determined by the Bayesian statistics software {\it BEHR} \footnote{http://hea-www.harvard.edu/AstroStat/BEHR/} by \citet{Park+2006}. 
\end{enumerate}

The \swift's UV-Optical Telescope (UVOT, \citealt{Roming+2005}) was performed as follows:

\begin{enumerate}
\item In each filter each observation was co-added using the UVOT task {\it uvotimsum}.
\item Region files had a circular region with radii of 5$^{''}$ and 20$^{''}$ for the source and background data, respectively.	
\item Fluxes and magnitudes were measured using the uvottool {\it uvotsource} which is based on the most recent calibration as described in \citet{Poole+2008} and \citet{Breeveld+2010}.
\item All magnitudes and fluxes used in this publication are corrected for Galactic reddening ($E_{\rm B-V}=0.035$; \citealt{Schlegel+1998}), using equation (2) in \citet{Roming+2009} applying the reddening curves given in \citet{Cardelli+1989}. The correcting magnitudes corresponding to each UVOT filter are shown in Table~\ref{corr_mag}.
\end{enumerate}
\begin{table}
 \centering
 \caption{Correcting magnitudes of UVOT filters.   }
  \begin{tabular}{cc}
  \hline
  \hline
UVOT filter & Correction   \\
\\
\hline 
V-corr &  0.1166 \\ 
B-corr & 0.1519 \\ 
U-corr & 0.1908 \\ 
UVW1-cor & 0.2394 \\ 
UVM2-corr & 0.3413 \\ 
UVW2-corr & 0.2867 \\ 

\hline
\hline
\label{corr_mag}
\end{tabular}
\end{table}

\begin{table}
 \centering
 \caption{Average yearly \swift\ XRT observations of \mrk335. 
 }
  \begin{tabular}{cc}
  \hline
  \hline
Epoch & Time interval   \\
\\
\hline 
Year 0 &  $<$2007 \\ 
Year 1 & 2007-05 to 2008-01 \\ 
Year 2 & 2008-06 to 2009-02 \\ 
Year 3 & 2009-05 to 2010-02 \\ 
Year 4 & 2010-05 to 2011-01 \\ 
Year 5 & 2011-05 to 2012-02 \\ 
Year 6 & 2012-05 to 2013-02 \\ 
Year 7 & 2013-05 to 2014-02 \\ 
Year 8 & 2014-05 to 2015-02 \\ 
Year 9 & 2015-05 to 2016-02 \\ 
Year 10 & 2016-05 to 2017-02 \\ 
Year 11 & 2017-05 to 2018-02 \\ 
Year 12 & 2018-05 to 2019-02 \\ 
Year 13 & 2019-05 to 2019-09 \\   
\hline
\hline
\label{swift_log0}
\end{tabular}
\end{table}
\begin{table*}
 \centering
 \caption{\swift\ XRT and UVOT observations of \mrk335\ since February 2018. All exposure times are given in seconds. Previous observations are listed in \citet{Grupe+2008, Grupe+2012} and \citet{Gallo+2018}. All observations had the target ID 33420. The full list of observations is shown online. The MJD is given in the middle of an observation. 
 }
  \begin{tabular}{ccccrrrrrrr}
  \hline
  \hline 
Segment & MJD$^1$ &  $T_{\rm start}$ (UT) & $T_{\rm end}$ (UT)  & $T_{\rm exp, XRT}$ 
& $T_{\rm exp, V}$ & $T_{\rm exp, B}$ & $T_{\rm exp, U}$ & $T_{\rm exp, W1}$ & $T_{\rm exp, M2}$ & $T_{\rm exp, W2}$ \\
\\
\hline 
151 & 58163.203  & 2018-02-14 04:41 & 2018-02-14 05:01 & 1226 & --- & --- & --- & --- & --- & 1226 \\ 
153  & 58267.401 & 2018-05-29 09:30 & 2018-05-29 09:45 &  869  & --- & --- & --- & --- & --- &  858 \\
154  & 58269.007 & 2018-05-31 00:03 & 2018-05-31 00:18 &  829  & ---  & --- & --- & --- & --- & 963 \\
155 & 58280.691  & 2018-06-11 16:28 & 2018-06-11 16:41 &  727 & ---  & --- & --- & --- & --- & 716 \\
158 & 58293.588  & 2018-06-24 14:02 & 2018-06-24 14:12 &  564 & --- & --- & --- & --- & --- & 557 \\
159 & 58301.417 &  2018-07-02 09:53 & 2018-07-02 10:07 &  812 & --- & --- & --- & --- & --- & 812 \\
160 &  58307.761 &  2018-07-08 17:26 & 2018-07-08 19:05 & 1039 &  82 &  82 &  82 &  165 & 236 & 330 \\
161 &  58309.244 & 2018-07-10 05:43 & 2018-07-10 06:00 & 1026 &   83 &  83 & 83 & 166 & 240 & 333 \\
162 & 58310.038 & 2018-07-11 09:01 & 2018-07-11 09:15 &   829 &   83 &  83 &  83 & 167 &  45 & 334 \\
163 & 58311.107 & 2018-07-12 02:26 & 2018-07-12 02:44 & 1044 &   84 &  84 &  84 & 168 & 250 & 336 \\

\hline
\hline
\label{swift_log}
\end{tabular}

\end{table*}

\begin{table*}
 \centering
 \caption{\swift\ XRT and UVOT measurements of \mrk335\ since February 2018. The XRT Count rates are given in units of counts s$^{-1}$. The hardeness ratios are defined as $HR = \frac{hard-soft}{hard+soft}$ where soft and hard are the counts in the 0.3-1.0 keV and 1.0-10.0 keV bands, respectively. The reddening corrected UVOT magnitudes are in the Vega system. The full list of observations is shown online.
 }
  \begin{tabular}{ccccccccc}
  \hline
  \hline
 MJD & XRT CR & XRT HR & V & B & U & UV W1 & UV M2 & UV W2 \\
\\
\hline 
58163.203 &   0.055$\pm$0.008  &   -0.04$\pm$0.12   &    ---         &    ---         &    ---         &    ---         &    ---         & 13.51$\pm$0.03  \\ 
58267.401 &   0.044$\pm$ 0.008 &  -0.10$\pm$0.12   &    ---         &    ---         &    ---         &    ---         &    ---         & 13.65$\pm$0.04 \\
58269.007 &   0.049 $\pm$0.008  & -0.18 $\pm$0.14   &    ---         &    ---         &    ---         &    ---         &    ---         &  13.67$\pm$0.04 \\
58280.691 &   0.052 $\pm$0.009  & -0.27$\pm$0.18    &    ---         &    ---         &    ---         &    ---         &    ---         &  13.65$\pm$0.04 \\		
58293.588 &   0.056 $\pm$0.011  & -0.18$\pm$0.20    &    ---         &    ---         &    ---         &    ---         &    ---         &  13.80$\pm$0.04 \\		
58301.417  &  0.052$\pm$0.009  & -0.39$\pm$0.16     &    ---         &    ---         &    ---         &    ---         &    ---         &  13.86$\pm$0.04 \\	
58307.761 &   0.054$\pm$0.008 &  -0.28$\pm$0.14     & 14.62$\pm$0.04  & 14.91$\pm$0.04  & 13.74$\pm$0.04   & 13.68$\pm$0.04  & 13.73$\pm$0.04 &  13.83$\pm$0.04 \\
58309.244  &  0.063$\pm$0.008 &  -0.07$\pm$0.15     & 14.56$\pm$0.04  & 14.89$\pm$0.04  & 13.77$\pm$0.04   & 13.67$\pm$0.04  & 13.67$\pm$0.04  & 13.81$\pm$0.04 \\
58310.038  &  0.043$\pm$0.008 &  -0.13$\pm$0.18     & 14.58$\pm$0.04  & 14.88$\pm$0.04  & 13.72$\pm$0.04   & 13.71$\pm$0.04  & 13.65$\pm$0.04  & 13.82$\pm$0.04 \\
58311.107  &  0.051$\pm$0.008 &  -0.21$\pm$0.16     & 14.55$\pm$0.04  & 14.88$\pm$0.04  & 13.72$\pm$0.04   & 13.68$\pm$0.04  & 13.71$\pm$0.04  & 13.81$\pm$0.04 \\

\hline
\hline
\label{swift_uvlog}
\end{tabular}
\end{table*}

\subsection{Optical Spectroscopy}
Optical spectral data used in this work are listed in Table~\ref{opt_log0}. Data covering the wavelength range 1152-6072 \AA\ were obtained from the HST Faint Object Spectrograph (FOS) and are representative of the bright state prior to 2007. We have 78 spectra obtained with MDM Observatory 1.3m McGraw-Hill telescope on Kitt Peak and the data processing is described in \citet{Grier+2012}. The target was observed multiple times with the 1.22~m telescope of the Asiago Astrophysical Observatory (Italy) using the Boller \& Chivens spectrograph with a 300 mm$^{-1}$ grating. The spectra have a dispersion of 2.6 \AA/px and an instrumental resolution of $\sim$700. The spectra were reduced using standard IRAF tools. The data were first corrected for bias and flat-field, and then calibrated in wavelength and flux. 
\begin{table}
 \centering
 \caption{Summary of optical observations of \mrk335.  
 }
  \begin{tabular}{ccc}
  \hline
  \hline
Observatory & Number of & Date  \\
&   observations& \\
\hline 
HST/FOS & 1  &1994-12-16 \\ 
MDM & 78 & 2010-08-31 to 2010-12-28 \\ 
ASIAGO & 4 & 2014-09-23 to 2018-07-07 \\  

\hline
\hline
\label{opt_log0}
\end{tabular}
\end{table}

\subsection{Ultra-violet Spectroscopy}
Ultra-violet observations of \mrk335\ representing the bright state are from the Far Ultra-violet Spectroscopic Explorer (FUSE) and HST Space Telescope Imaging Spectrograph (STIS). HST cosmic origins spectrograph (COS) data between 2010-2018 are used for the dim state. The detailed analysis of these observations have been done by \citet{Longinotti+2013, Longinotti+2019} and \citet{Parker+2019}. The UV observation log is shown in Table ~\ref{uv_log0}.

\begin{table}
 \centering
 \caption{Ultra-violet observations of \mrk335.  
 }
  \begin{tabular}{cccc}
  \hline
  \hline
Instrument & Grating/Tilt & Date & Wavelength studied \\
\\
\hline 
FUSE & -- &1999 & 912-1186 \AA\\ 
FUSE & -- &2000 & 912-1186 \AA\\
\hline 
HST STIS & E140M  & 2004-07-01 & 1125-1425 \AA\\ 
\hline
HST COS  & G160M & 2009-10-31 & 1421-1810 \AA\\ 
\hline
HST COS  & G160M & 2010-02-08 & 1421-1795 \AA\\
HST COS    & G130M & 2010-02-08 & 1135-1425 \AA\\ 
\hline
HST COS  & G160M & 2016-01-04 & 900-1150 \AA\\
HST COS  & G140L & 2016-01-04/-07 & 900-2150 \AA\\
HST COS  & G130M & 2016-01-04 & 1110-1425 \AA\\ 
\hline
HST COS  & G160M & 2018-07-24 & 1390-1750 \AA\\
HST COS  & G130M & 2018-07-23 & 1110-1390 \AA\\ 
\hline
\hline
\label{uv_log0}
\end{tabular}
\end{table}

\section{Optical and UV variability}
\label{sect:vary}
The available optical and ultra-violet observations of \mrk335\ have been studied in the past (e.g. \citealt{Grier+2012, Longinotti+2013, Longinotti+2019}). In this section, we quantify the average behaviour of flux variations in the optical/UV data of the source over several years. 
\subsection{Optical variability}
\label{sect:vary1}
\begin{figure}
   \centering
   \advance\leftskip-0.cm
   {\scalebox{0.6}{\includegraphics[trim= 0.5cm 0cm 0cm 0cm, clip=true]{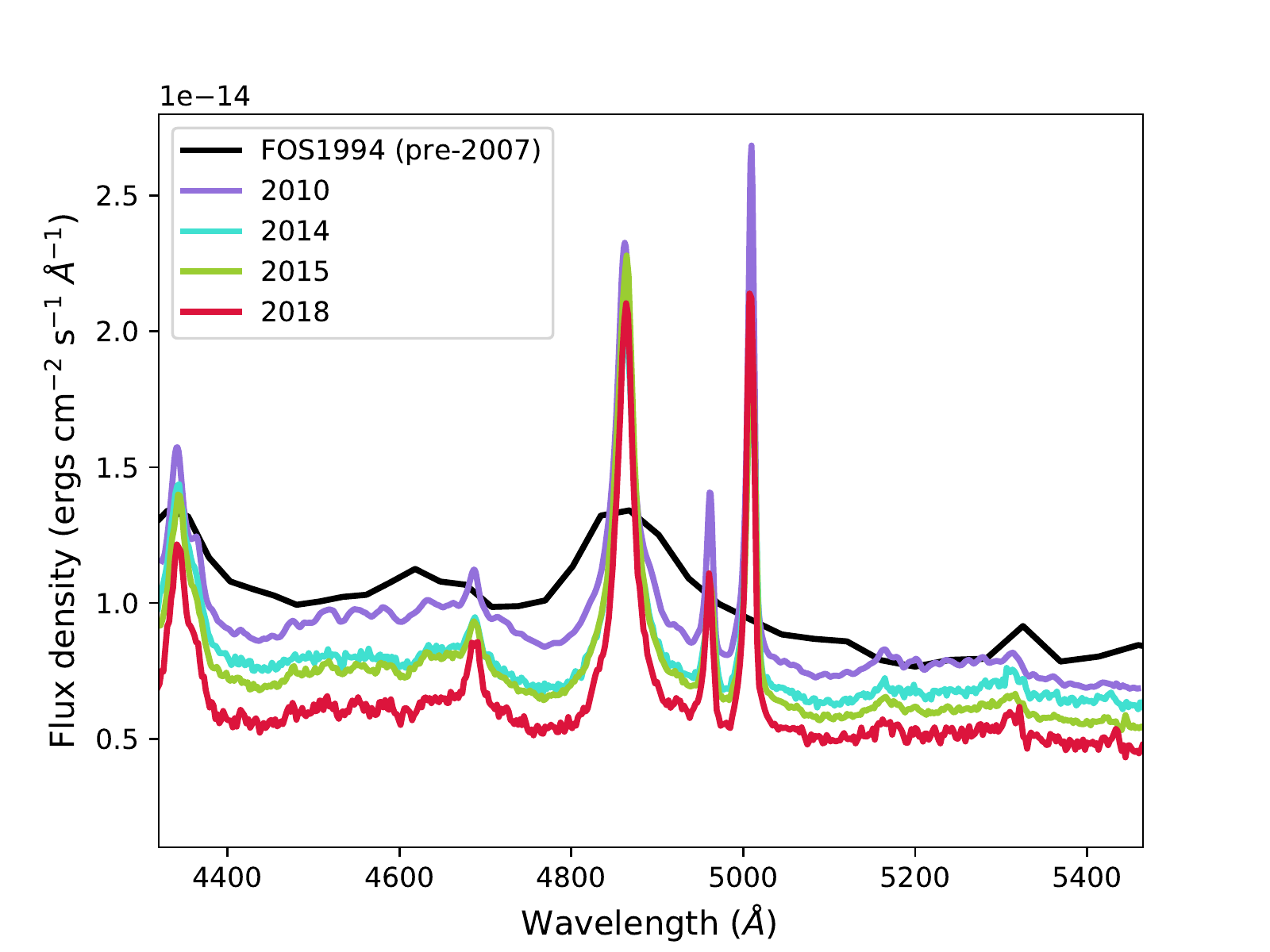}}}      
   \caption{Variations in the H$\beta$ spectral region between 1994-2018 are on the same order as the more rapid changes \citep{Grier+2012}, but show a general downward trend, from 1994 (bright state) to 2018. The 1994 FOS data are of lower resolution compared to the other spectra.}
\label{fig:opt2}
\end{figure}

\begin{figure}
   \centering
   {\scalebox{0.45}{\includegraphics[trim= 1.0cm 0cm 0cm 0cm, clip=true]{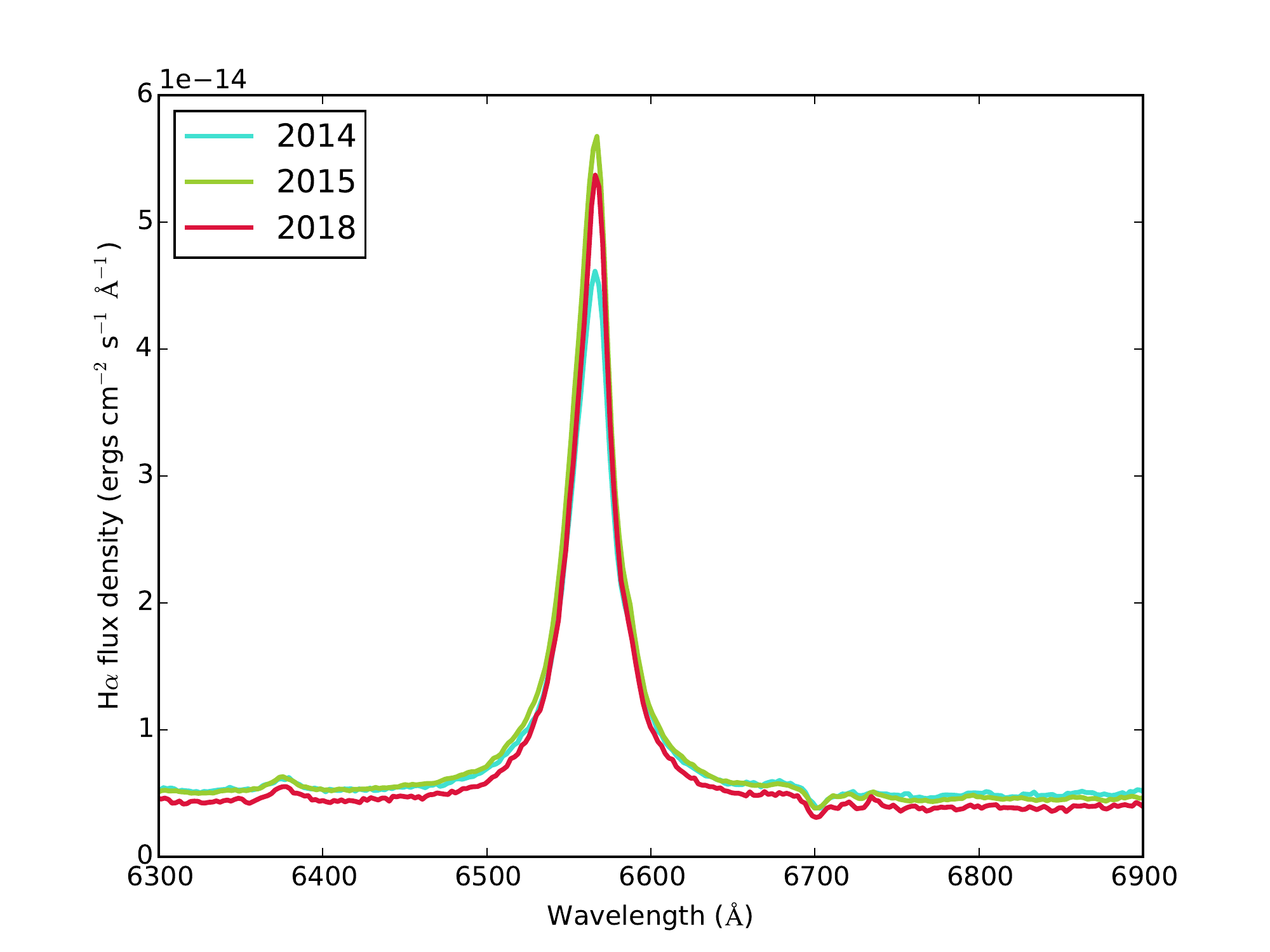}}}
   \caption{Variations in the H$\alpha$ line from 2014-2018 appear to be less significant than in H$\beta$. }
\label{fig:opt4}
\end{figure}

The spectra of 78 observations show a general decrease in the flux with time \citep{Grier+2012}. An average change in the flux density is noted at 20 $\%$ level over a shorter time span of 4 months ($\sim$ 120 days). In Figure~\ref{fig:opt2}, the variations in the flux density for all the optical spectra in the H$\beta$ spectral region are presented which span longer timescales ($\sim$ 20 yr). For these observations, we notice a general downward trend in the flux consistently, from the bright state in 1994 to the low flux state in 2018. The spectra were also plotted in the H$\alpha$ spectral region for the dim 4-year period (2014-18) which suggest modest variations in the flux as shown in Figure~\ref{fig:opt4}. 

\subsection{UV spectral variability}
\label{sect:vary2}

The ultra-violet spectra in the Ly$\alpha$ region ($\sim$ 1190-1225 \AA) from 2004 (bright state) to 2018 span longer timescales ($\sim$ 14-year). Measuring the average change in the flux density within the regions devoid of emission lines, it varies at less than 10 $\%$ level. Similarly, the average flux density in the OVI region ($\sim$ 1015-1050 \AA) and in the CIV region ($\sim$ 1525-1575 \AA) vary within 25$\%$ and 20 $\%$, respectively. The spectra show a general downward trend in brightness over time. It is to be noted that the optical photometric studies of \mrk335\ in 1995-2004 also revealed that the brightness of the source has decreased systematically over those years \citep{Doroshenko+2005}. The detailed investigation of the variations in these line fluxes have been performed by \citet{Longinotti+2013, Longinotti+2019} and \citet{Parker+2019}.

\subsection{Measured optical-UV parameters}
\label{sect:vary3}
We also measured optical and UV spectral parameters with the help of available spectroscopic and photometric data. 

\indent The UV spectral index $\alpha_{uv}$ is measured from the powerlaw fitting of photometric data with six \swift\ UVOT filters. The typical value of $\alpha_{uv}$ from our measurements is $\sim$ -0.4, which is comparable to the index value -0.44 found for the SDSS spectra \citep{VandenBerk+2001}. 

\indent Flux densities at 2500~\AA\ ($F_{uv}$) were obtained with the UVW1 band centred at $\sim$~2600~\AA. Flux densities at 5100~\AA\ ($F_{opt}$) were obtained both using spectroscopic and photometric data. With photometry, $F_{opt}$ was obtained using the V ($\sim$~5468~\AA) or B ($\sim$~4392~\AA) filters when available, otherwise by extrapolating the powerlaw of $\alpha_{uv}$. 

\indent The root-mean-squared fractional variability for the UVW2 (1928~\AA) light curve (F$_{\rm{varUV}}$) during all epochs of the dim 13-year period were computed using the equation of F$_{\rm{var}}$ described in \citet{Edelson+2002}, that quantifies the intrinsic variability of the lightcurve while accounting for uncertainty on the data points. That is,
\begin{equation}
\label{eqn:fvar}
F_{\rm{var}} = \sqrt{ \frac{S^{2} -
\overline{\sigma_{\rm{err}}^{2}}}{\bar{x}^{2}}}.
\end{equation}
where $S^{2} - \overline{\sigma_{\rm{err}}^2}$ is the excess variance normalised by the mean count rate $\bar{x}$.  $S^2$ is the light curve variance and $\overline{\sigma_{\rm{err}}^{2}}$ is the mean square error due to flux measurements.
This parameter allows to compare the amplitude of any variations present in a given energy band by calculating the standard deviation of counts to the average in that energy band. 

\begin{figure*}
   \centering
   \advance\leftskip-0.cm
      {\scalebox{0.5}{\includegraphics[trim= 0cm 0cm 0cm 0cm, clip=true]{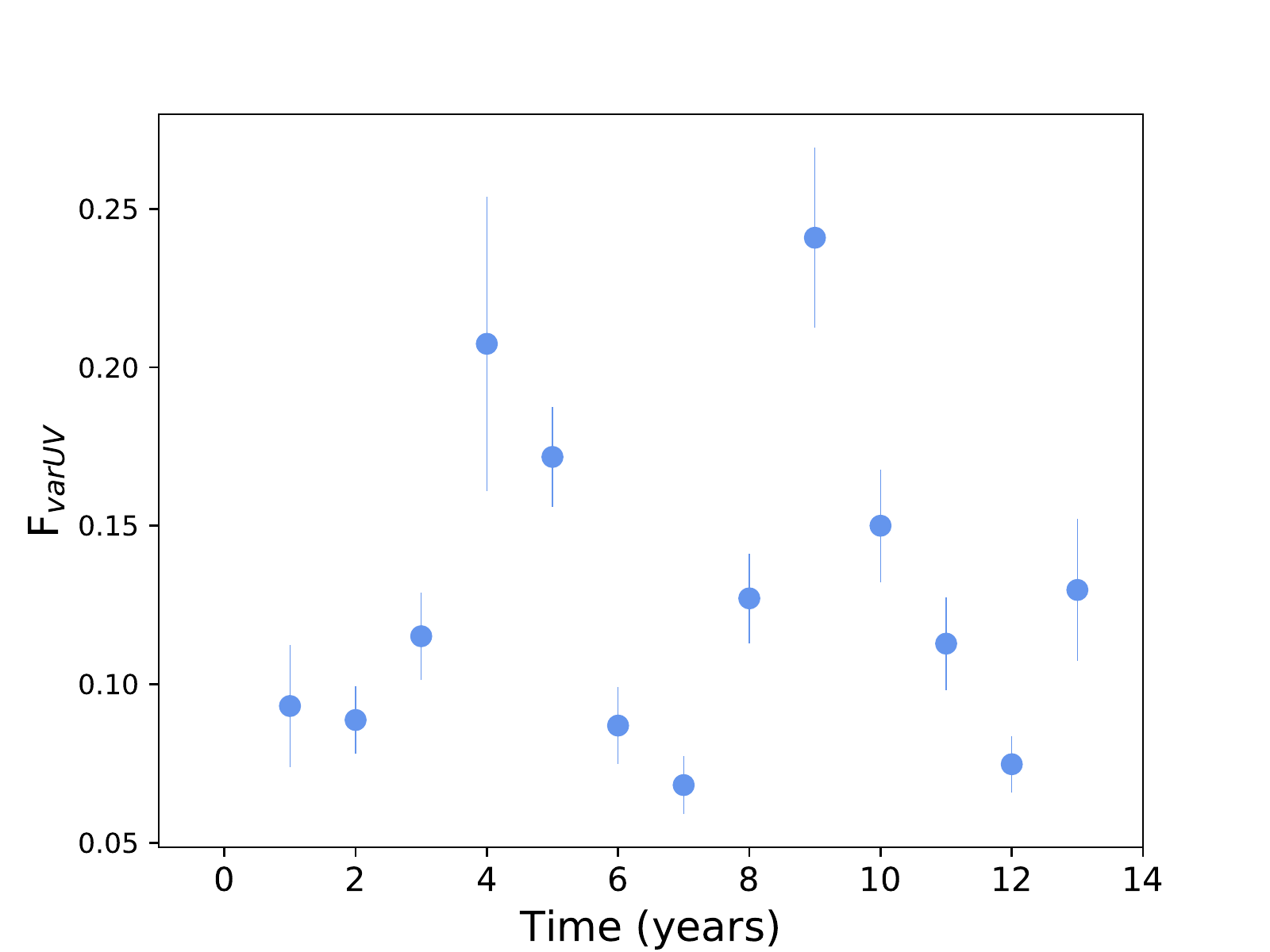}}}  
\hfill
     {\scalebox{0.5}{\includegraphics[trim= 0cm 0cm 0cm 0cm, clip=true]{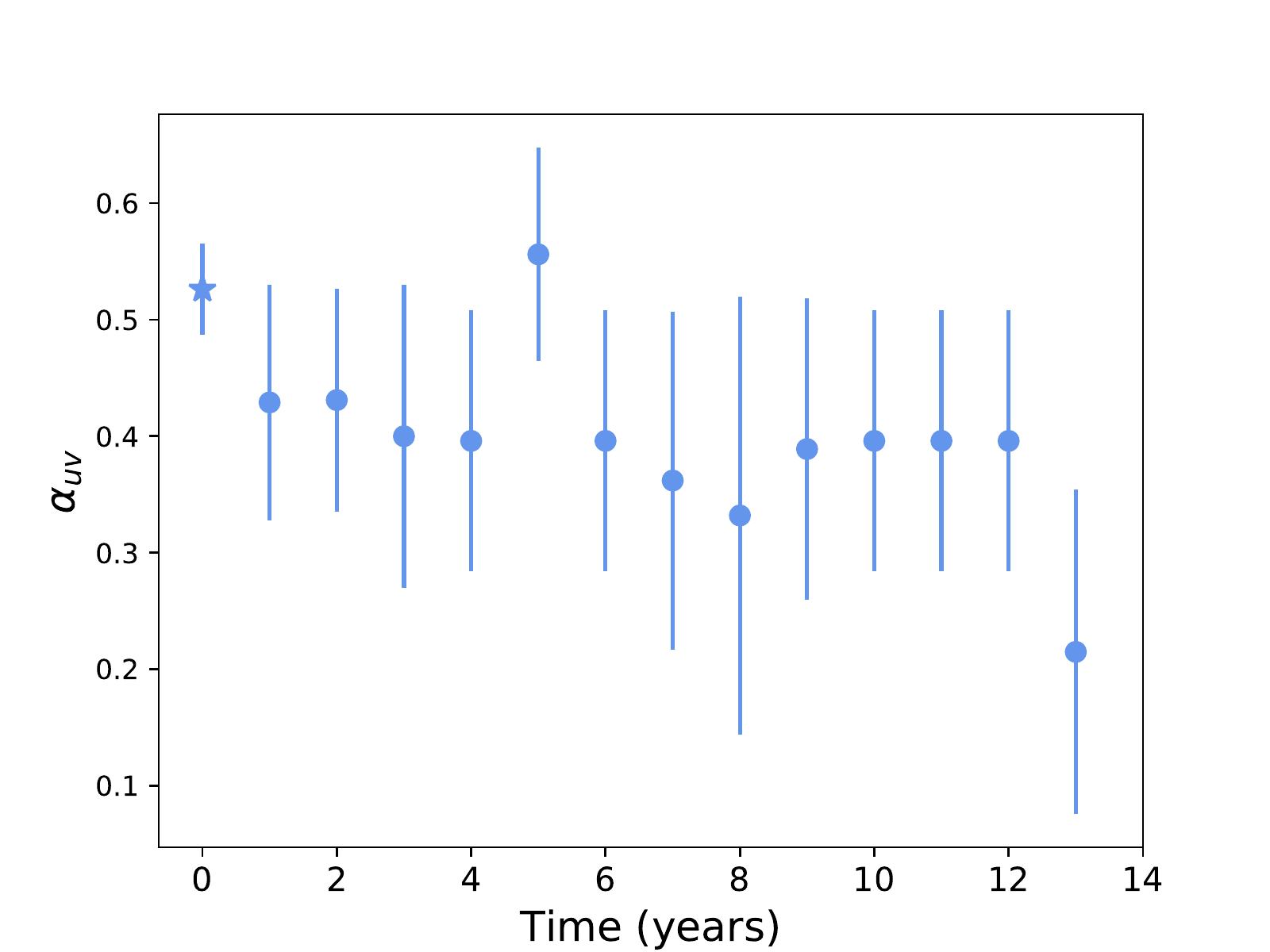}}}  

      {\scalebox{0.5}{\includegraphics[trim= 0cm 0cm 0cm 0cm, clip=true]{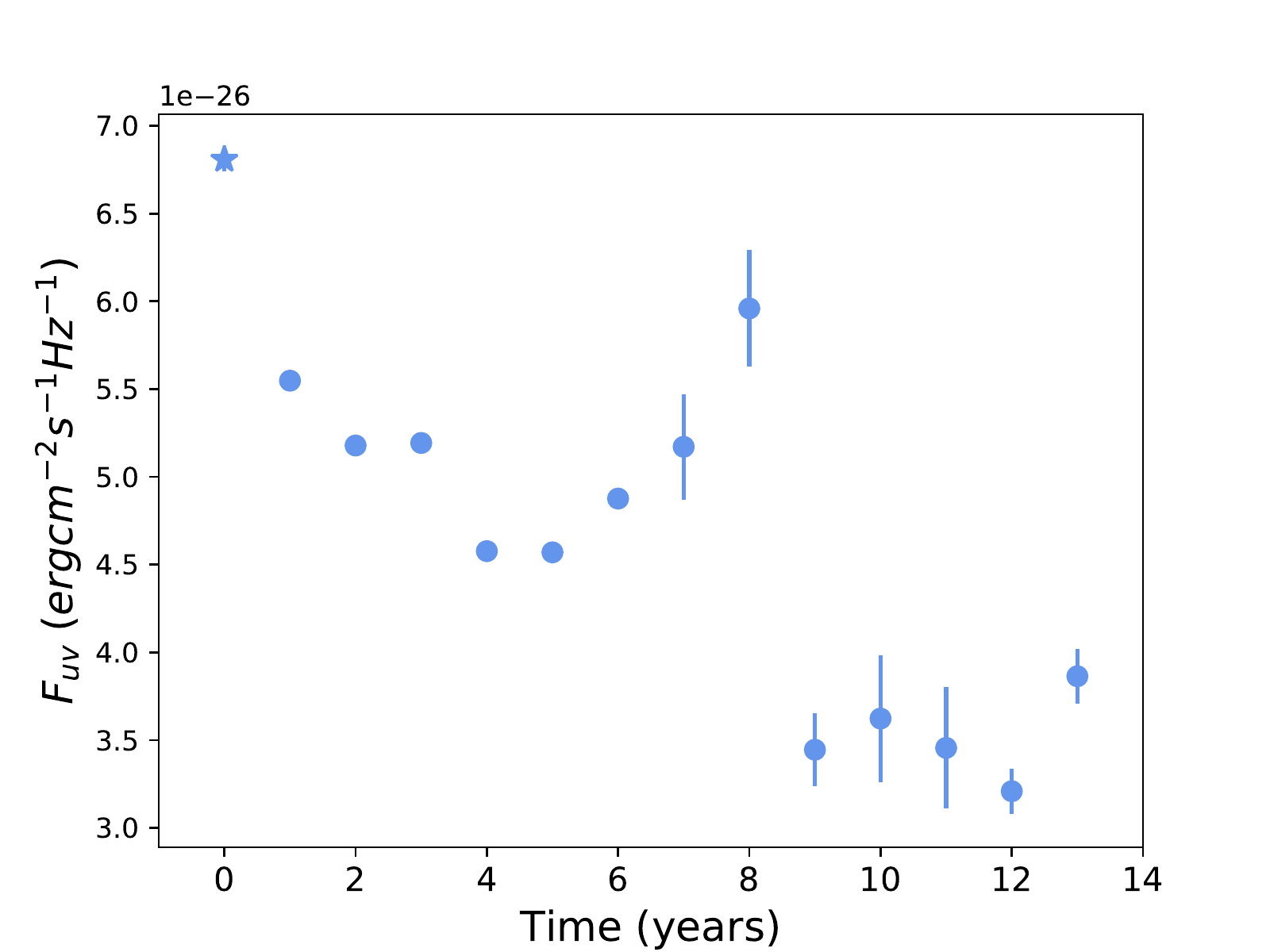}}} 
\hfill 
     {\scalebox{0.4}{\includegraphics[trim= 0cm 0cm 0cm 0cm, clip=true]{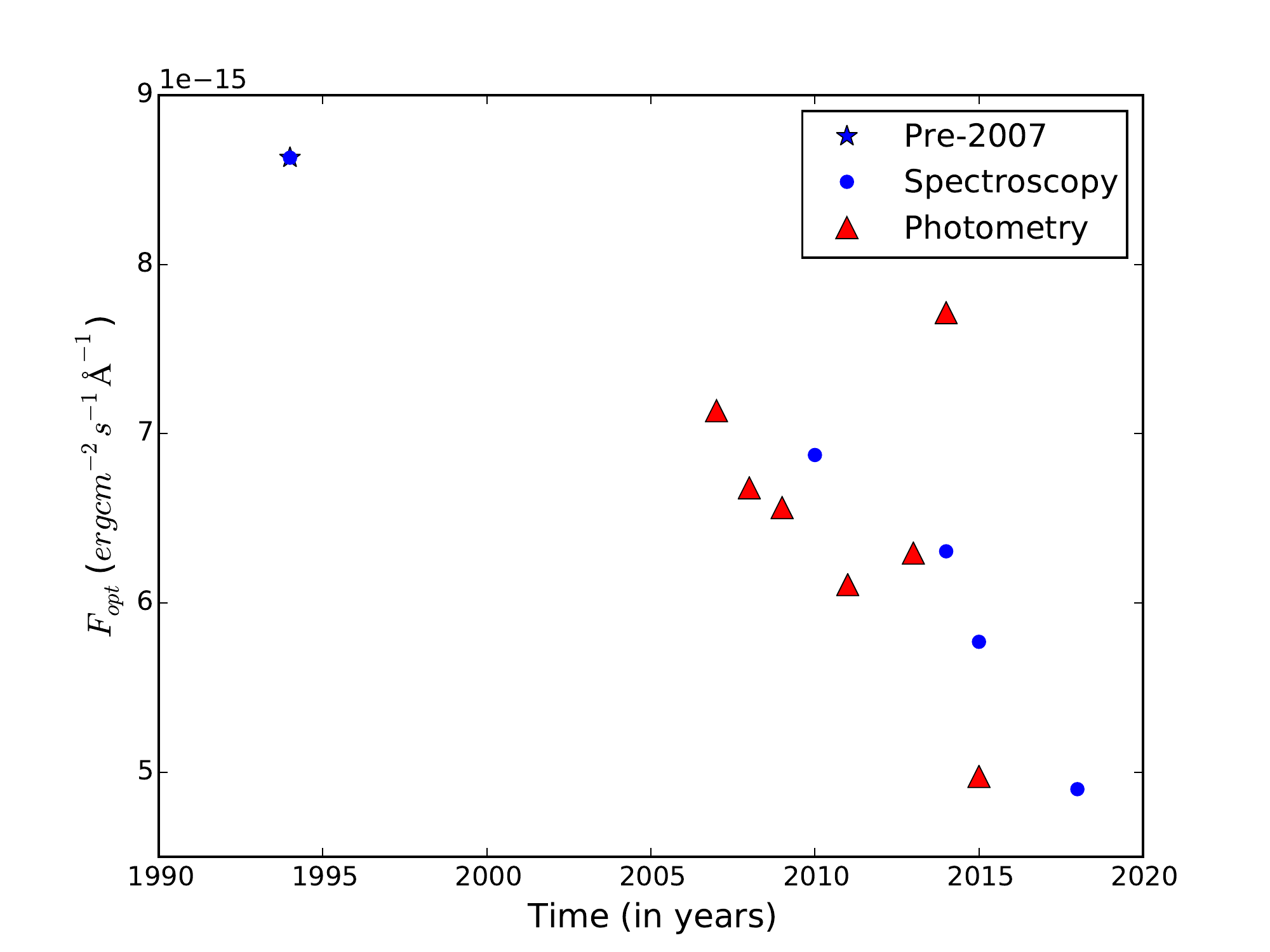}}}  

   \caption{Top left: Variations of F$_{varUV}$; Top right: UV spectral slope $\alpha_{UV}$; Bottom left: Flux density at 2500 \AA\ ($F_{uv}$); Bottom right: Flux density at 5100 \AA\ ($F_{opt}$) both obtained with spectroscopy and photometry over the years. Year 0 represents bright state of \mrk335\ and is represented by a `star' symbol.}
\label{fig:corr3}
\end{figure*}

\indent Table~\ref{tab:tab1} lists the measured UV parameters corresponding to each year. Average variations in these parameters with time are shown in Figure~\ref{fig:corr3}.  The F$_{\rm{varUV}}$ exhibits a range of variability amplitudes.  Years 4, 5, 9, and 10 display large values of $\sim$~15-25 $\%$.  Relatively low values of $\sim$~7 $\%$ are measured in Years 7 and 12.  The remaining years exhibit intermediate values.  The UV spectral slope $\alpha_{uv}$ is statistically constant over the observing period, but there is noticeable flattening from the high state (Year 0) to Year 1 and additional flattening during Year 13 (Figure~\ref{fig:corr3}). $F_{uv}$ and $F_{opt}$ trace a general downward trend with time. Note the significant down step in these values for the recent 2-3 years. 

\indent To show the spread of parameter values within each year as well as year-to-year variations, it is more informative to employ the boxplot diagrams\footnote{Boxplot displays variation in the data without making any assumptions of the underlying statistical distribution} that show the range of possible values for a parameter as well as its median value. Figure~\ref{fig:Box} (bottom panel) represent the boxplots of the flux densities corresponding to the UVW1 and UVW2 filters. These diagrams suggest that \mrk335\ has become fainter over the last decade with respect to median values (marked by horizontal brown lines in the boxes). Year 12 shows smaller median and variance values, however for Year 13, the AGN is getting slightly brighter and more variable.


\section{X-ray Variability: Measured parameters}
\label{sect:xray}
\begin{figure*}
   \centering
   \advance\leftskip-0.cm
   {\scalebox{0.42}{\includegraphics[trim= 0cm 0cm 0cm 0cm, clip=true]{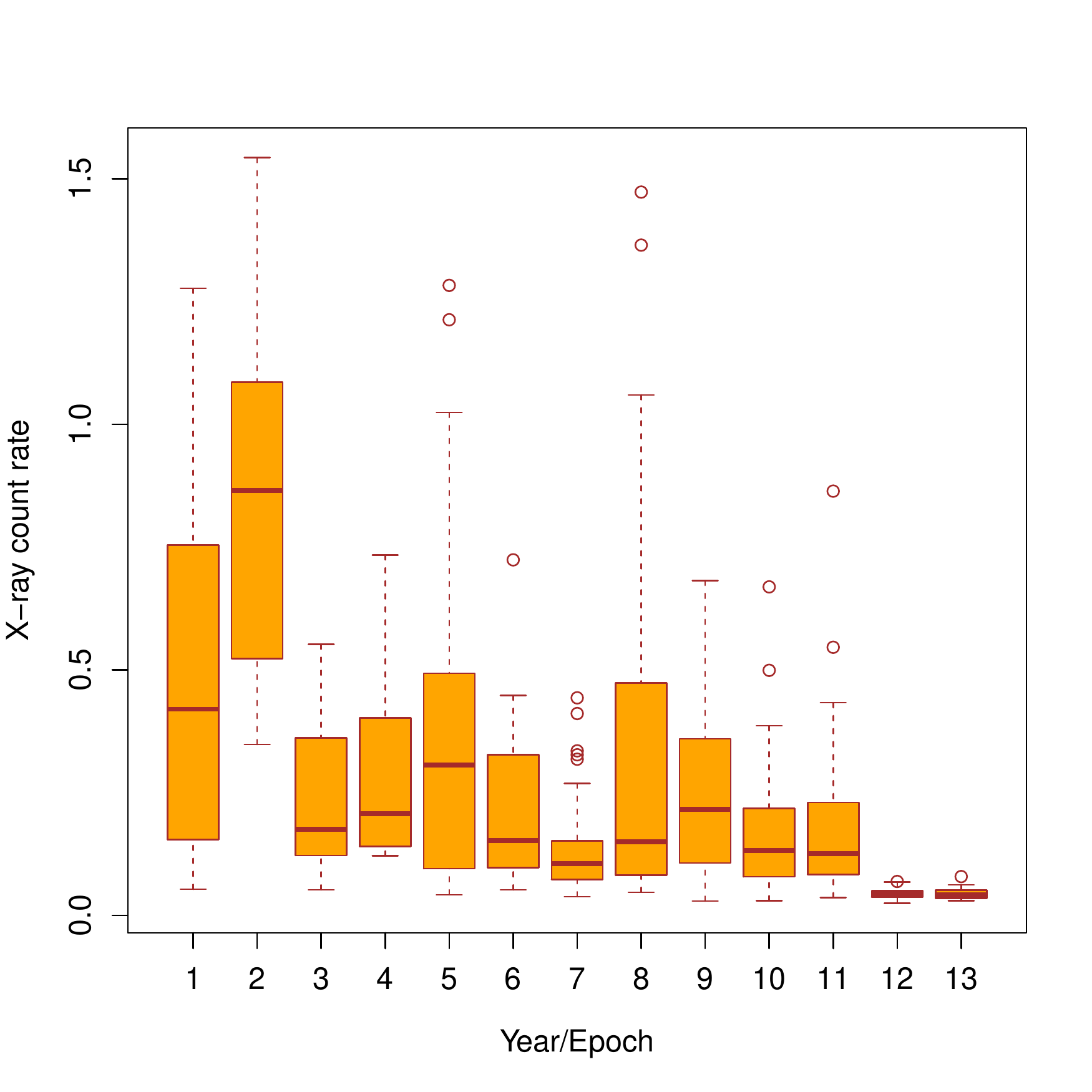}}} 
\hfill     
   {\scalebox{0.42}{\includegraphics[trim= 0cm 0cm 0cm 0cm, clip=true]{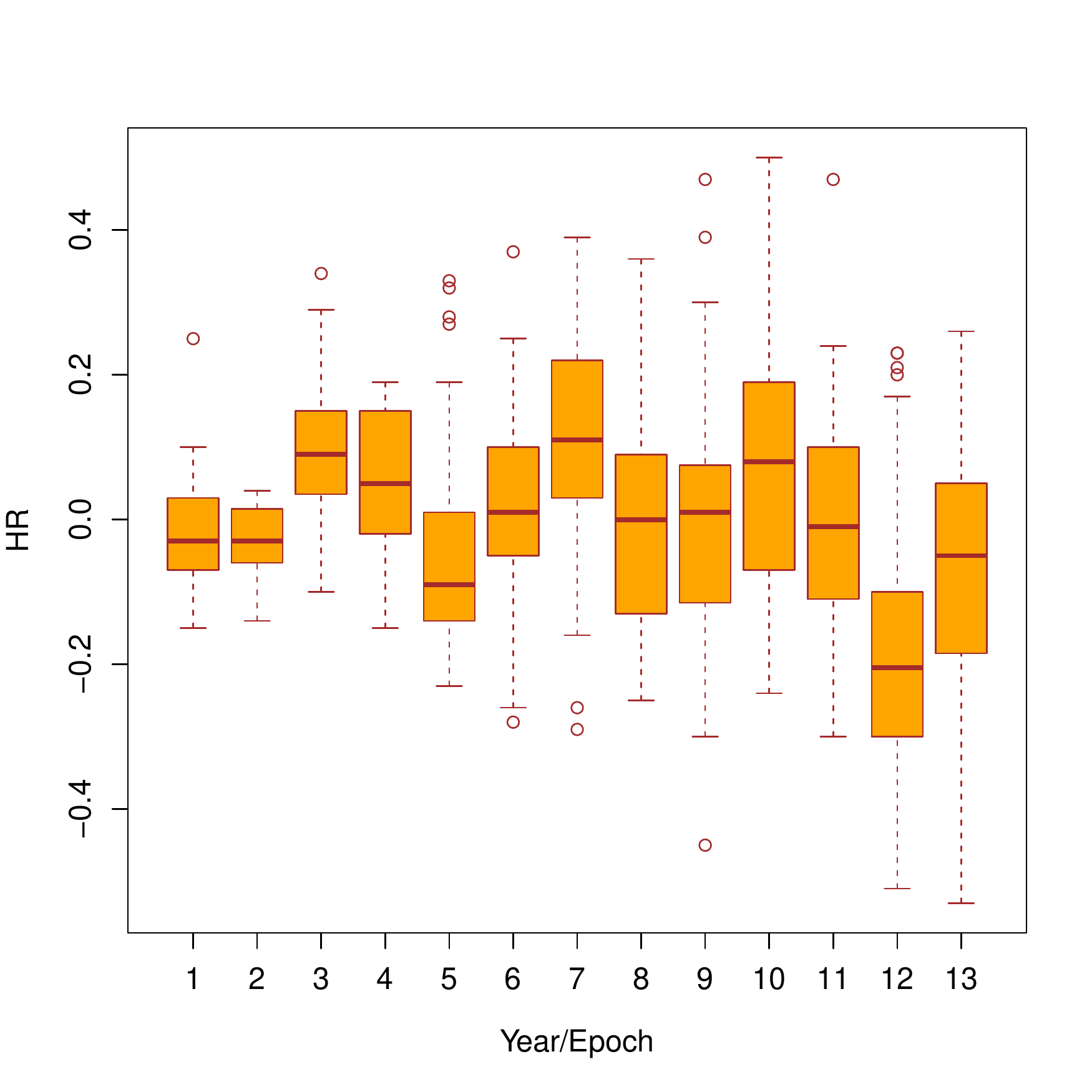}}}
   {\scalebox{0.34}{\includegraphics[trim= 0cm 0cm 0cm 0cm, clip=true]{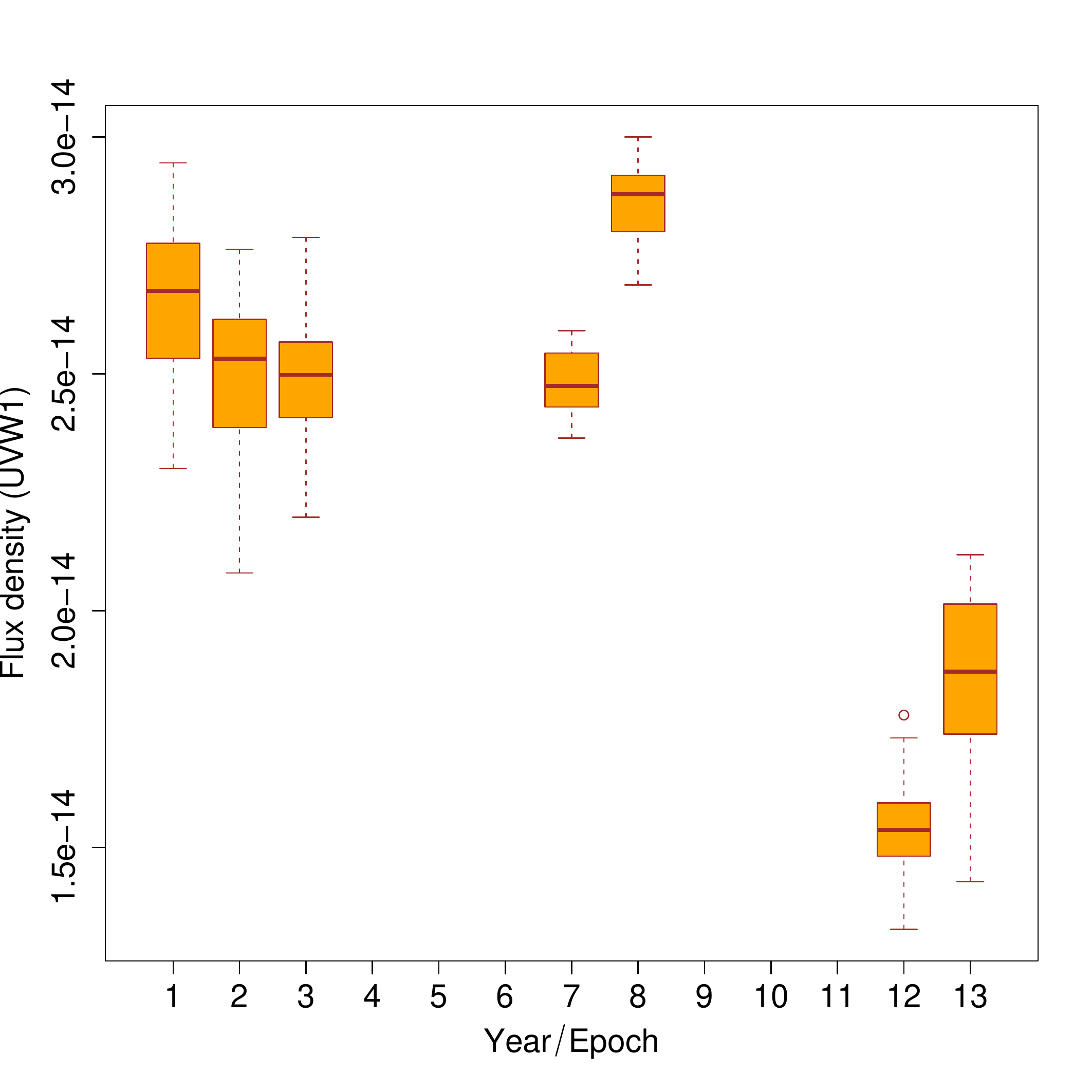}}}
\hfill
   {\scalebox{0.34}{\includegraphics[trim= 0cm 0cm 0cm 0cm, clip=true]{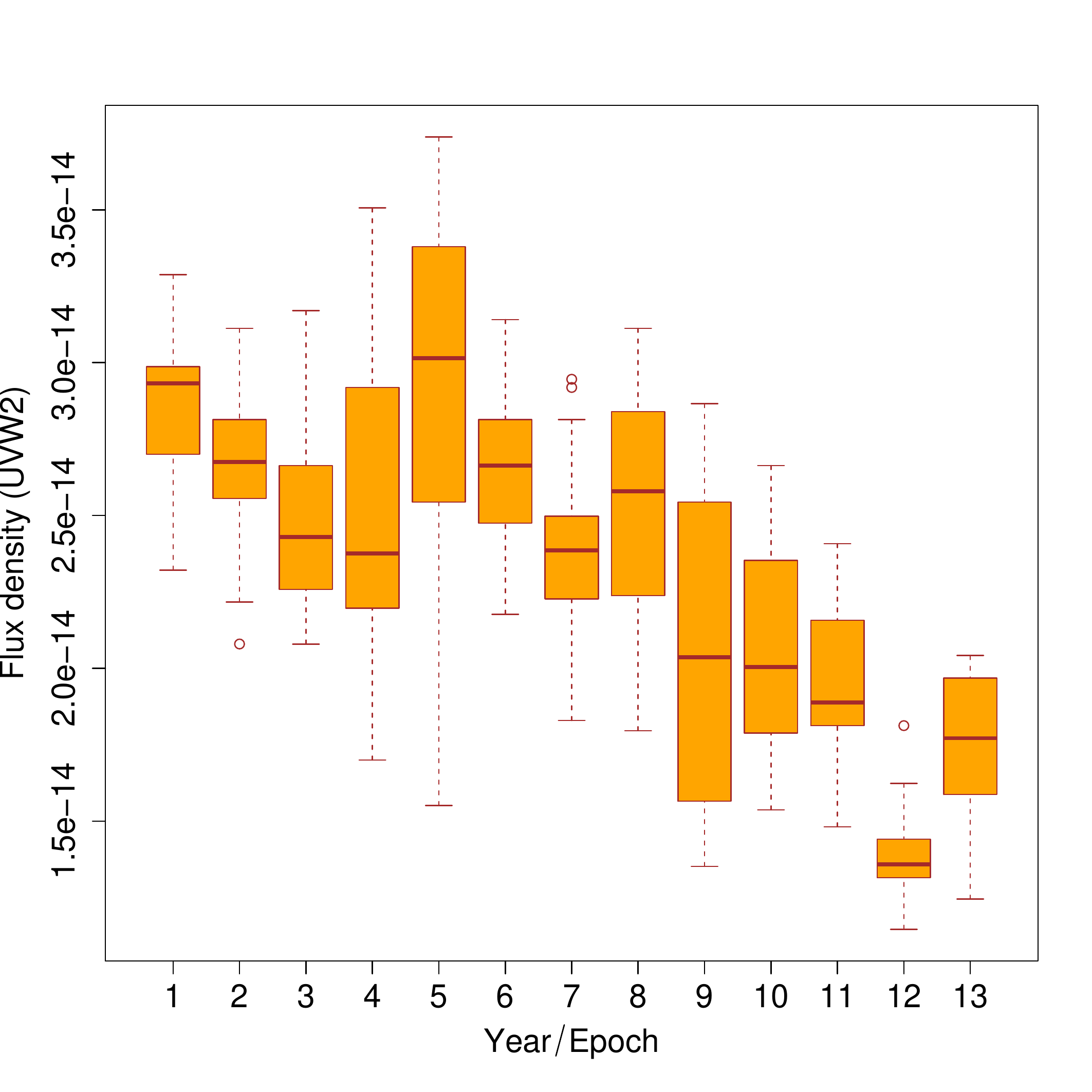}}} 
       
   \caption{Box plots of various parameters to examine the degree of variability during a year. The width of the box describes the variations or spread in the values of any parameter. The brown horizontal line that divides the box into two parts marks the median value. The ends of the box show the upper and lower quartiles. The dashed vertical lines are whiskers that connect to the highest and lowest value of the parameter excluding the outliers shown as open circles. Top panels: Variability in the X-ray count rate and the hardess ratio at each epoch.  Bottom panels: Variations of flux densities (in units of erg~cm$^{-2}$~s$^{-1}$\AA$^{-1}$) in UVW1 ($\sim$~2600 \AA) and UVW2 ($\sim$~1928 \AA) for each epoch. There are no observations in the UVW1 filter for Year 4, Year 6, Year 10 and Year 11. Also, Year 5 and Year 9 are removed due to limited statistics in UVW1.}
\label{fig:Box}
\end{figure*}

\indent  As a first step, we used the 13-year monitoring data (Table~\ref{swift_uvlog}), to show the X-ray variability pattern of count rates (CR) and the hardness (HR) parameter through boxplot diagrams (Top panel: Figure~\ref{fig:Box}). Besides the year-to-year variations, boxplot for the X-ray CR again emphasize that \mrk335\ has fallen inactive over the last two years. HR boxplot conveys spectral variability over years. However, there is a small caution to exercise in the interpretation of the HR boxplot. When we take a look at the last two years, it suggests that \mrk335\ was highly variable in the spectra. However, for these years, the count rate was quite low and the HR was based on fewer counts compared to previous epochs.  

\indent The second step is to perform the spectral analysis to measure the X-ray parameters. The underlying idea is to examine the general shape of the broad-band continuum by simply fitting with a black body and a power law. To fit the spectrum of \mrk335\ properly, we would need to include warm absorbers/emission; blurred reflection and/or partial covering (e.g. \citealt{O'Neill+2007, Longinotti+2007, Parker+2014, Gallo+2015}). To address the general shape of the spectrum, all the \swift\ XRT spectra during the low state between Year 1 to Year 13 are modelled with a powerlaw and a blackbody with absorption fixed to the Galactic value \citep{Willingale+2013} in the energy band 0.3 - 10 keV. This model fits all the XRT spectra sufficiently well. 

\begin{figure}
\minipage{0.48\textwidth}
{\scalebox{0.4}{\includegraphics[trim= 3.3cm 0cm 0cm 0cm, clip=true]{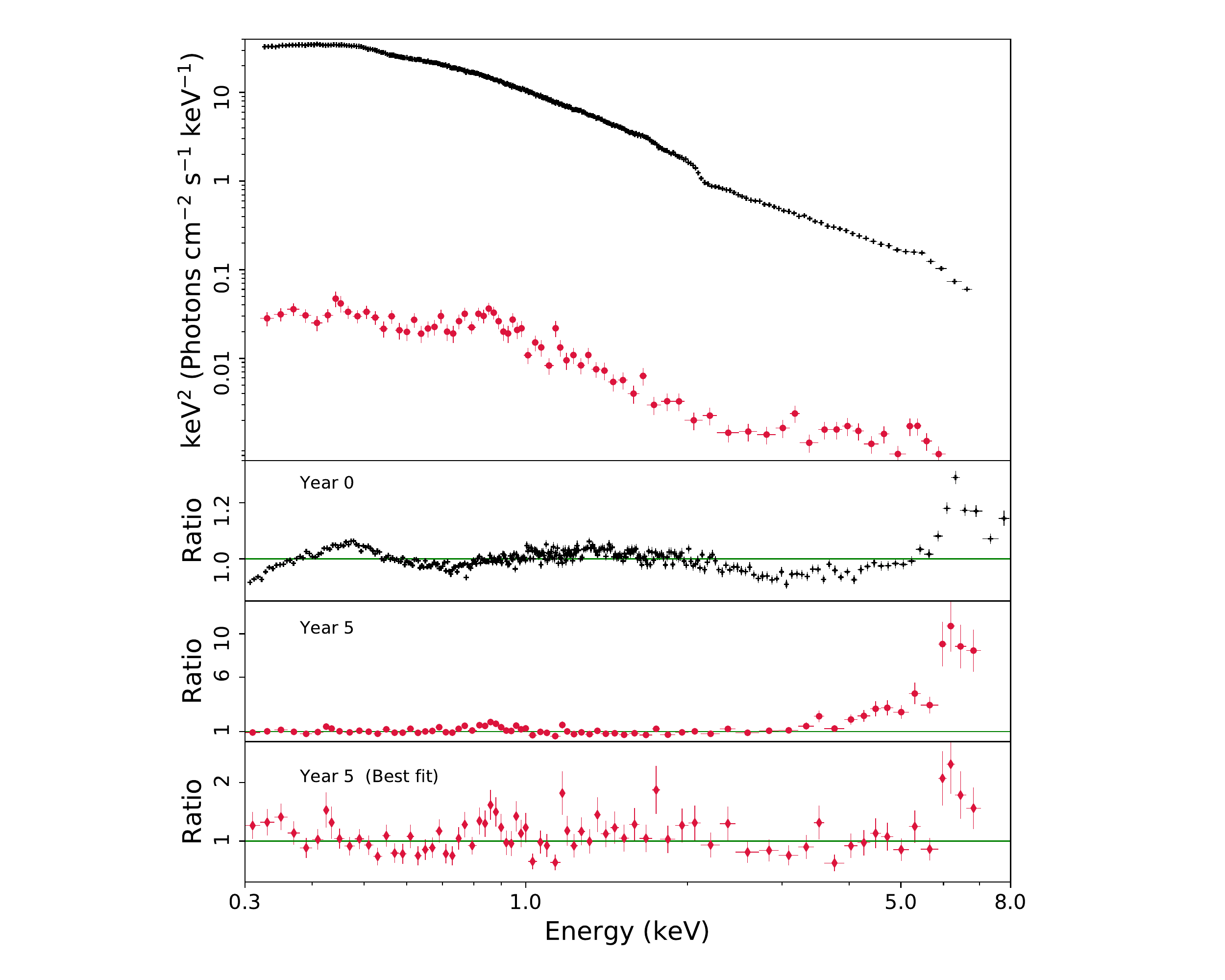}}}  
\vskip -0.2cm
\caption{Upper panel : The average X-ray spectrum of \mrk335 in the pre-2007 bright state (black, crosses) and in one of the low states (Year 5) (red, circles). Spectrum using \xmm\ deep observation in 2006 has been used to represent the pre-2007 bright state. Upper-middle panel shows the ratio plot for pre-2007 bright state by fitting the blackbody and powerlaw (black, crosses), Lower-middle panel shows the ratio plot using the same pre-2007 model but renormalized to Year 5 (red, circles) and the lower panel shows the best-fit ratio plot of Year 5 (red, diamonds).}\label{fig:xray}
\endminipage
\end{figure}

\indent Figure~\ref{fig:xray} shows the average high and low flux spectrum for Year 0 and 5 along with the ratio plots in the bottom panels. A more detailed modelling of these residuals is out of the scope of this paper and would not provide further insights on the physical origin of these features.

\begin{figure*}
   \centering
   \advance\leftskip-0.cm
   {\scalebox{0.42}{\includegraphics[trim= 0cm 0cm 0cm 0cm, clip=true]{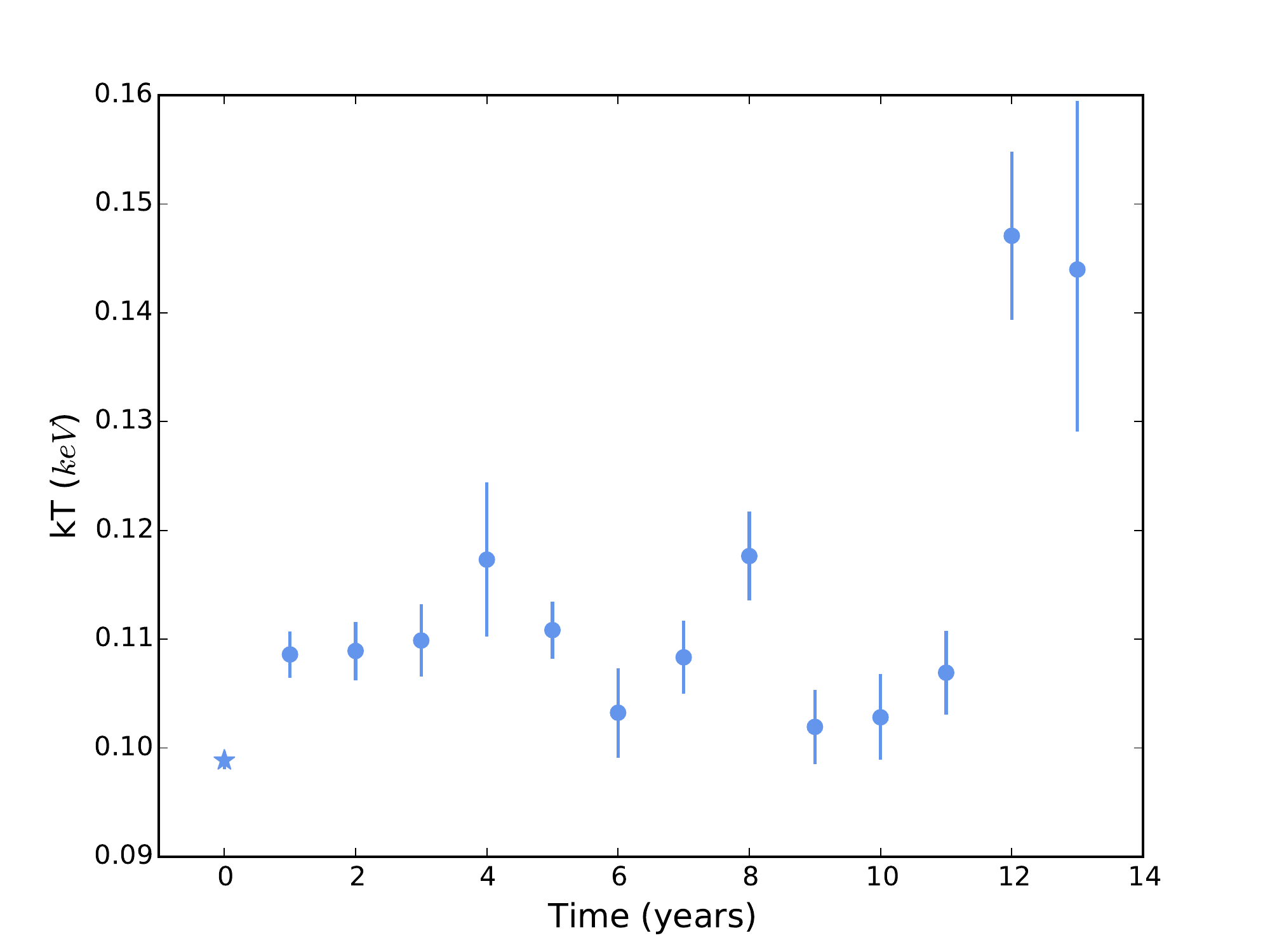}}} 
\hfill     
   {\scalebox{0.42}{\includegraphics[trim= 0cm 0cm 0cm 0cm, clip=true]{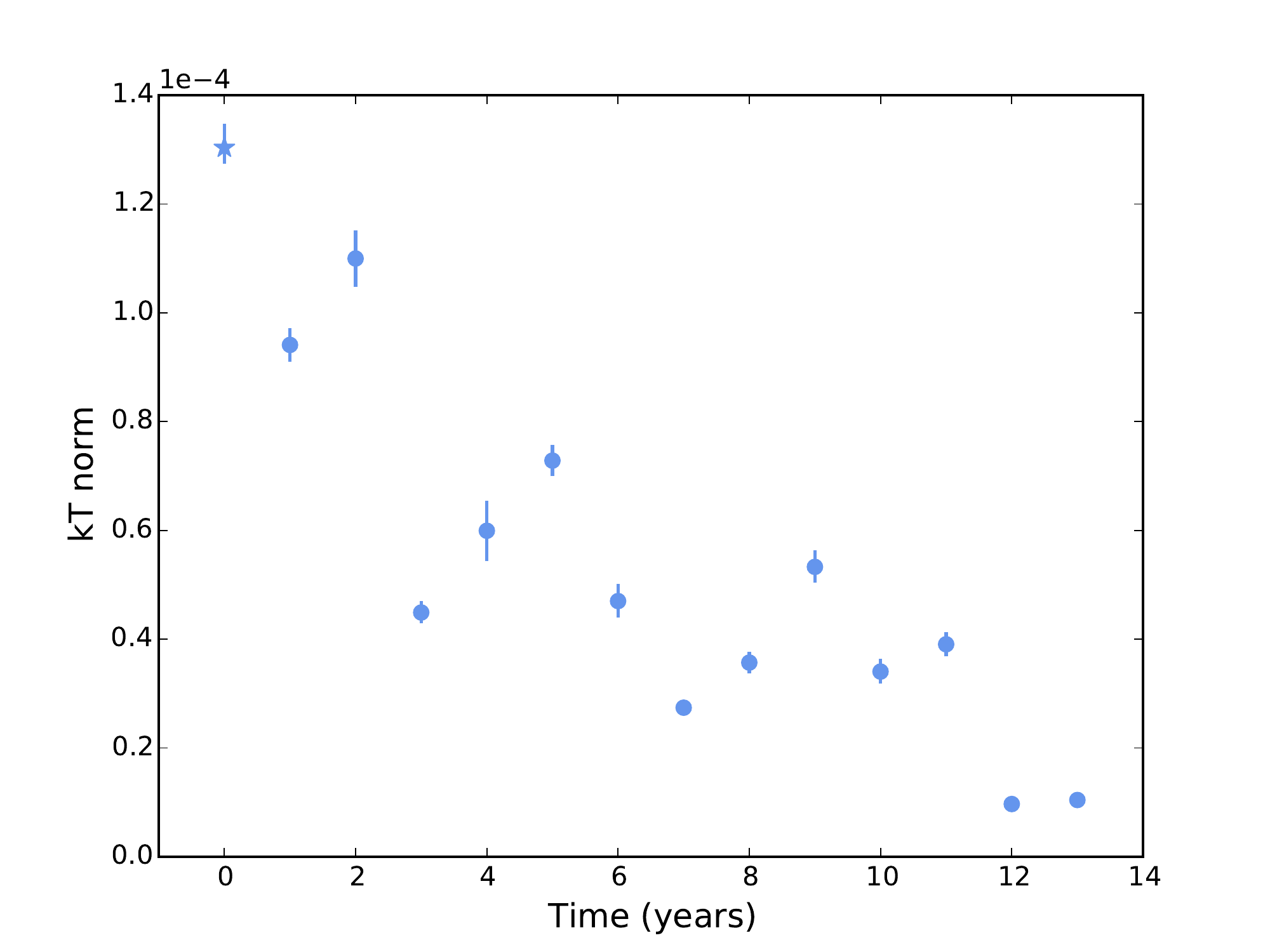}}}
   {\scalebox{0.42}{\includegraphics[trim= 0cm 0cm 0cm 0cm, clip=true]{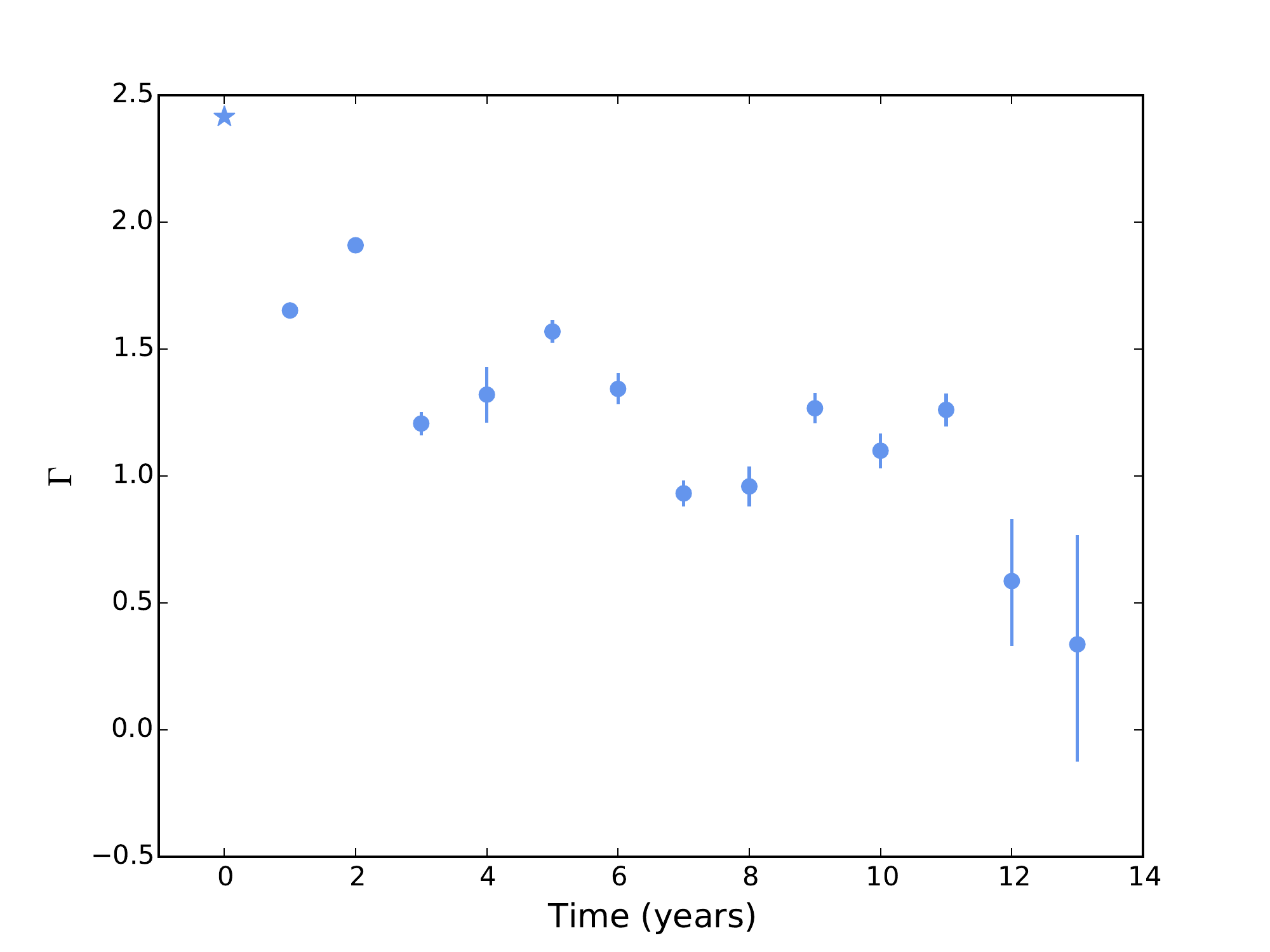}}}
\hfill
   {\scalebox{0.42}{\includegraphics[trim= 0cm 0cm 0cm 0cm, clip=true]{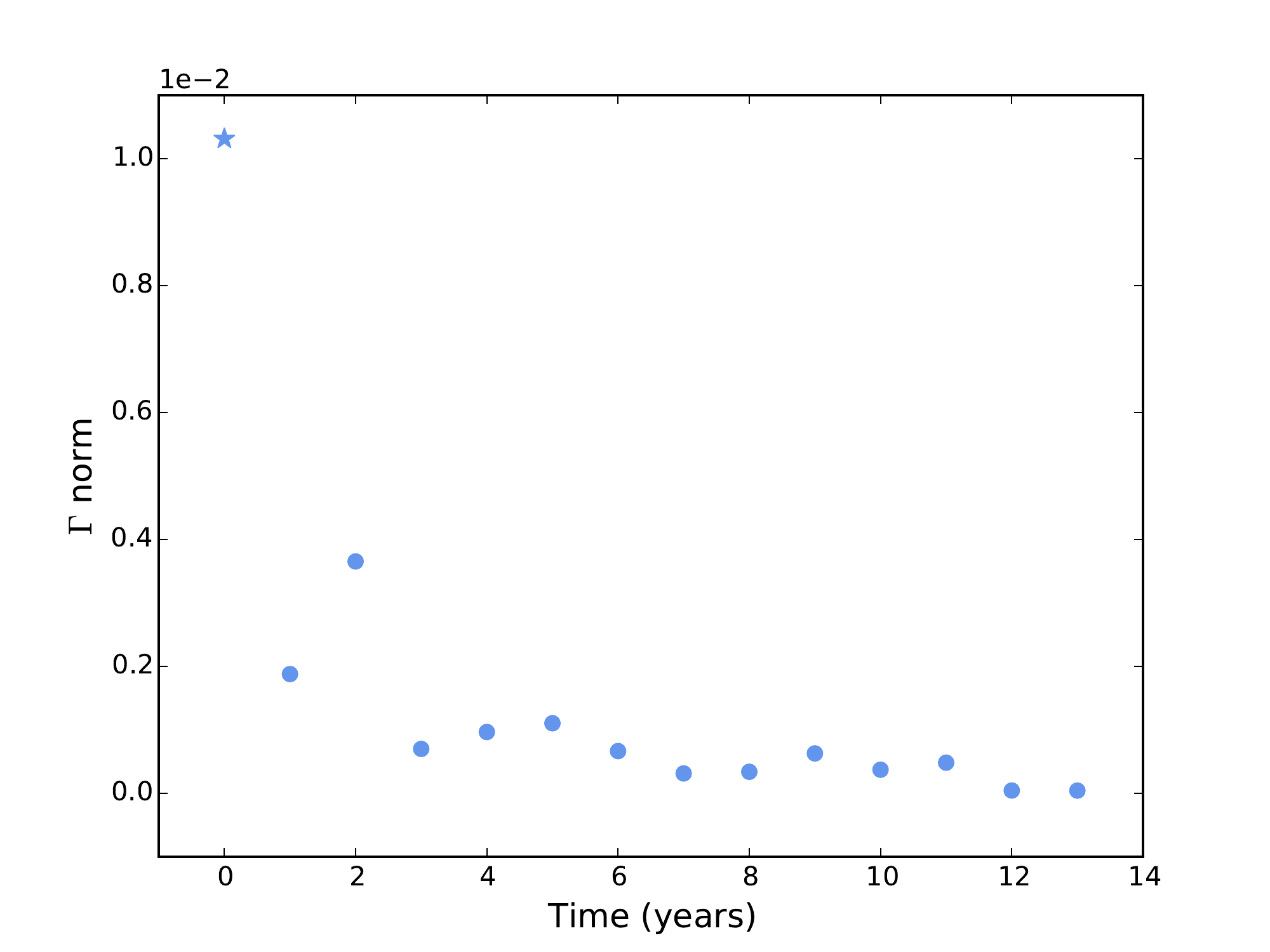}}} 
   {\scalebox{0.42}{\includegraphics[trim= 0cm 0cm 0cm 0cm, clip=true]{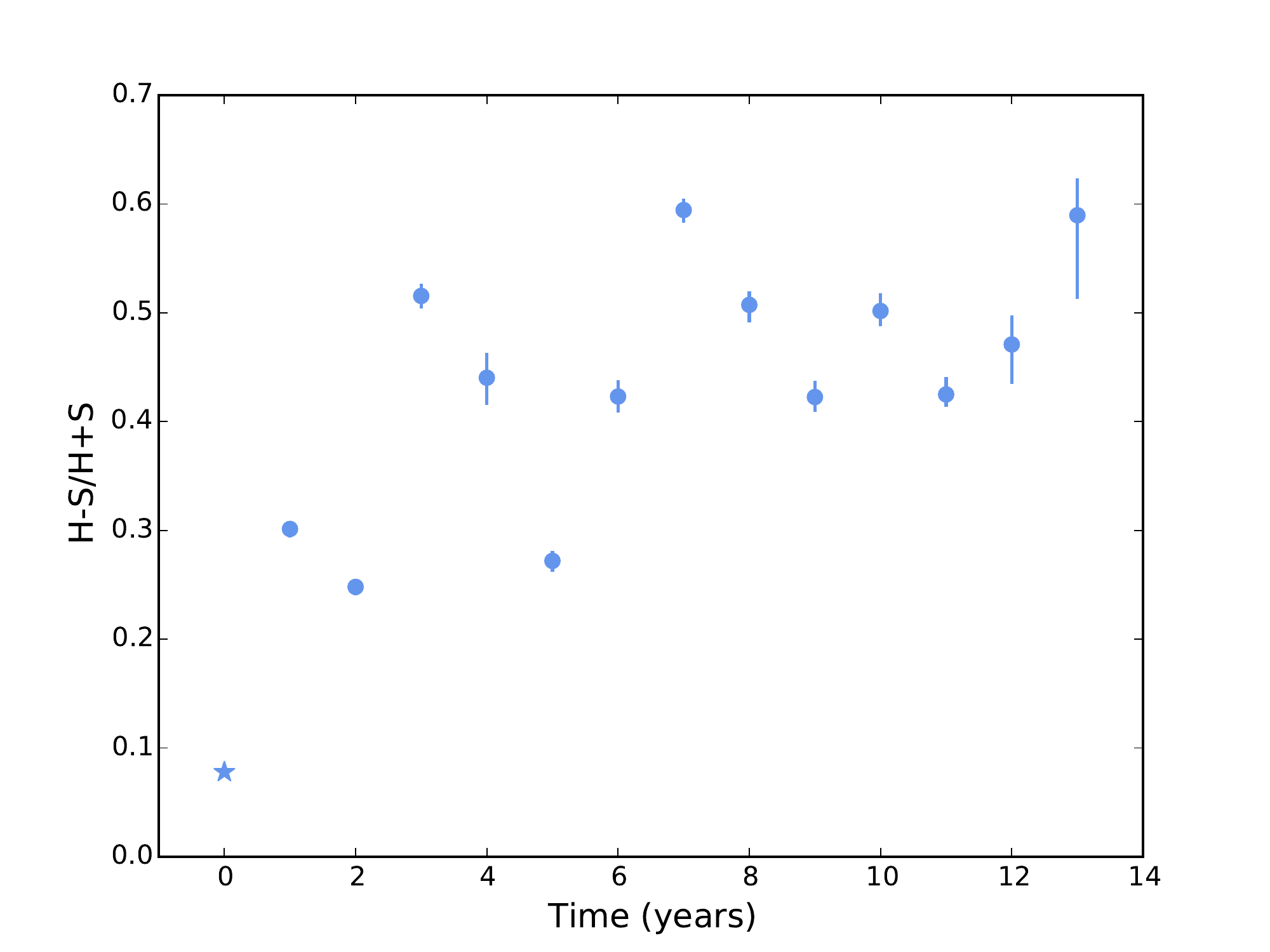}}}
\hfill
   {\scalebox{0.42}{\includegraphics[trim= 0cm 0cm 0cm 0cm, clip=true]{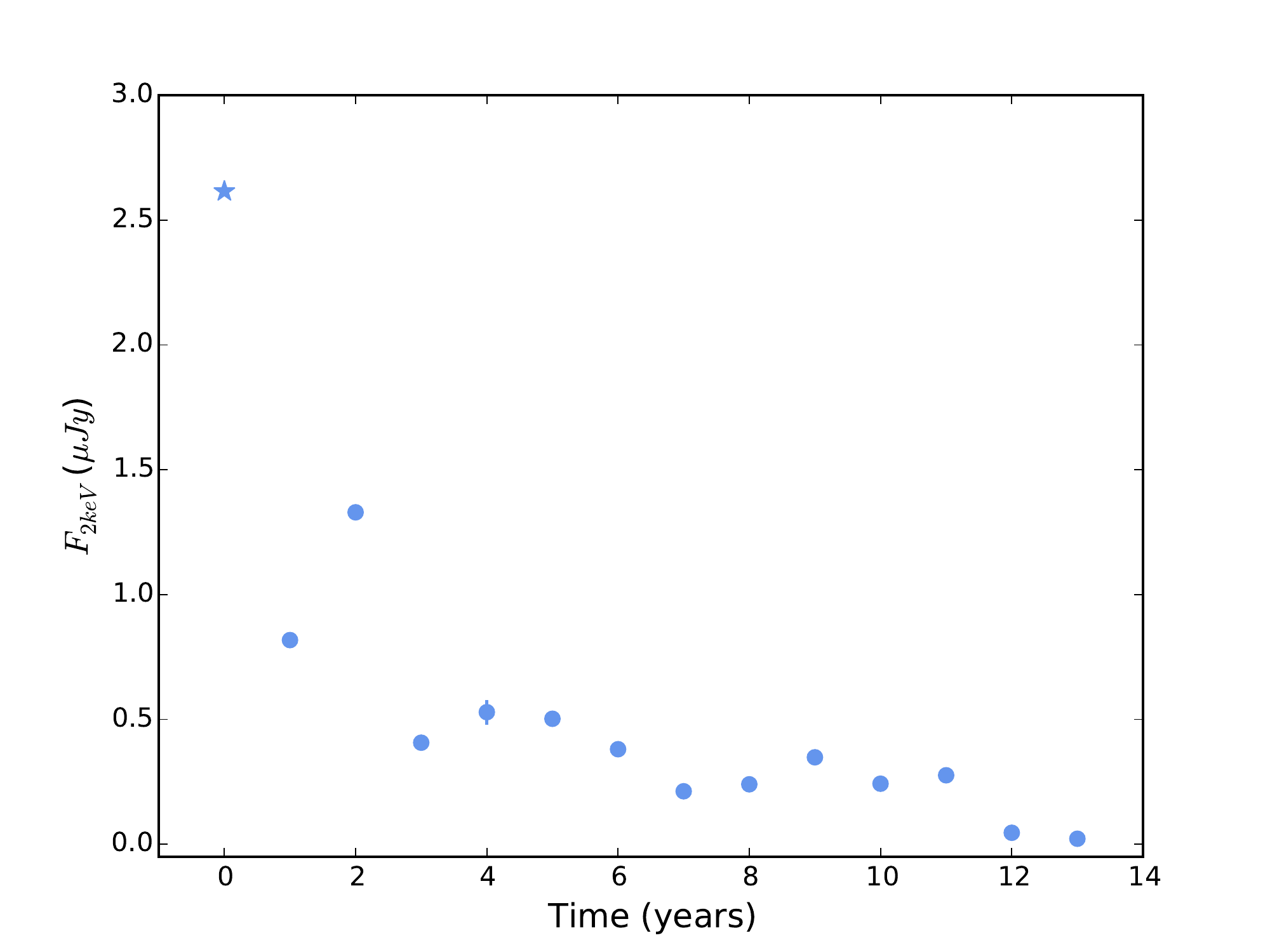}}} 
     
   \caption{Top panel (left to right): Variations of blackbody parameters(kT and kTnorm); Middle panel (left to right): powerlaw parameters ($\Gamma$ and $\Gamma$norm); Bottom panel: Hardness ratio and Flux density at 2 keV with respect to time. Year 0 represents bright state of \mrk335 and is represented by a `star' symbol. kT$_{norm}$ is in units of $(R_{in}/km)/(D/10~kpc)$, where $R_{in}$ is the inner radius and $D$ is the distance and $\Gamma_{norm}$ is in units of photons~keV$^{-1}$~cm$^{-2}$~s$^{-1}$ at 1 keV.}
\label{fig:corr1}
\end{figure*}

\indent From the spectral modeling of the yearly-averaged X-ray observations, we have computed model parameters and the hardness ratios (Table~\ref{tab:tab1}). All the uncertainties presented in the tables and figures correspond to $1\sigma$ error bars. The variability of these measured X-ray parameters as a function of time are shown in Figure~\ref{fig:corr1}. The photon index or the spectral shape changes with time, complying with an overall downward trend with scatter; steeper values for the initial epochs, with subsequent flatter values in the later epochs. The blackbody temperature is significantly higher for the years 12 and 13 when compared to other epochs. Hardness ratios were calculated from the annual spectra using the 0.3-1.0 keV for the soft band and 1.0-10 keV for the hard band, to optimize S/N. For consistency, we have checked that the trend of hardness ratios with time do not change if the soft and hard energy bands are defined as 0.3-2 keV and 3-10 keV respectively. In few instances, year 4 stands out from the other low flux epochs possibly due to changes in the underlying physical processes. 

\indent In addition to measuring the spectral parameters, at each epoch, we also quantify the fractional variability $F_{var}$, soft excess strength and the optical-to-X-ray spectral slope $\alpha_{ox}$.

\begin{figure*}
   \centering
   \advance\leftskip-0.cm
      {\scalebox{0.54}{\includegraphics[trim= 0cm 0cm 0cm 0cm, clip=true]{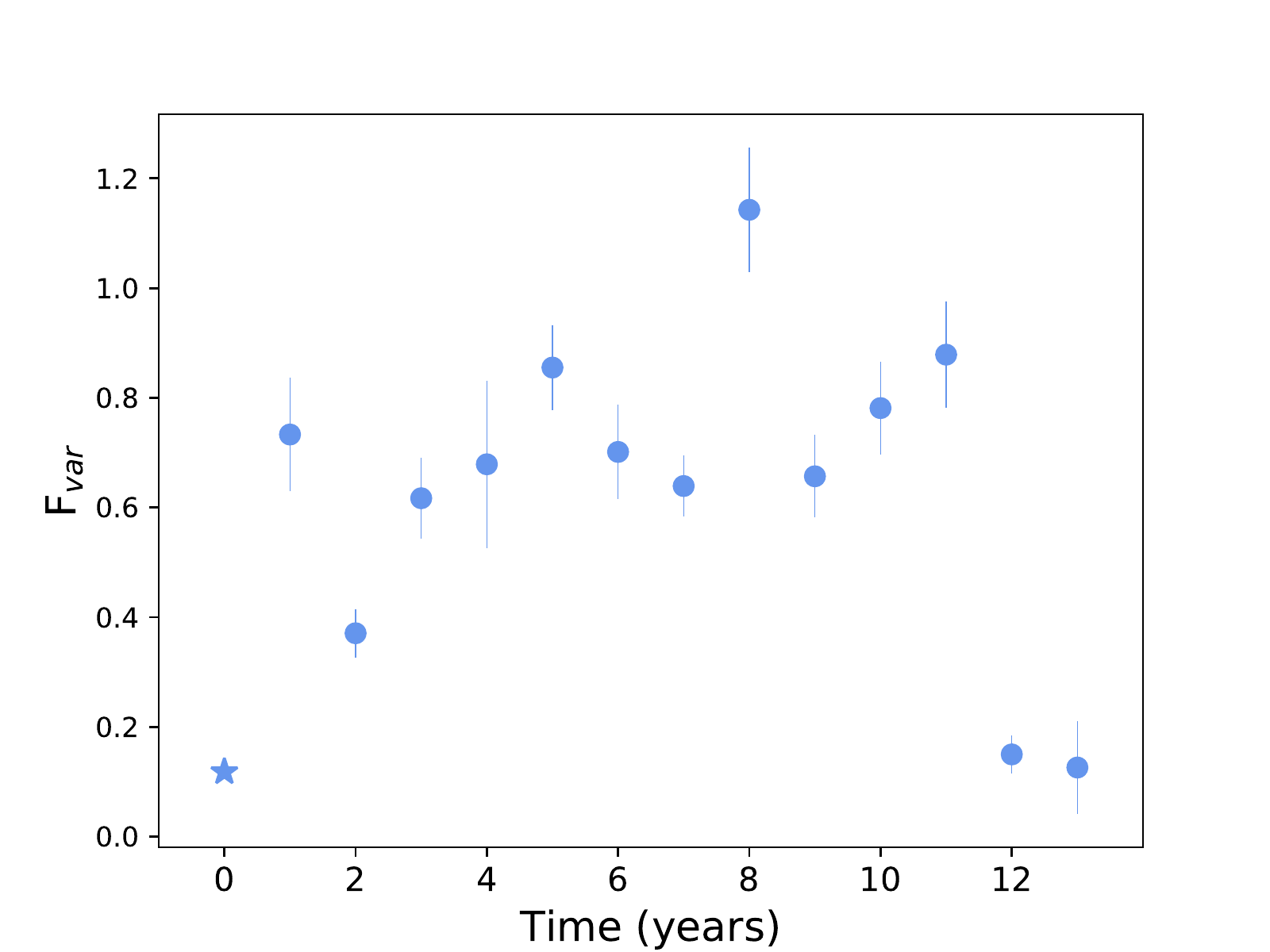}}} 
\hfill 
      {\scalebox{0.42}{\includegraphics[trim= 0cm 0cm 0cm 0cm, clip=true]{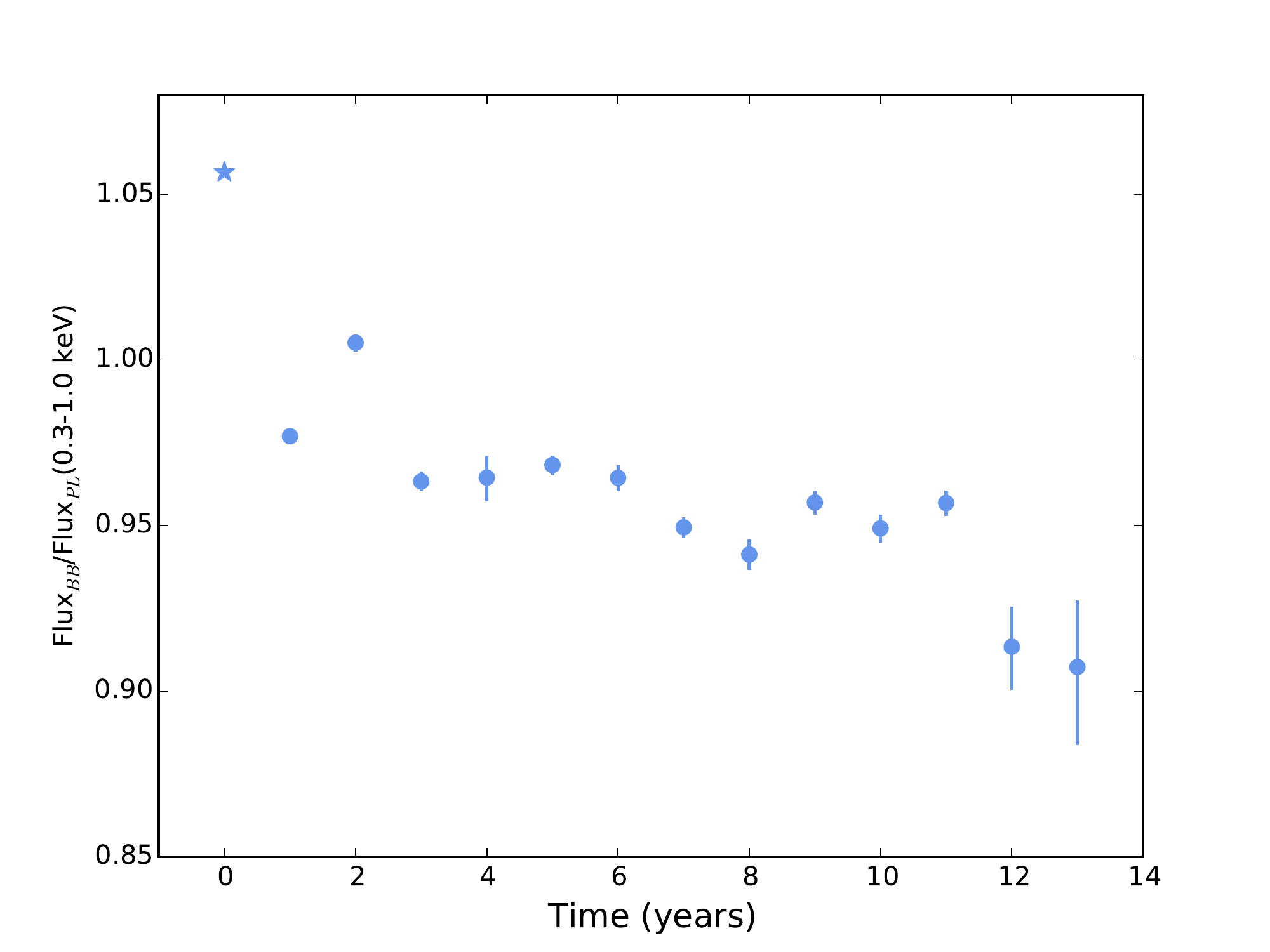}}} 
   
   \caption{Left: Fractional variability over time. Right: Soft excess determined from the ratio of fluxes computed with `cflux' in 0.3-1 keV for the model component blackbody and powerlaw. Year 0 represents high state of \mrk335\ and is represented by a `star' symbol.}
\label{fig:corr2}
\end{figure*}

\indent A fractional variability analysis ($F_{var}$) determining the relative strengths of observed variability across the energy band has been computed for each year. The yearly variation of the $F_{var}$ computed in the 0.3-10 keV is represented (left, Figure~\ref{fig:corr2}). $F_{var}$ values vary within 70 $\%$ for all years except for Year 8 where it records fractional variability of 114 $\%$. Year 8 is marked by the high X-ray flaring event studied in detail through deep observations \citep{Wilkins++2015}, and shows the value of $F_{var}$ significantly higher than the average value. 

\indent Soft excess is determined from the ratio of blackbody and powerlaw fluxes computed using the convolution model `cflux' (within \texttt{XSPEC v12.9.1}) in the energy band 0.3-1.0 keV for all the years. The time variability of soft excess shows a gradual downward trend (right, Figure~\ref{fig:corr2}). This shows that the source is consistently exhibiting lesser contribution from the soft X-ray band (0.3-1.0) over the years.

\begin{figure*}
   \centering
   \advance\leftskip-0.cm
     
{\scalebox{0.54}{\includegraphics[trim= 0cm 0cm 0cm 0cm, clip=true]{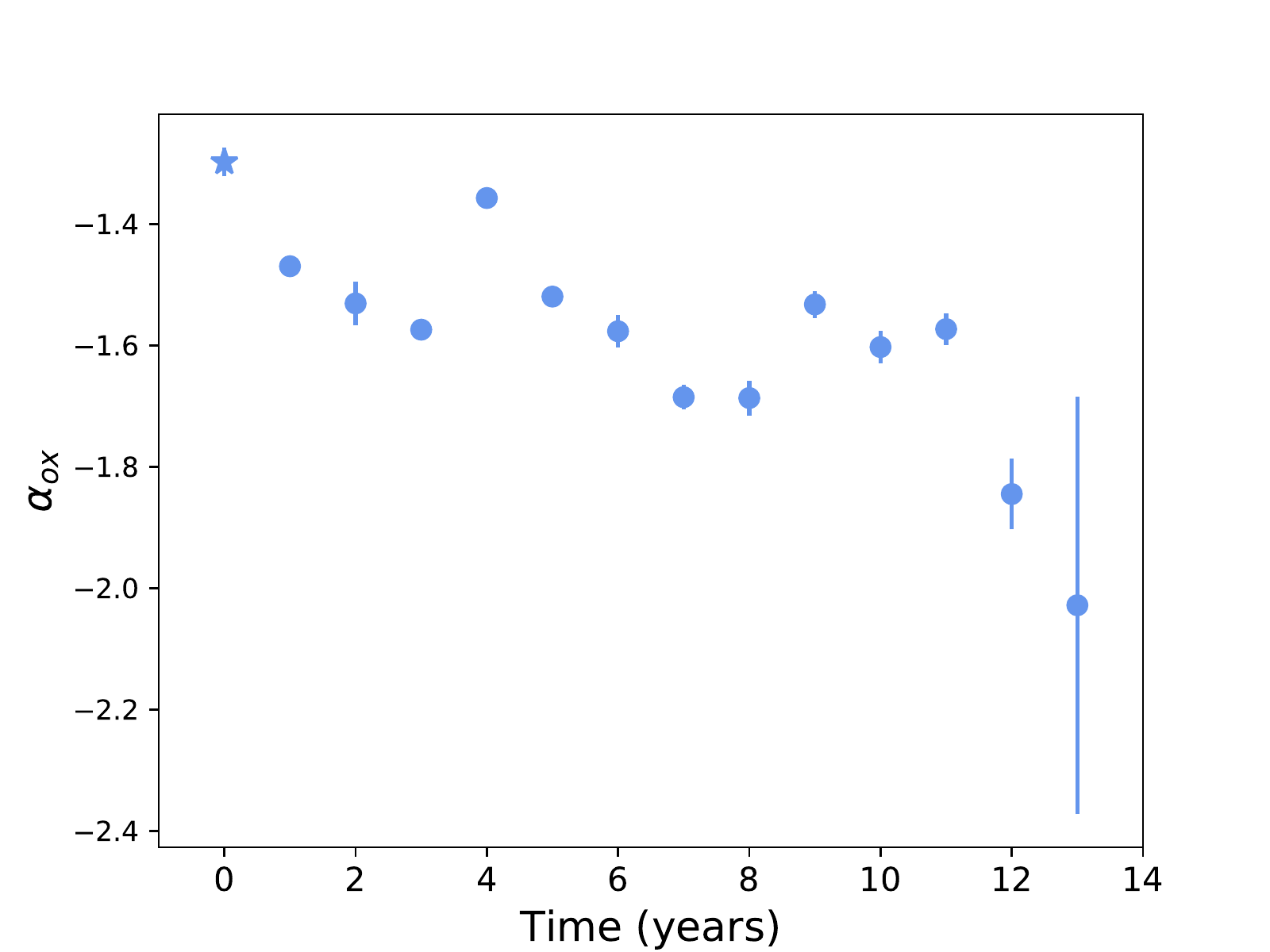}}}
\hfill
{\scalebox{0.54}{\includegraphics[trim= 0cm 0cm 0cm 0cm, clip=true]{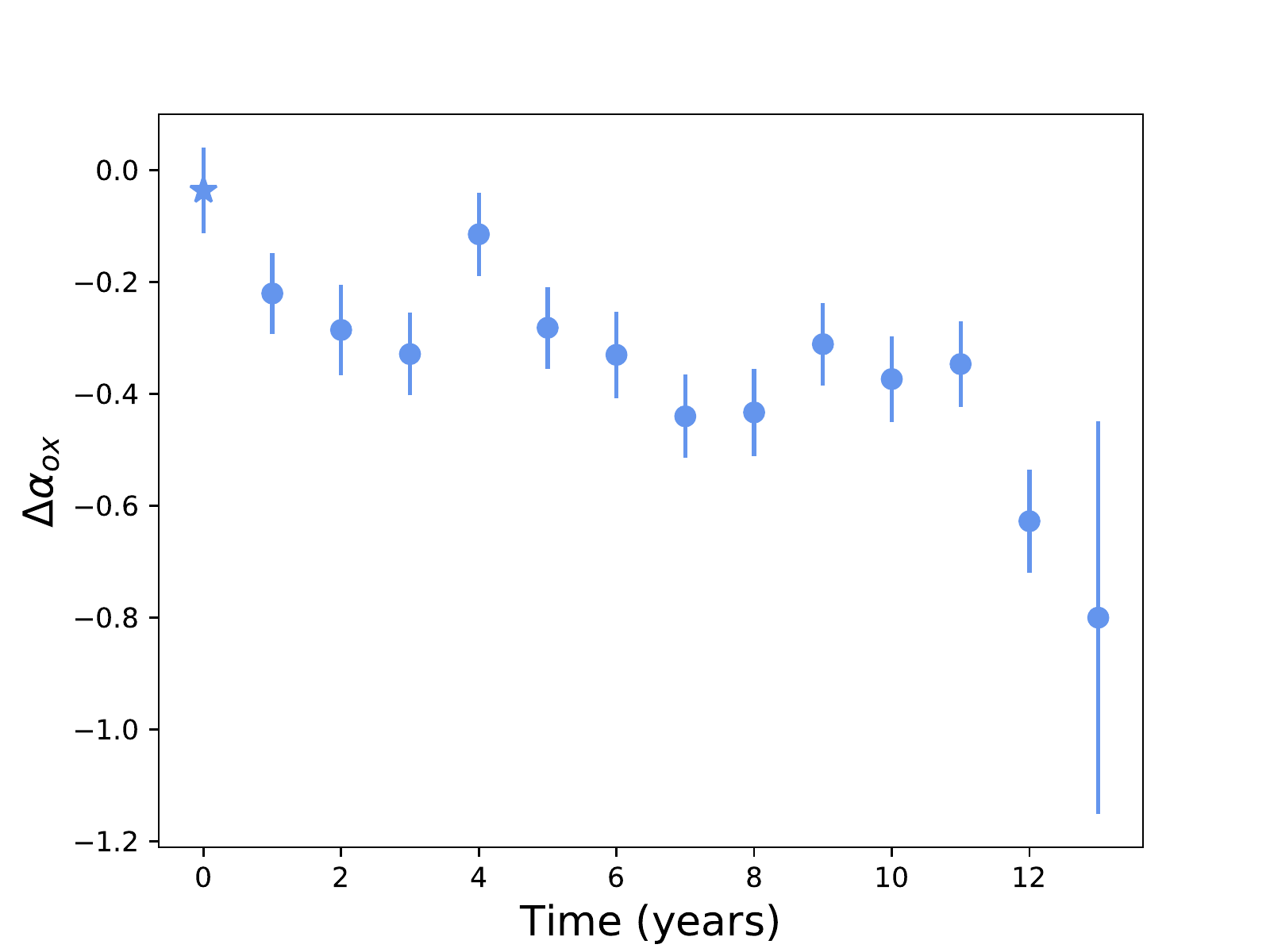}}}      
   \caption{Left: The time variability of the optical-to-X-ray spectral slope $\alpha_{ox}$ is shown. Right: Variation of $\Delta\alpha_{ox}$ with time. Year 0 represents high state of \mrk335\ and is represented by a `star' symbol.}
\label{fig:corr2-3}
\end{figure*}

\indent The availability of simultaneous X-ray and UV measurements allow us to gauge the broadband spectral variability through parameter $\alpha_{ox}$. It is a hypothetical power law between 2500 \AA\ and 2 keV \citep{Tananbaum+1979} measured from the simultaneous X-ray and UV observations.  We find that $\alpha_{ox}$ takes steeper values in general during the low-flux extended period indicating that the X-rays have dimmed more than the UV (left, Figure~\ref{fig:corr2-3}).

\indent Given the UV luminosity of \mrk335\, measurements of $\alpha_{ox}$ were compared to the expected values (e.g. \citealt{Strateva+2005}). This gives us with a measure of $\Delta\alpha_{ox}(\Delta\alpha_{ox} = \alpha_{ox} - \alpha_{ox}(L_{2500\A}))$ and is an indicator of the X-ray strength compared to the UV. 

\indent \mrk335\ displays values of $\Delta\alpha_{ox}\approx0$ during the bright state, indicating that it appears like a typical or ``normal'' AGN. It is interesting to note that during the low extended period despite large uncertainties, the source show nearly constant behaviour, at least, for a few epochs, suggesting a new `normal' state. Physically, it might imply that after acquiring the X-ray dim state, X-rays and UV both are possibly varying in similar strength for those epochs as shown in Figure~\ref{fig:corr2-3}.

\section{Spectral Energy Distribution (SED): Analysis and Measurements}
\label{sect:sed}
In this work, the spectral energy distributions of \mrk335\ over the 13-year \swift\ monitoring period are constructed with the simultaneous multi-wavelength optical/UV to X-ray data. The X-ray emission gives valuable insight into the processes in the innermost regions of the supermassive black hole, while the UV radiation emanating from the accretion disc dominates the bolometric emission. Therefore, studying the simultaneous broadband SED of \mrk335\ makes it possible to investigate the link between the accretion disc and corona. For the preparation of SEDs, we have made use of the \swift\ XRT, UVOT data and the optical spectra (from other instruments), if available. Also, the number of observations of \mrk335\ by \swift\ over the 13-year period provide an opportunity to study the variation of these SEDs in a systematic manner. In addition, we have made average, flux-resolved SEDs on which the analysis and measurements were performed as described in the following subsections.
\subsection{The 13-year average dim SED compared to the bright state SED prior to 2007}
\label{sect:sed1}
\begin{figure*}
\includegraphics[width=8.5cm, trim= 0cm 0cm 0cm 0cm, clip=true]{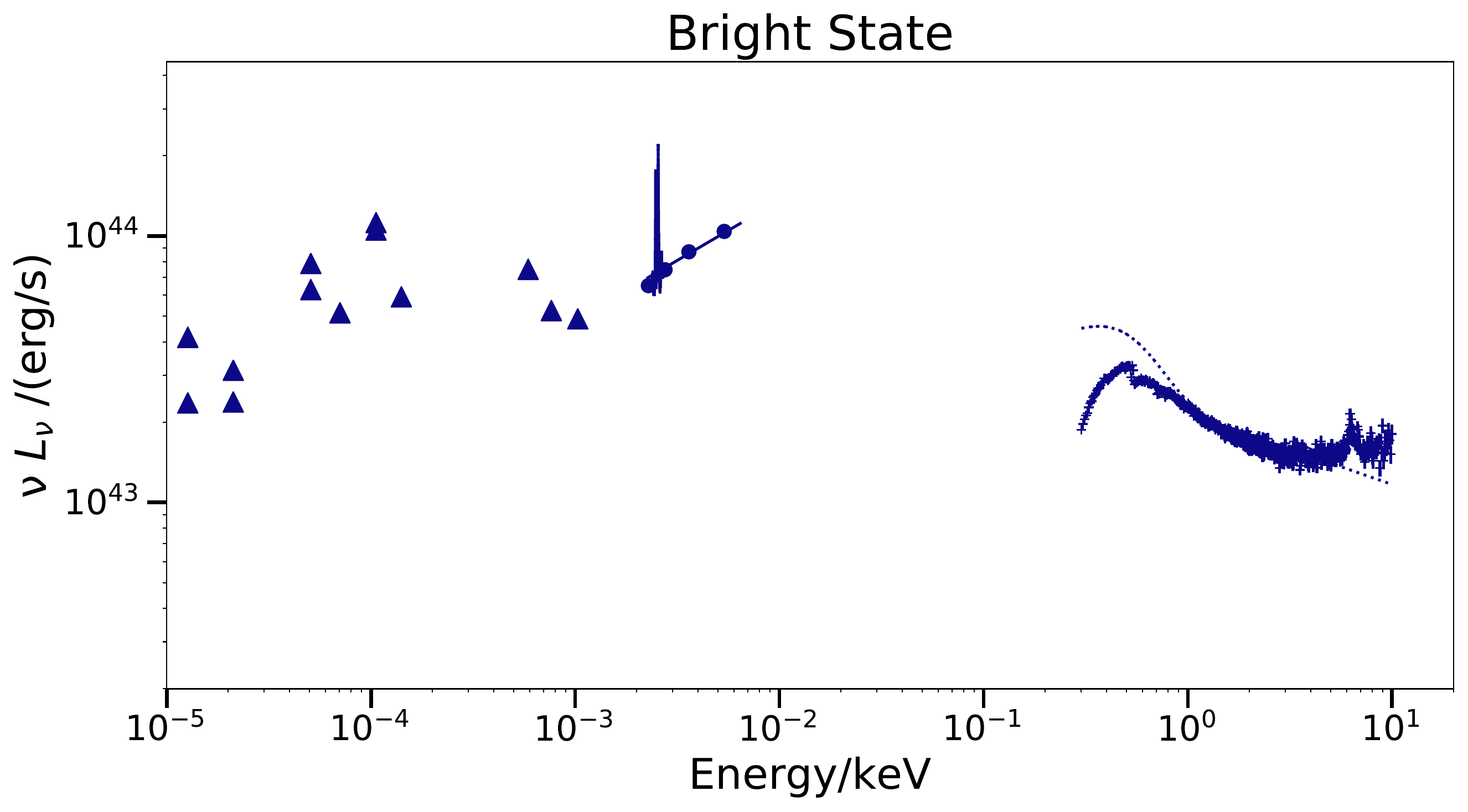}
\hfill
\includegraphics[width=8.5cm, trim= 0cm 0cm 0cm 0cm, clip=true]{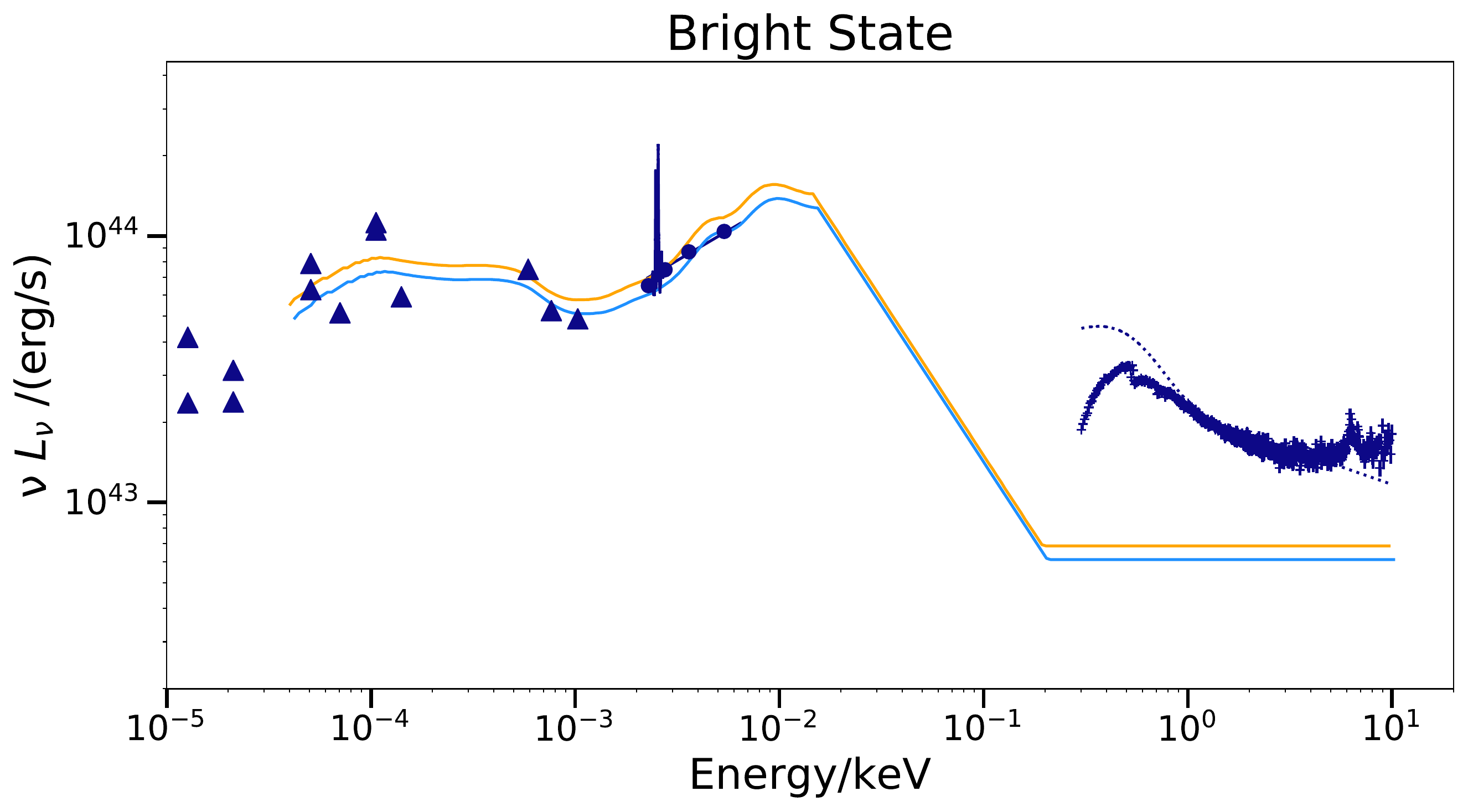}

\vspace{2.00mm}

\includegraphics[width=8.5cm, trim= 0cm 0cm 0cm 0cm, clip=true]{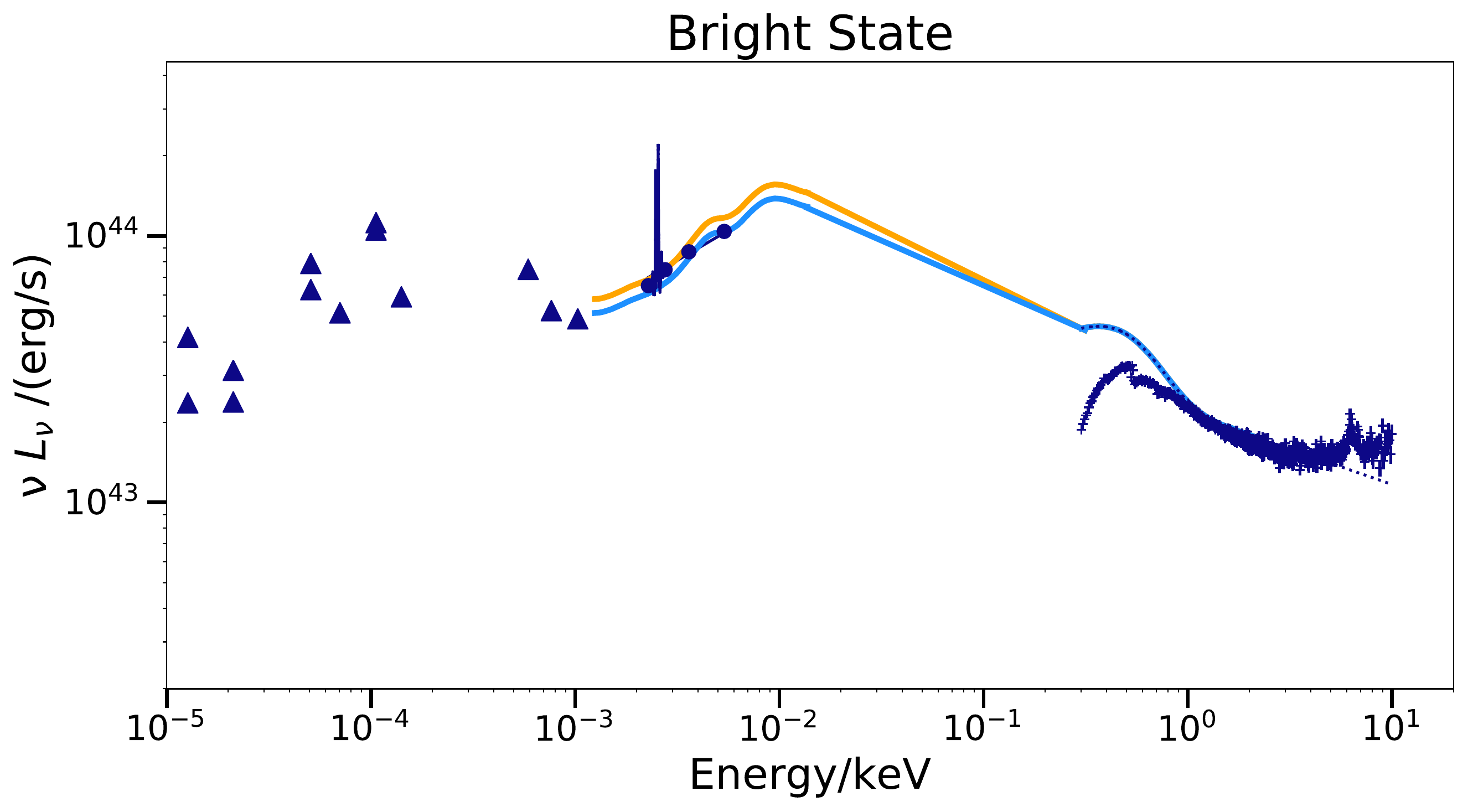}
\hfill
\includegraphics[width=8.5cm, trim= 0cm 0cm 0cm 0cm, clip=true]{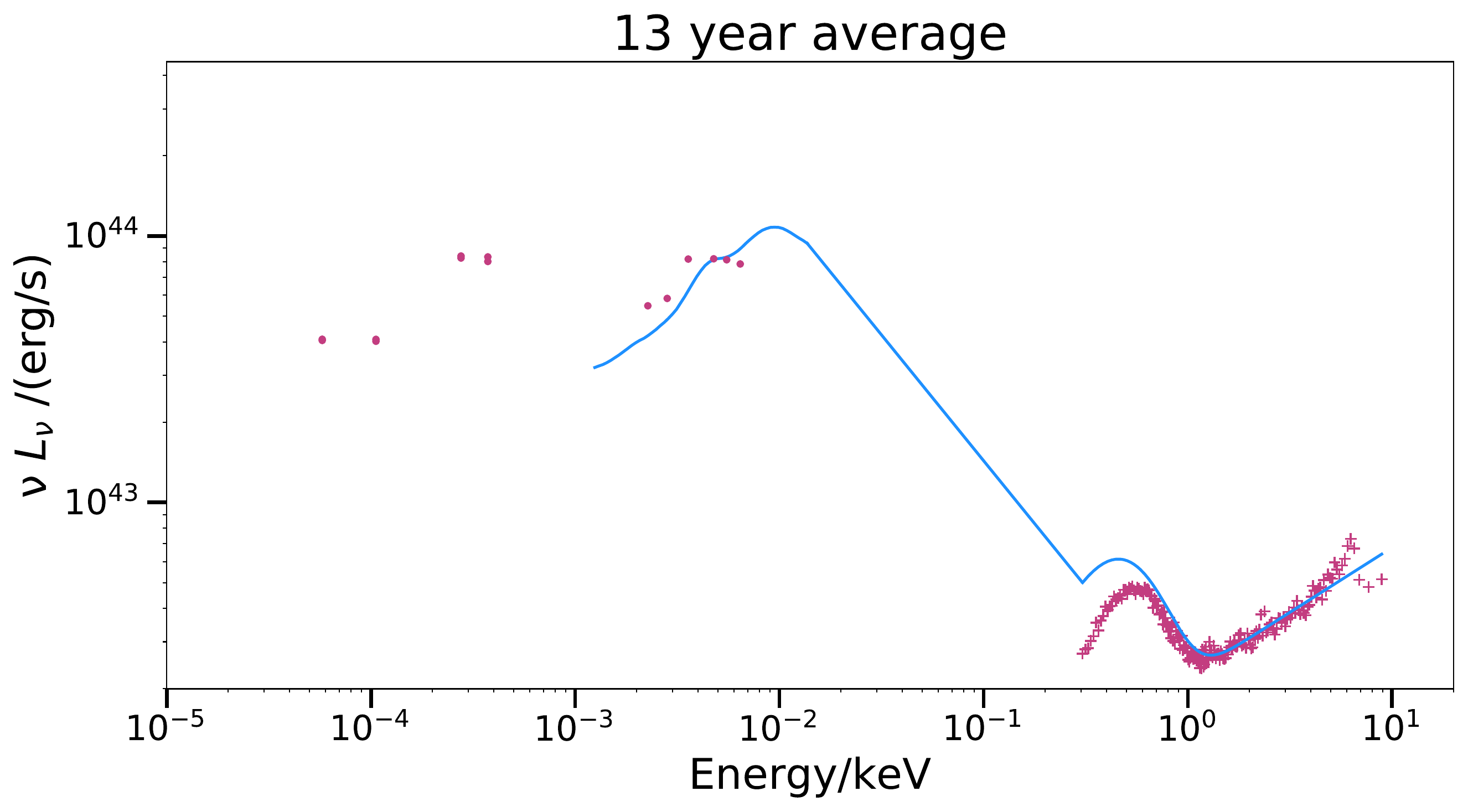}
\caption{First panel (upper, left): shows data for the bright state of \mrk335. Archival infrared data represented in triangles are shown for visual purpose only. Dotted line depicts the unabsorbed model over the X-ray data and the solid line connecting the UV data defines the spectral slope $\alpha_{UV}$. Second panel (upper, right): shows the Krawczyk model (derived for the low-luminosity quasars in the SDSS-DR7 catalog in \citet{Krawczyk+2013}) by solid lines in blue and yellow corresponding to 2500 \AA\ and 5100 \AA\ normalizations respectively, in addition to the data for bright state. This empirical model spans the mid-infrared (1$\mu$m) to the far ultra-violet FUV (912 \AA) region in the rest-frame. It is clear that the template does not accurately predict the X-ray emission in \mrk335. Third panel (lower, left): shows the modified Krawczyk SED model as described in this work. Specifically, the 1 $\mu$m - 912 \AA\ SED is connected via a power law to the unabsorbed X-ray spectrum at 0.3 keV. Fourth panel (lower, right): Similar to the third panel, but for the 13-year average. Only the model normalized to 2500 \AA\ is shown for clarity. Also, notice the steepness of the power law connecting the point at 912 \AA\ and the point at $\sim$~0.3 keV from the bright state to 13-year average SED of the source. }
\label{fig:newsed}
\end{figure*}

The SED of \mrk335\ in the bright state and the 13-year average dim state are constructed using the simultaneous optical-UV to X-ray data with \swift\ as shown in Figure~\ref{fig:newsed}. 

\indent First of all, we plotted the simultaneous optical/UV and X-ray data in the $keV$$-$$\nu L_{\nu}$ rest-frame plane for each year/epoch (First panel, Figure~\ref{fig:newsed}). To facilitate the computation of SED, we have employed an empirical SED template generated for the low-luminosity AGNs in the SDSS-DR7 catalog in a recent study by \citet{Krawczyk+2013}.

\indent \citealt{Krawczyk+2013} studied the SEDs for 119,652 AGNs using mid-IR data from \spitzer\ and \wise, near-infrared data from the Two Micron All Sky Survey and UKIDSS, optical data from the Sloan Digital Sky Survey, and UV data from \galex. Regarding our work, we have used the mean SED for low-luminosity AGNs (log($\nu L_{\nu})|_{\lambda=2500}$~\AA$\leqslant~45.41$). We will refer to this SED template as `Krawczyk model' in the remainder of this paper. 

We performed the measurement of SED by splitting the data into three parts. In the first part, we consider data that roughly spans the frequencies between 1$\mu$m$-$912 \AA. Then the Krawczyk model is used for the data that covers the frequencies between $\sim$~1$\mu$m$-$912 \AA. The Krawczyk model was not extended for the frequencies in the X-rays as their measurements were done through extrapolation on limited X-ray data while we have the available simultaneous X-ray data to make reliable estimates. The full Krawczyk model on the bright state data are plotted (Second panel (upper, right), Figure~\ref{fig:newsed}). Now, Krawczyk model is modified for the data at normalisation wavelengths of 2500 \AA\ and 5100 \AA. This is simply done by scaling the Krawczyk SED template with the ratio of rest-frame luminosity ($\nu L_{\nu}$) point at 2500 \AA\ in our data and Krawczyk model $\frac{\nu L_{\nu (data)}}{\nu L_{\nu (model)}}$. Similar calculation was performed to modify the Krawczyk model at the normalization wavelength 5100 \AA.

\indent In the second part, we need to extrapolate the data in the UV to X-ray gap (912 \AA$-$0.3 keV) to compute the bolometric luminosity. In each SED, we assume a power-law spectrum to the modified Krawczyk model at its rest-frame $\nu L_{\nu}$ point at 912 \AA. We then linearly connect this luminosity at 912 \AA\ to the luminosity corresponding to the frequency of 0.3 keV.     

\begin{equation}
L_\nu = (L_{1 kev}) \times (\nu / 1 keV)^{b}
\label{eq:formula2}
\end{equation}
where $b$ is the spectral index of the extreme UV region. 

\indent For the third part spanning the X-ray data, we deduced the total unabsorbed luminosity through spectral modelling between the energy range 0.3-2 keV only. The data beyond 2 keV has not been used to avoid the double counting of photons resulting due to reprocessing processes \citep{Krawczyk+2013}. Finally, the bolometric luminosities are computed for each year/epoch by integrating\footnote{Python package (`simps') was used to perform integration.} luminosities from three parts in the $keV$$-$$\nu L_{\nu}$ rest-frame plane as discussed above.  

\indent The measured bolometric luminosities for the bright and 13-year average dim states are $L_{Bol}$=$5.53\times10^{44}$ and $2.43\times10^{44}$ erg~s$^{-1}$, respectively.
\subsection{Flux-resolved SED}
\label{sect:sed2}
\begin{figure*}
   \centering
   \advance\leftskip-0.cm
            {\scalebox{0.5}{\includegraphics[trim= 0cm 0cm 0cm 0cm, clip=true]{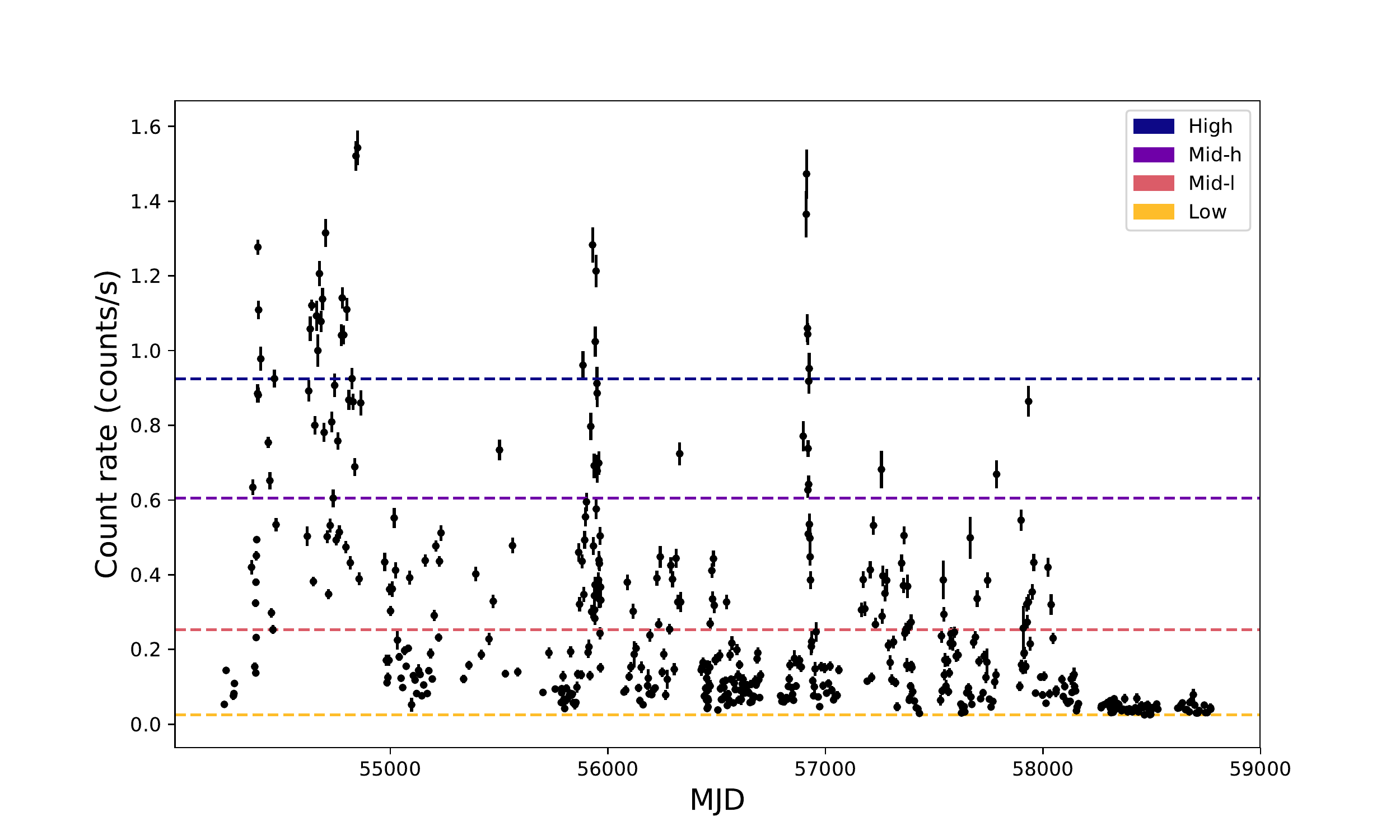}}}
\hfill  
              {\scalebox{0.5}{\includegraphics[trim= 0cm 0cm 0cm 0cm, clip=true]{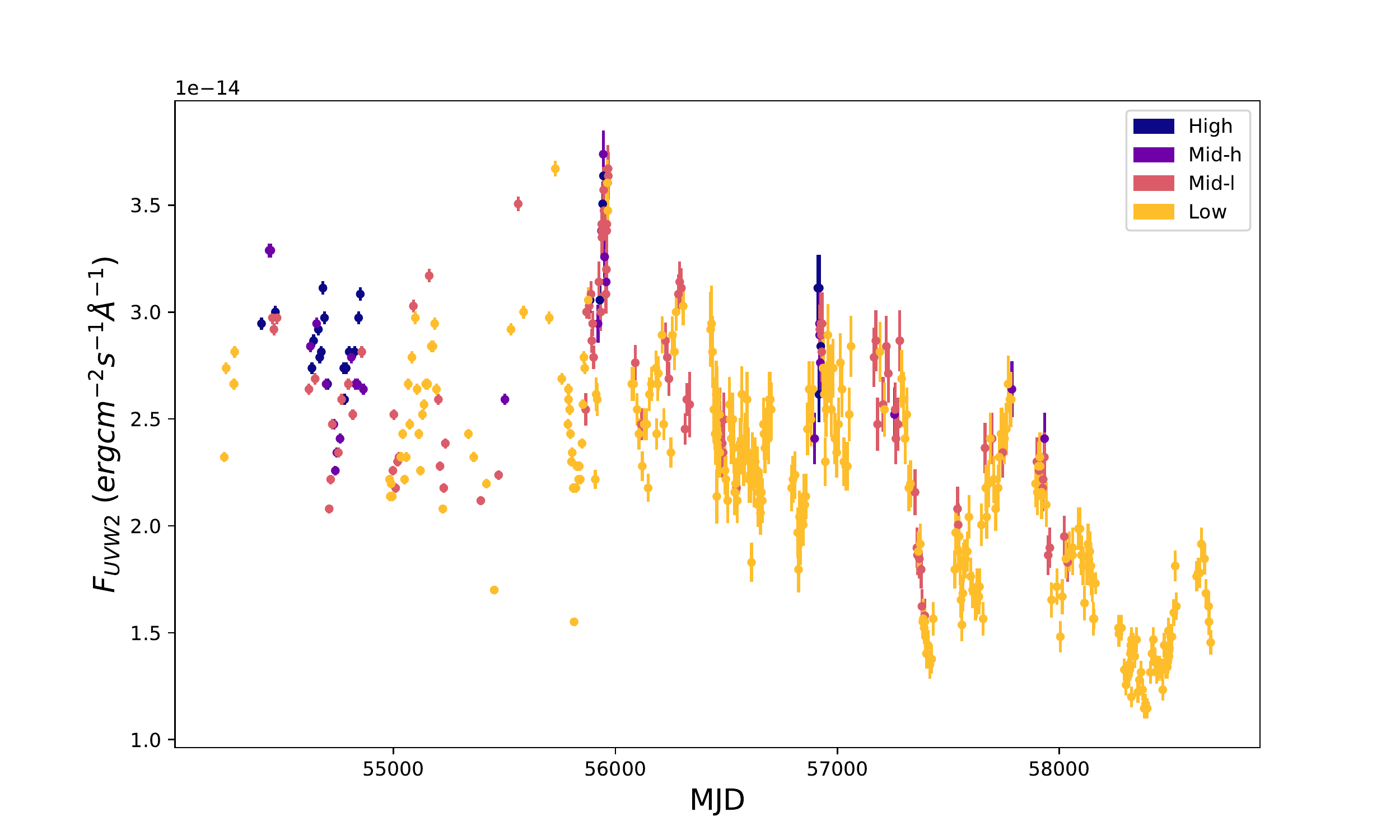}}}  
   \caption{Top to bottom : X-ray lightcurve of \mrk335. The cuts in the count rates are shown as horizontal lines whereas the UVOT lightcurve is color-coded, defined by the X-ray flux cuts.}
\label{fig:lcxrayuv}
\end{figure*}

\begin{figure*}
   \centering
   \advance\leftskip-0.cm
   {\scalebox{0.42}{\includegraphics[trim= 0cm 0cm 0cm 0cm, clip=true]{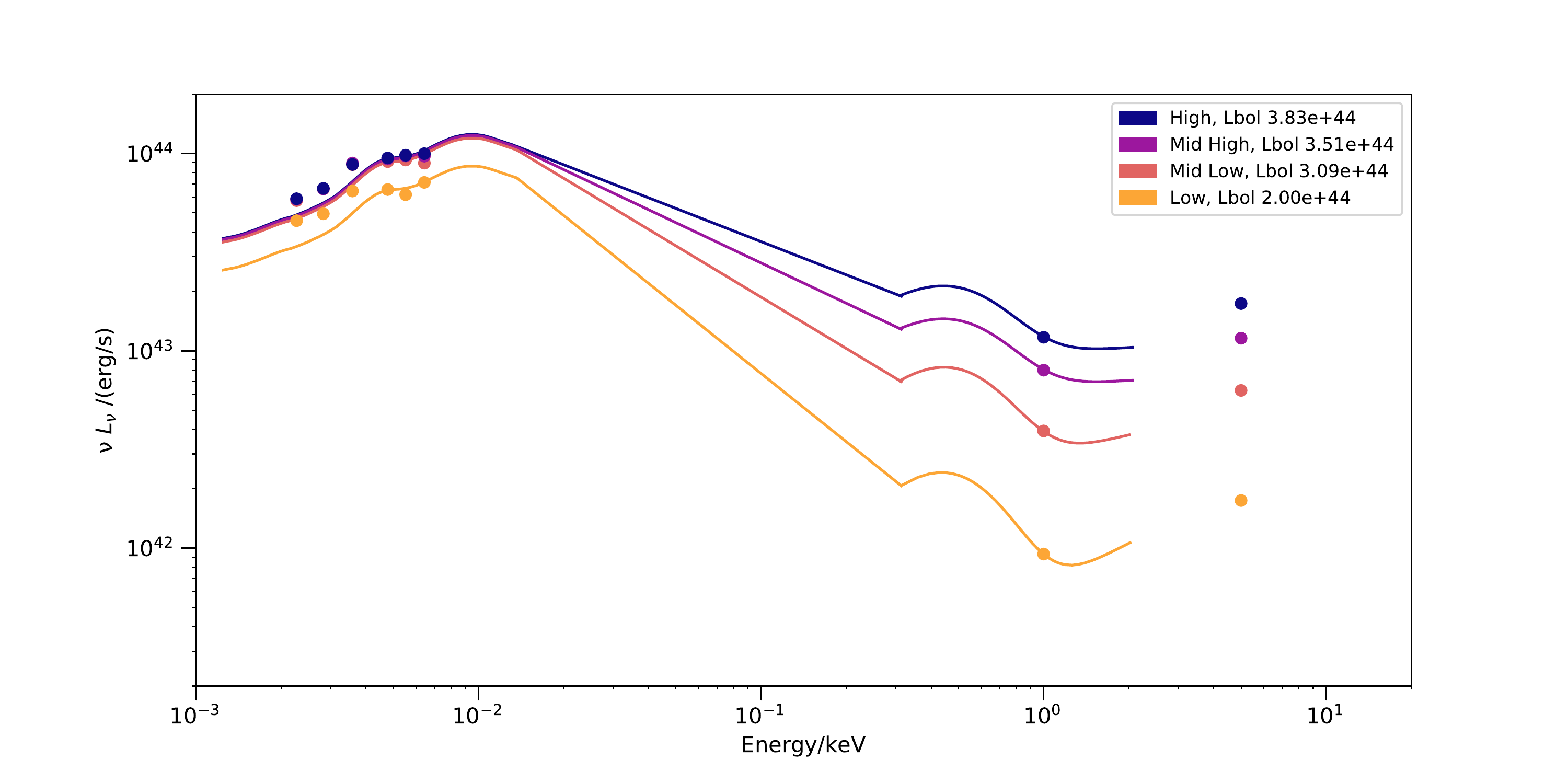}}}      
   
   \caption{SED plot for the flux-resolved data of \mrk335. }
\label{fig:fluxsed}
\end{figure*}

To perform the flux-resolved SED analysis, first the X-ray count rates in the soft band (0.3-1 keV) and the hard band (1-10 keV) were calculated using the hardness ratios. These count rates were sorted into four flux intervals defined as high, mid-high, mid-intermediate and low flux represented by horizontal dashed lines in the top plot of Figure~\ref{fig:lcxrayuv}. Subsequently, the data corresponding to these four intervals are plotted in the SED diagram along with the corresponding UVOT data in the $keV$$-$$\nu L_{\nu}$ rest-frame plane. These are multiplicatively scaled with the yearly SED model template using the same procedure as described in the subsection~\ref{sect:sed1} and then integration is performed to yield the corresponding bolometric luminosities (Figure~\ref{fig:fluxsed}). The bolometric luminosities computed for the four intervals i.e. high, mid-high, mid-low and low flux are $L_{Bol}$=$3.83\times10^{44}$, $3.51\times10^{44}$, $3.09\times10^{44}$ and $2.00\times10^{44}$ erg~s$^{-1}$, respectively.\\

\indent X-ray spectral analysis were also performed for these flux intervals and the measured spectral parameters are presented in Table~\ref{tab:xflux}.

\begin{table}
 \centering
\caption{Variations of the X-ray spectral parameters over flux epochs.}
  \begin{tabular}{ccccc}
  \hline
  \hline
 Flux-epoch & kT (eV) & kT$_{norm}$$^{a}$ & $\Gamma$ & $\Gamma_{norm}$$^{b}$\\
\\
\hline 
High  &  109$\pm$3    &   1.42$\pm$0.07   & 1.95$\pm$0.03            & 4.40$\pm$0.12 \\ \\
Mid-High &  110$\pm$2.7    &   1.18$\pm$0.06   & 1.85$\pm$0.03            & 3.12$\pm$0.10 \\ \\
Mid-Low &  105$\pm$1.6    &   0.84$\pm$0.01   & 1.53$\pm$0.02            & 1.42$\pm$0.03 \\ \\
Low   &  114$\pm$1.6    &   0.26$\pm$0.01   & 0.91$\pm$0.03            & 0.26$\pm$0.01 \\ \\
\hline
\hline
\label{tab:xflux}
\end{tabular}

\vskip -0.1cm~\footnotesize{$^{a}$ kT$_{norm}$ is in units of 10$^{-4}$~$(R_{in}/km)/(D/10~kpc)$, where $R_{in}$ is the inner radius and $D$ is the distance};~\footnotesize{$^{b}$ $\Gamma_{norm}$ is in units of 10$^{-3}$~photons~keV$^{-1}$~cm$^{-2}$~s$^{-1}$ at 1 keV}.
\end{table}

\subsection{Year-averaged SED}
\label{sect:sed3}
The individual yearly SEDs are generated using the same procedure described in the Section~\ref{sect:sed1}. For comparison, the SEDs were normalised at 2500 \AA ~and 5100 \AA.  Both produced comparable results. These models were then used to compute the bolometric luminosity $L_{Bol}$  each year. These SEDs are compiled in the Appendix~\ref{sect:app2}. For Years 0, 4, 8, 9, 10 and 12, the optical spectroscopic data were  available to include in the SEDs. 

\indent $L_{Bol}$ is often divided by $L_{Edd}$ to yield a parameter that is proportional to the accretion rate, which is generally called as Eddington ratio and commonly described by the symbol $\lambda$=$L_{Bol}$/$L_{Edd}$, where $L_{Edd}$~\footnote{$L_{Edd}=1.26\times~10^{38}~(M_{BH}/M_{\odot}$) erg~s$^{-1}$, where M$_{\odot}$ is the mass of the Sun.} is the maximal theoretical luminosity to account for the equilibrium between the radiation pressure and the gravitational force.\\

\begin{figure*}
   \centering
   \advance\leftskip-0.cm
   {\scalebox{0.5}{\includegraphics[trim= 0cm 0cm 0cm 0cm, clip=true]{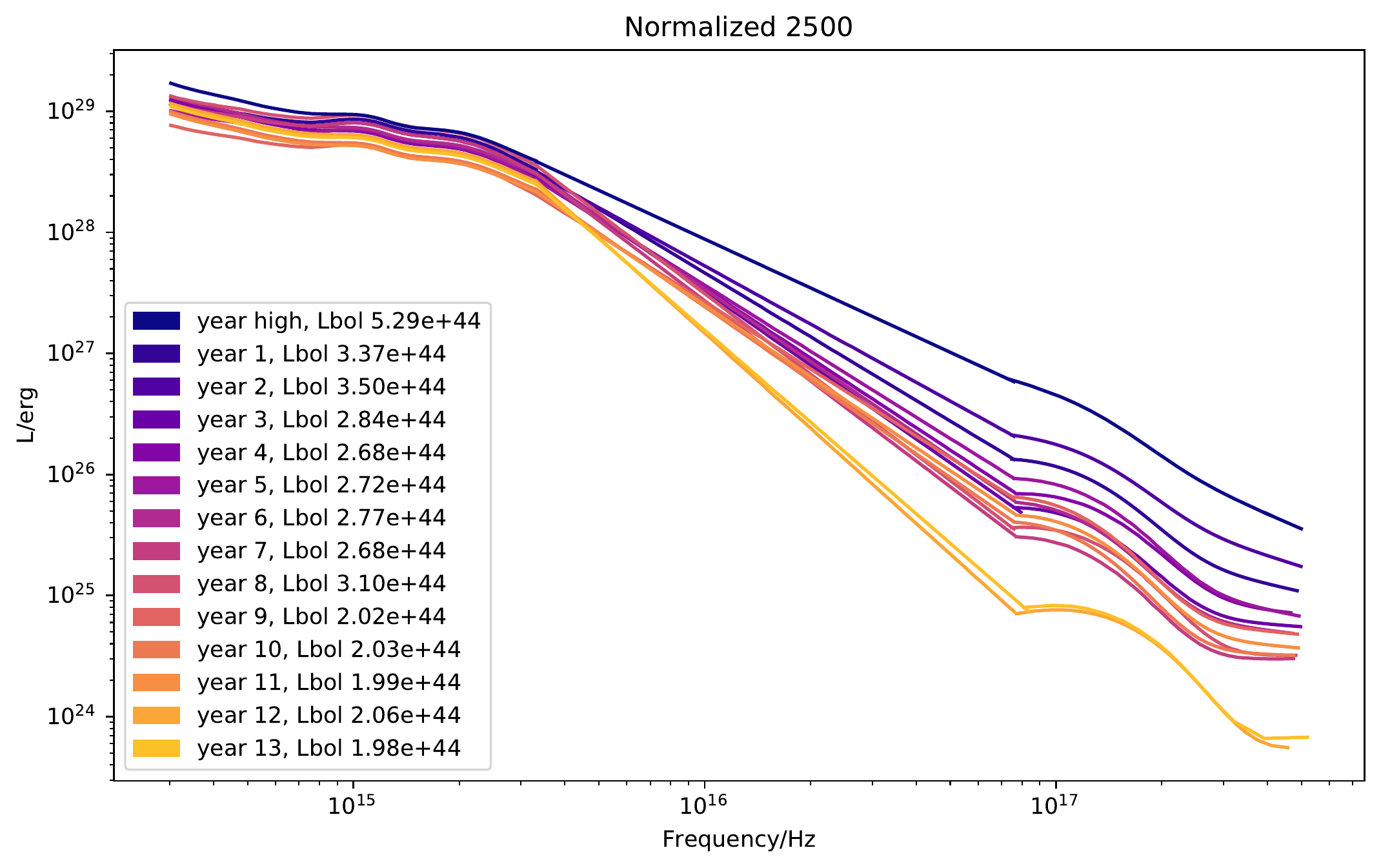}}}      
  
   \caption{SED measurements for all years with normalization at 2500 \AA. 
}
\label{fig:sed0}
\end{figure*}

\indent The variation of the SED models of \mrk335\ for all the years with respect to normalization wavelength at 2500 \AA\ are shown in Figure~\ref{fig:sed0}. 

\begin{figure*}
   \centering
   \advance\leftskip-0.cm
            {\scalebox{0.42}{\includegraphics[trim= 0cm 0cm 0cm 0cm, clip=true]{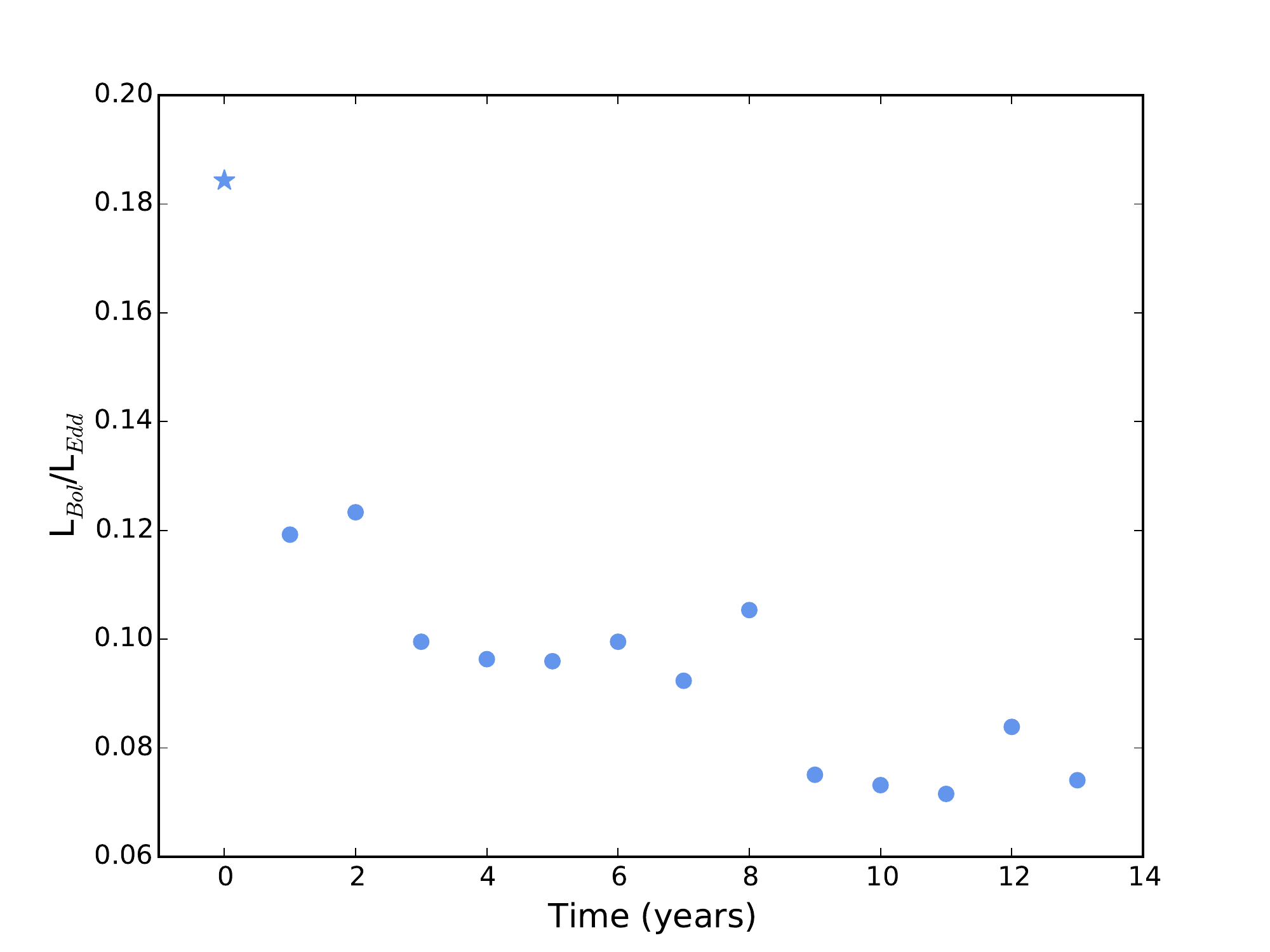}}}  
              {\scalebox{0.42}{\includegraphics[trim= 0cm 0cm 0cm 0cm, clip=true]{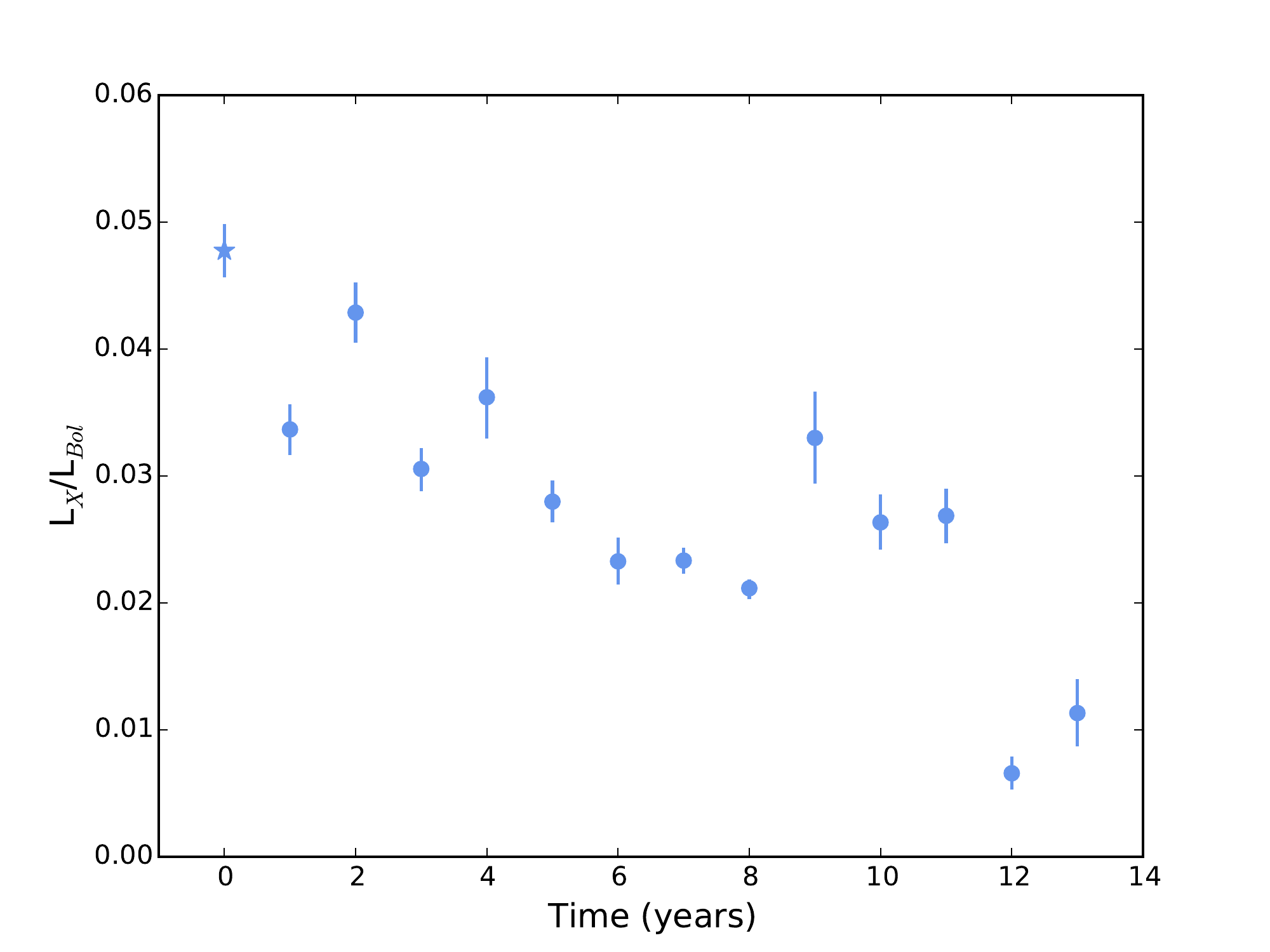}}}  
   \caption{From left to right : Bolometric luminosity Eddington ratio with time and the X-ray to Bolometric Luminosity ratio, assuming the Black hole mass of $(25\pm~3)\times~10^{6}~M\odot$ \citep{Grier+2012}. Year 0 represents bright state of \mrk335\ and is represented by a `star' symbol.}
\label{fig:corr4}
\end{figure*}

\indent We have measured the yearly variability of bolometric Luminosity Eddington ratio $L_{Bol}/L_{Edd}$ (decrease at 11 $\%$ level) and the X-ray to bolometric luminosity ratio $L_{X}/L_{Bol}$ (decrease at 4 $\%$ level) in Figure~\ref{fig:corr4}. $L_{Bol}/L_{Edd}$ follow a significant downward trend from the pre-2007 epoch to the low state epochs.

\begin{landscape}
\begin{table}
 \centering
\caption{Year-wise measurements of \mrk335\ since January 2007. The XRT values are in units of counts s$^{-1}$. The hardness ratio is defined as $HR = \frac{hard-soft}{hard+soft}$ where soft and hard are the counts in the 0.3-1.0 keV and 1.0-10.0 keV bands, respectively.}
  \begin{tabular}{ccccccccccccccc}
  \hline
  \hline
 Year & kT (eV) & kT$_{norm}$$^a$ & $\Gamma$ & $\Gamma_{norm}$$^b$& HR & F$_{var}$ & F$_{2keV}$$^c$ & $L_{x}/L_{bol}$ & $\alpha_{ox}$ & SE$\times$10$^{-1}$ & L$_{bol}$/L$_{Edd}$ & $\alpha_{uv}$ &F$_{varUV}$ & F$_{2500}$$^d$
\\
\\
\hline 
Year 0  &  99$\pm$1    &   1.30$\pm$0.04   & 2.42$\pm$0.01            & 10.31$^{+0.04}_{-0.05}$ & 0.08$\pm$0.01             & 0.12$\pm$0.01   & 2.61$\pm$0.01    & 0.05$\pm$0.02 & -1.30$\pm$0.02     & 10.57$\pm$0.01  & 0.18  & 0.5$\pm$0.04 & ---& 6.8$\pm$0.4   \\ \\
Year 1  &  109$\pm$2   &   0.94$\pm$0.03   & 1.65$\pm$0.03            & 1.88$\pm$0.06             & 0.30$\pm$0.01             & 0.73$\pm$0.10   & 0.82$\pm$0.02    & 0.03$\pm$0.02 & -1.47$\pm$0.01     & 9.77$\pm$0.02   & 0.12  & 0.4$\pm$0.1 &0.09$\pm$0.02  & 5.5$\pm$0.01   \\ \\
Year 2  &  109$\pm$3   &   1.10$\pm$0.05   & 1.91$\pm$0.03            & 3.65$\pm$0.09             & 0.25$\pm$0.01             & 0.37$\pm$0.15   & 1.33$\pm$0.03    & 0.04$\pm$0.01 & -1.53$\pm$0.01     & 10.05$\pm$0.03  & 0.12  & 0.4$\pm$0.1 &0.09$\pm$0.05 & 5.2$\pm$0.1   \\ \\
Year 3  &  110$\pm$3   &   0.45$\pm$0.02   & 1.21$\pm$0.05            & 0.70$^{+0.04}_{-0.03}$  & 0.52$\pm$0.01             & 0.62$\pm$0.07   & 0.41$\pm$0.02    & 0.03$\pm$0.03 & -1.57$\pm$0.02     & 9.63$\pm$0.03   & 0.10  & 0.4$\pm$0.1 &0.12$\pm$0.01&5.2$\pm$0.1  \\ \\
Year 4  &  117$\pm$7   &   0.60$\pm$0.06   & 1.32$\pm$0.11            & 0.97$\pm$0.11             & 0.44$\pm$0.02             & 0.68$\pm$0.04   & 0.53$\pm$0.05    & 0.04$\pm$0.02 & -1.36$\pm$0.04     & 9.65$\pm$0.07   & 0.10  & 0.4$\pm$0.1  &0.21$\pm$0.01& 4.6$\pm$0.01  \\ \\
Year 5  &  111$\pm$3   &   0.73$\pm$0.03   & 1.57$\pm$0.04            & 1.10$\pm$0.05             & 0.27$\pm$0.01             & 0.86$\pm$0.08   & 0.50$\pm$0.02    & 0.03$\pm$0.01 & -1.52$\pm$0.01     & 9.68$\pm$0.03   & 0.10  & 0.6$\pm$0.1 &0.17$\pm$0.02 & 4.6$\pm$0.01  \\ \\
Year 6  &  103$\pm$4   &   0.47$\pm$0.03   & 1.34$\pm$0.06            & 0.67$\pm$0.04             & 0.42$\pm$0.01             & 0.70$\pm$0.09   & 0.38$\pm$0.02    & 0.02$\pm$0.01 & -1.58$\pm$0.02     & 9.64$\pm$0.04   & 0.10  & 0.4$\pm$0.1 &0.09$\pm$0.01 & 4.9$\pm$0.1   \\ \\
Year 7  &  108$\pm$3   &   0.27$\pm$0.01   & 0.93$\pm$0.05            & 0.31$\pm$0.02             & 0.59$\pm$0.01             & 0.64$\pm$0.06   & 0.21$\pm$0.01    & 0.02$\pm$0.01 & -1.68$\pm$0.02     & 9.49$\pm$0.03   & 0.10  & 0.4$\pm$0.2 &0.07$\pm$0.01 & 5.2$\pm$0.01  \\ \\
Year 8  &  118$\pm$4   &   0.36$\pm$0.02   & 0.96$\pm$0.08            & 0.34$\pm$0.03             & 0.51$^{+0.01}_{-0.02}$  & 1.14$\pm$0.11   & 0.24$\pm$0.02    & 0.02$\pm$0.01 & -1.69$\pm$0.03     & 9.41$\pm$0.05   & 0.11  & 0.3$\pm$0.2 &0.13$\pm$0.01 & 5.9$\pm$0.01  \\ \\
Year 9  &  102$\pm$3   &   0.53$\pm$0.03   & 1.27$\pm$0.06            & 0.63$\pm$0.04             & 0.42$\pm$0.01             & 0.66$\pm$0.08   & 0.35$\pm$0.02    & 0.03$\pm$0.03 & -1.53$\pm$0.02     & 9.57$\pm$0.04   & 0.08  & 0.4$\pm$0.1  &0.24$\pm$0.03& 3.5$\pm$0.01   \\ \\
Year 10 &  103$\pm$4   &   0.34$\pm$0.02   & 1.10$\pm$0.07            & 0.37$\pm$0.03             & 0.50$^{+0.02}_{-0.01}$  & 0.78$\pm$0.08   & 0.24$\pm$0.02    & 0.03$\pm$0.02 & -1.60$\pm$0.03     & 9.49$\pm$0.04   & 0.07  & 0.4$\pm$0.1  &0.15$\pm$0.02& 3.6$\pm$0.01   \\ \\
Year 11 &  107$\pm$4   &   0.39$\pm$0.02   & 1.26$\pm$0.07            & 0.48$\pm$0.03             & 0.42$^{+0.02}_{-0.01}$  & 0.88$\pm$0.10   & 0.28$\pm$0.02    & 0.03$\pm$0.02 & -1.57$\pm$0.03     & 9.57$\pm$0.04   & 0.07  & 0.4$\pm$0.1  &0.11$\pm$0.01& 3.5$\pm$0.1   \\ \\
Year 12 &  147$\pm$8   &   0.10$\pm$0.01   & 0.59$^{+0.26}_{-0.24}$ & 0.04$\pm$0.01             & 0.47$^{+0.03}_{-0.04}$  & 0.15$\pm$0.03   & 0.05$\pm$0.01    & 0.01$\pm$0.01 & -1.84$\pm$0.06     & 9.13$\pm$0.13   & 0.08  & 0.4$\pm$0.1 &0.07$\pm$0.01 & 3.2$\pm$0.01  \\ \\
Year 13 &  144$\pm$15  &   0.10$\pm$0.01   & 0.34$^{+0.46}_{-0.43}$ & 0.04$^{+0.03}_{-0.02}$  & 0.59$^{+0.03}_{-0.08}$  & 0.13$\pm$0.08   & 0.02$\pm$0.02    & 0.01$\pm$0.02 & -2.03$\pm$0.3     & 9.07$\pm$0.20   & 0.07  & 0.2$\pm$0.1 &0.13$\pm$0.02 & 3.9$\pm$0.07  \\ \\
\hline
\hline
\label{tab:tab1}
\end{tabular}
\vskip -0.1cm~\footnotesize{$^a$ kT$_{norm}$ is in units of 10$^{-4}$~$(R_{in}/km)/(D/10~kpc)$, where $R_{in}$ is the inner radius and $D$ is the distance};~\footnotesize{$^b$ $\Gamma_{norm}$ is in units of 10$^{-3}$~photons~keV$^{-1}$~cm$^{-2}$~s$^{-1}$ at 1 keV};~\footnotesize{$^c$ F$_{2keV}$ is in units of $\mu$Jy.};~\footnotesize{$^d$ F$_{2500}$ is in units of 10$^{-26}$~erg~cm$^{-2}$~s$^{-1}$~Hz$^{-1}$}.
\end{table}
\end{landscape}
\section{Investigating correlations between the SED parameters}
\label{sect:corsed}
\subsection{Correlation Analysis}
\label{sect:cor}
\begin{figure*}
   \centering
   \advance\leftskip-0.cm
{\scalebox{1.0}{\includegraphics[trim= 0cm 0cm 0cm 0cm, clip=true]{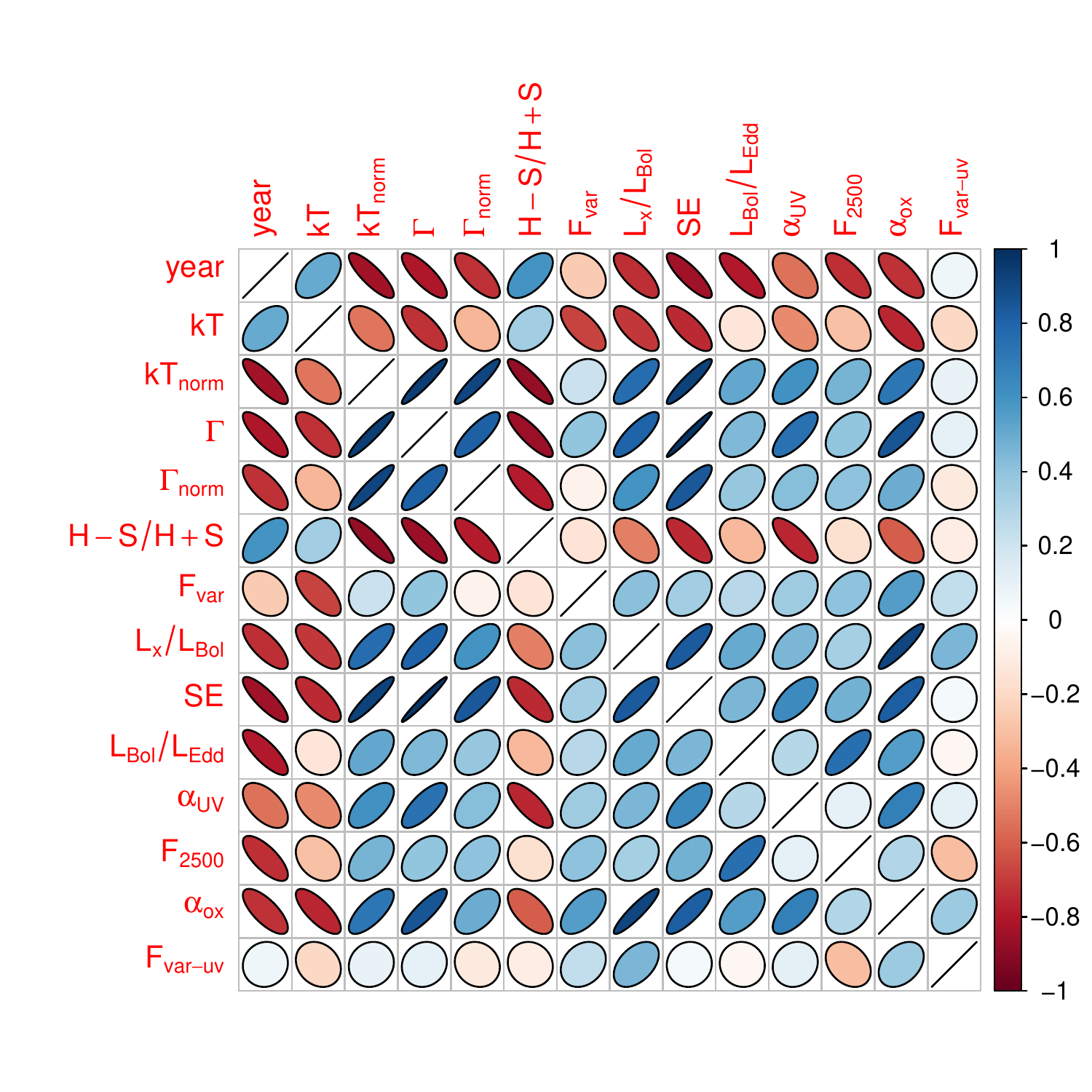}}}
      
   \caption{Correlogram shows all the correlations among the variables presented in Table~\ref{tab:tab1}. Correlation coefficients is coloured according to the value such that the colours of the ellipse indicate the sign of the correlation, and their shapes indicate the strength (narrower ellipses = higher correlations)}
\label{fig:pca1}
\end{figure*}

We have searched for the correlations among the measured parameters. To present the results of the correlation analysis in one plot, a correlation matrix function `corrplot' \citep{corrplot2017} in `R' \citep{R} is used as shown in Figure~\ref{fig:pca1}. This correlogram is used here to highlight the most correlated variables in Table~\ref{tab:tab1}. We have excluded the flux density at 2 keV from the analysis as by definition, this parameter is redundant with $\alpha_{ox}$. The colours (as well as the orientation) of ellipses in the correlogram show positive or negative correlations while the width of the ellipses implies the strength of correlation. We discuss the important findings of this analysis below.

\indent The blackbody temperature (kT) (i.e. the shape of the soft excess) and the hardness ratio (HR) show a moderate correlation with time (i.e. year). There is no obvious relation between time and either F$_{varUV}$ or F$_{var}$. All other parameters show an anti-correlation with the time.

\indent The soft excess emission (SE) shows significant correlations with several parameters. We find a very strong correlation between the soft excess and $L_{x}/L_{bol}$. This could be due to the contribution of primary X-ray emission in the soft excess strength, as defined in our work. Some studies have indicated that the lack of inverse correlation between the soft excess and 2-10 keV $L_{x}/L_{bol}$ could possibly disfavour the reflection and absorption scenarios, and could be naturally explained by the warm Comptonization scenario (e.g. \citealt{Gliozzi+2019}). There is a clear correlation between the soft excess and the steepness of the X-ray spectrum ($\Gamma$), however, we found only a marginal relationship with L$_{bol}$/L$_{Edd}$. Some studies have found that L$_{bol}$/L$_{Edd}$ is positively correlated to soft excess emission (SE) (e.g. \citealt{Boissay+2016}). Both the soft excess and $L_{x}/L_{bol}$ are correlated with $\alpha_{ox}$ indicating as the X-ray brightens so to must the UV.   Interestingly, there is also a tight correlation between SE and $\alpha_{uv}$  possibly suggesting a common mechanism in the soft X-ray and UV bands.

\indent The parameter L$_{bol}$/L$_{Edd}$ is significantly correlated with time, implying a decrease in its strength successively each year. L$_{bol}$/L$_{Edd}$ shows weak correlation with photon index $\Gamma$. This dependance has been studied for \mrk335\ in the previous works (e.g. \citealt{Sarma+2015, Keek+2016}) and a positive correlation has been found in the previous studies on large samples that have shown that L$_{bol}$/L$_{Edd}$ is indeed the primary parameter driving the conditions in the corona, giving rise to $\Gamma$ (e.g. \citealt{Wang+2004, Grupe2004, Porquet+2004, Bian+2005, Kelly+2007, Shemmer+2006, Shemmer+2008, Risaliti+2009, Jin+2012, Fanali+2013, Brightman+2013, Gliozzi+2019}). The weaker correlation found in this work possibly suggest that we may not be necessarily measuring the true shape of the power law since the spectrum is modified by absorption or reflection that we do not include. L$_{bol}$/L$_{Edd}$ appears only moderately correlated with X-ray normalisation parameter of blackbody (kT$_{norm}$) and  $L_{x}/L_{bol}$.

\indent An interesting significant correlation is found between L$_{bol}$/L$_{Edd}$ and F$_{2500}$. This is rather a surprising result as F$_{2500}$ is not found to be correlated with any other parameter except with year. Statistically, this correlation could be driven due to their strong dependance on time. Otherwise, physically, it could suggest the possible influence of conditions driving the changes in the accretion rate on 2500 \AA\ emission. 

\indent L$_{bol}$/L$_{Edd}$ is seen to be positively correlated to $\alpha_{ox}$ which is in consistence with the previous studies on the sample of luminous AGNs (L$_{bol}$/L$_{Edd}$ $>$ $10^{-3}-10^{-2}$) that have derived this positive relationship with the slope of such as 0.397 \citep{Lusso+2010}, and 0.11 \citep{Grupe+2010}. This is interesting as we did not find any correlation between $\alpha_{ox}$ and F$_{2500}$ and this could probably mean that the Eddington ratio (or accretion rate) is possibly playing role in driving the changes in the X-ray strength. Previous studies on the optically-selected samples and X-ray samples have found strong dependance between $\alpha_{ox}$ and L$_{2500}$ (e.g. \citealt{Vignali+2003,Strateva+2005,Lusso+2010}).

\indent Next, concentrating on  $L_{x}/L_{bol}$, it is positively correlated with the photon index $\Gamma$, blackbody parameters and $\alpha_{ox}$. For high redshift sources, the positive dependance of $\Gamma$ with X-ray luminosity has been noted (e.g. \citealt{Dai+2004, Saez+2008}) but not observed in the sample of local AGN (e.g. \citealt{Brightman+2013}). 

\indent We checked the correlations of F$_{varUV}$ and F$_{var}$ with all the parameters. F$_{varUV}$ is not correlated with any parameter. However, we find some interesting behaviour from the parameter F$_{var}$. During the thirteen-year low state, it is found to be significantly anti-correlated with blackbody temperature (kT) and positively correlated with $\alpha_{ox}$; however, the correlations disappear if the bright state (Year 0) is included in the analysis. This suggests an important result that the 13-year monitoring period has been witnessing significant X-ray variability in the source although being in an X-ray low state. Also, if kT can be assumed to indicate the ionization of the absorbing or reflecting material, the anti-correlation between F$_{var}$ and kT might suggest the amplitude of the variations diminish when the ionization is high.

\indent One of the many interesting results that arise from this analysis is that overall behaviours in an individual AGN i.e. \mrk335, mirror the range of variations and their correlations as seen in large samples of different objects. Perhaps this suggest that the same physical process applies in all cases, and is scalable in some form. The physical scenarios that can predict/produce these relationships are discussed in detail in Section~\ref{sect:disc}.

\subsection{Principal Component Analysis}
\label{sect:pca}
\begin{figure*}
   \centering
   \advance\leftskip-0.cm
      {\scalebox{0.8}{\includegraphics[trim= 0cm 0cm 0cm 0cm, clip=true]{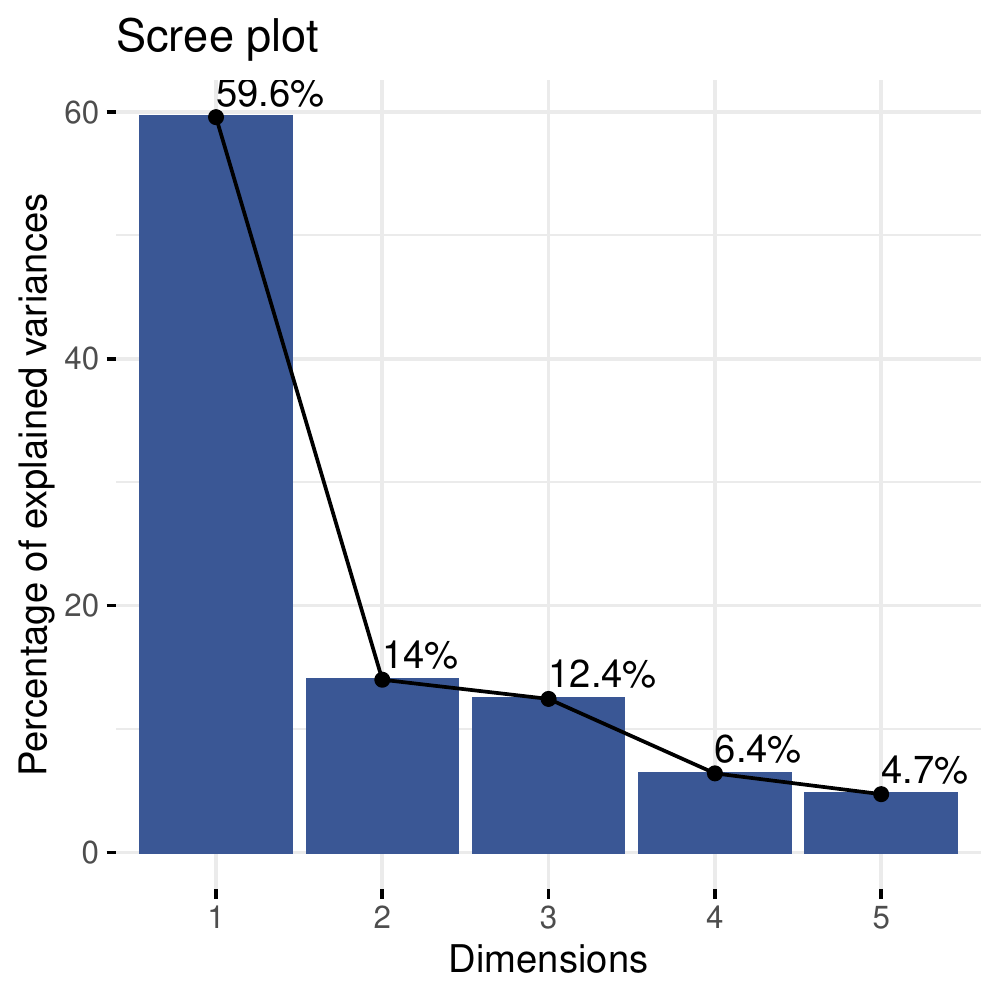}}} 
\hfill 
      {\scalebox{0.8}{\includegraphics[trim= 0cm 0cm 0cm 0cm, clip=true]{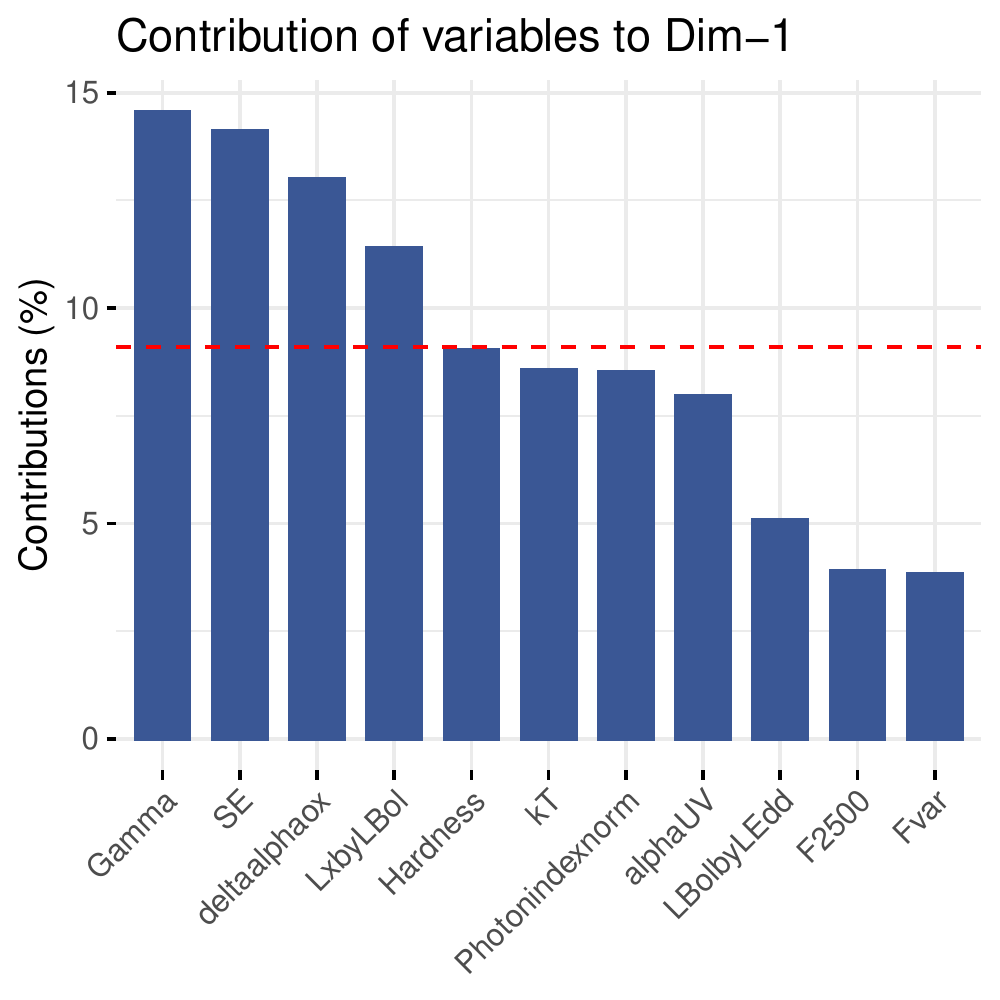}}} 
      {\scalebox{0.8}{\includegraphics[trim= 0cm 0cm 0cm 0cm, clip=true]{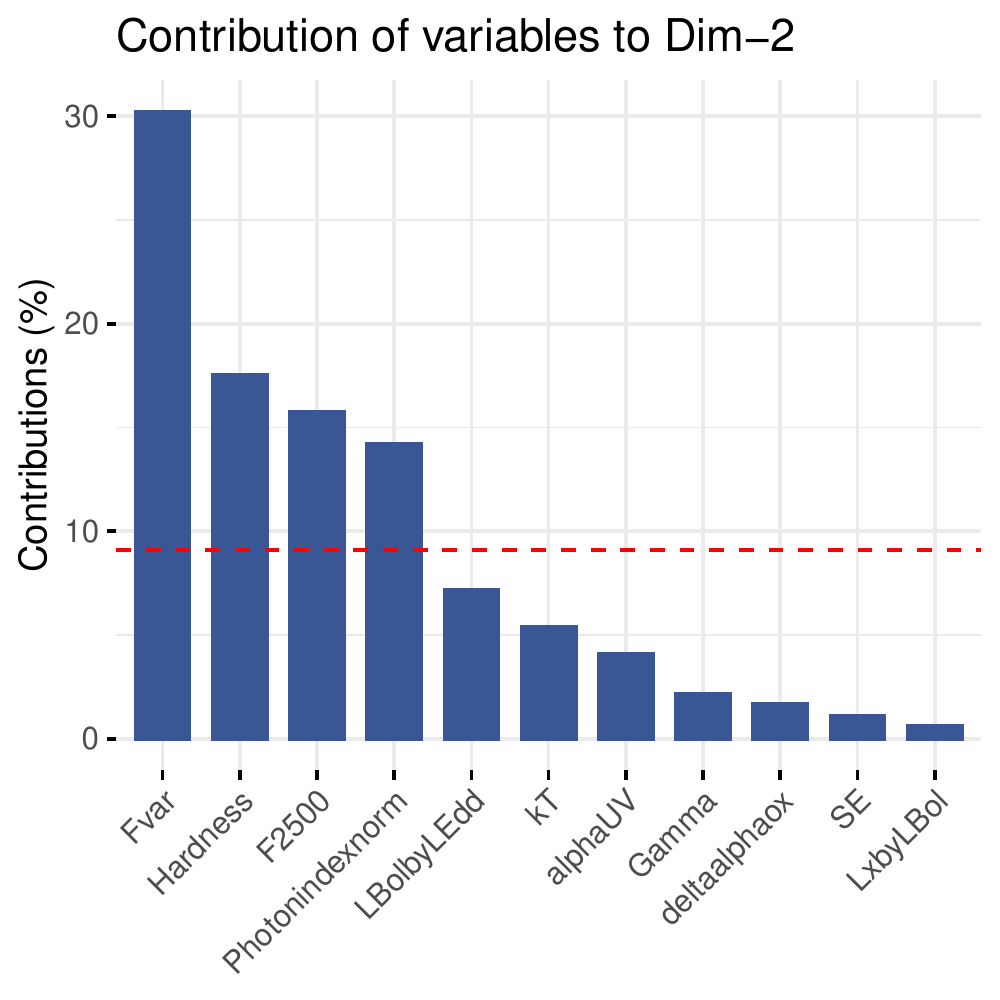}}} 
\hfill 
      {\scalebox{0.8}{\includegraphics[trim= 0cm 0cm 0cm 0cm, clip=true]{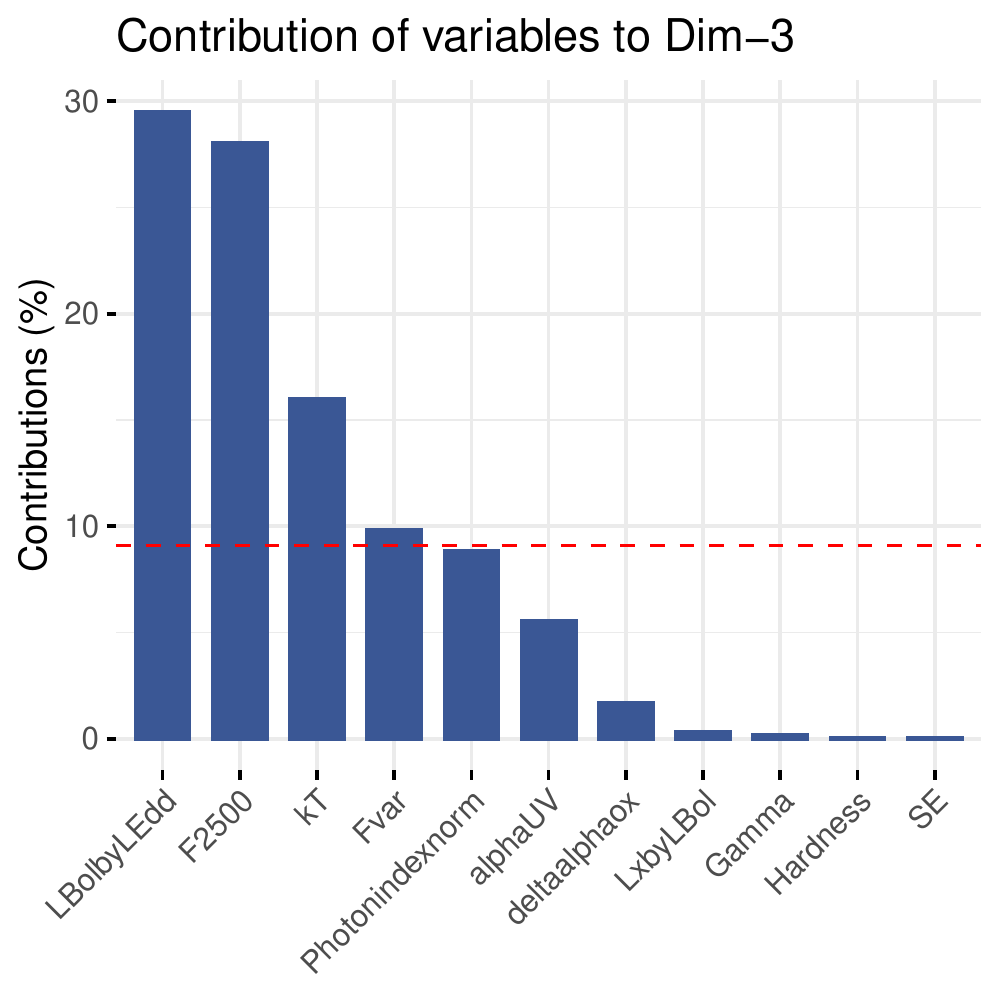}}}
\hfill 
   \caption{Upper Left: shows the screeplot from the Principal Component Analysis on all the parameters presented in Table~\ref{tab:tab1} which is the plot of eigenvalues ordered from largest to the smallest. The number of components are determined at the point, beyond which the remaining eigenvalues are all relatively small and of comparable size, Upper right figure shows the bar plot of contributions from all the variables to Principal Component 1. The red dashed line on the graph above indicates the average of contribution from all the parameters. Figures on the lower left and lower right represent the bar plots of contributions of variables to Principal Components 2 and 3, respectively.}
\label{fig:pca}
\end{figure*}

Principal component analysis (PCA) \citep{Francis+1999} is a statistical method to determine relevant properties that explain the maximum amount of variability in the dataset. It is often useful to visualize complicated processes involving multiple physical parameters through PCA analysis. We performed PCA using function `FactoMineR' \citep{F} in `R' on the multi-wavelength data of \mrk335\ presented in Table~\ref{tab:tab1}, and the results are shown in Figure~\ref{fig:pca}. We have excluded few parameters such as F$_{2keV}$, F$_{varUV}$ (uncorrelated with parameters) and kT$_{norm}$ (redundant with $SE$) from the analysis. This is necessary to eliminate noise in the PCA output from the uncorrelated/redundant variables. Also, $\alpha_{ox}$ is replaced with $\Delta\alpha_{ox}$. An important point to note is that we have carried out this analysis on the parameters for the low extended 13-year period (Table~\ref{tab:tab1}, Year 1 to Year13) in order to investigate the underlying cause of variability within the monitoring data. 

\indent The first step in the PCA analysis is to determine the number of significant principal components in the data. This is done by generating a screeplot (upper left, Figure~\ref{fig:pca}) that plots the variances (in percentages) against the number of the principal component. With our dataset, this resulted in the three significant principal components that cumulatively account for $\sim$ 86$\%$ of variance in the data. Next step is to investigate these principal components and get information on the set of parameters that give rise to each one of them. 

\indent It is interesting to see that the first principal component has the most significant contributions (above the marked line in `red') from the X-ray parameters i.e. photon index $\Gamma$, soft excess ($SE$), X-ray weakness ($\Delta\alpha_{ox}$), and normalised X-ray luminosity ($L_{x}/L_{bol}$). The data are well described by this principal component which accounts for more than $\sim$ 60$\%$ variability (upper left, Figure~\ref{fig:pca}) through the X-ray parameters.

\indent The second principal component accounting for $\sim14$ per cent of the variability is further influenced by the X-ray parameters. The most important parameters here are the fractional variability (F$_{var}$) ($\sim$ 30$\%$), hardness ratio (18$\%$). F$_{2500}$ and photon index normalisation ($\Gamma_{norm}$) have marginal contributions (15$\%$). 

\indent The third principal component is driven significantly ($\sim$ 30$\%$ equal contribution) by L$_{bol}$/L$_{Edd}$ and the optical-UV parameter F$_{2500}$. This principal component carries about $\sim$ 12$\%$ variability of the data.

\indent The fourth principal component has marginal contribution ($\sim$ 6$\%$) to explain the variability of data through ultraviolet parameter alone ($\alpha_{uv}$).

\indent This analysis is a statistical, model-independent approach to understand the significance of the multitude of parameters describing the dataset in terms of variability. However, it is subject to limitations in terms of physical meaning to the outcomes of this analysis. The analysis suggests clearly the dominance of X-ray properties of the source towards the primary variability ($\sim$ 70$\%$), as also suggested by correlation analysis.  Less important are the changes in the optical-UV properties of the AGN  ($\sim$ 8-10$\%$). 

\begin{figure*}
   \centering
   \advance\leftskip-0.cm
      {\scalebox{0.8}{\includegraphics[trim= 0cm 0cm 0cm 0cm, clip=true]{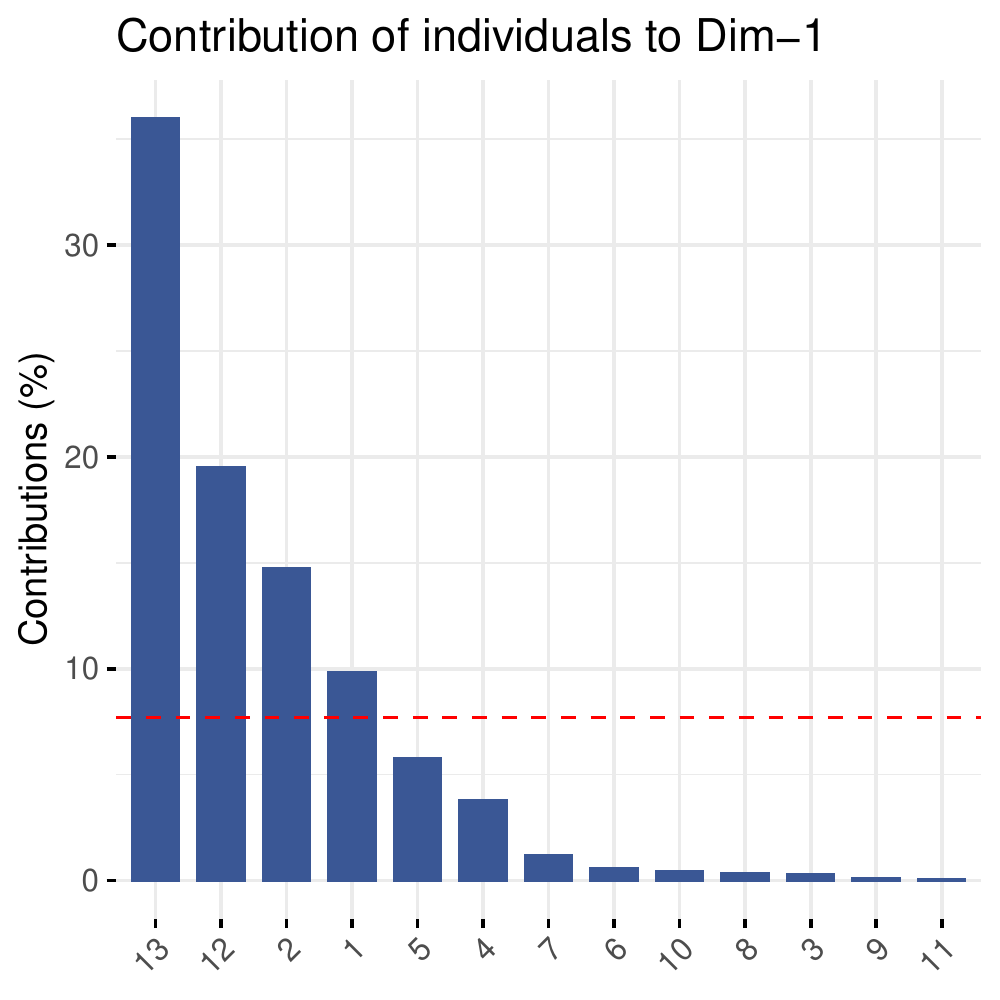}}}  
\hfill
      {\scalebox{0.8}{\includegraphics[trim= 0cm 0cm 0cm 0cm, clip=true]{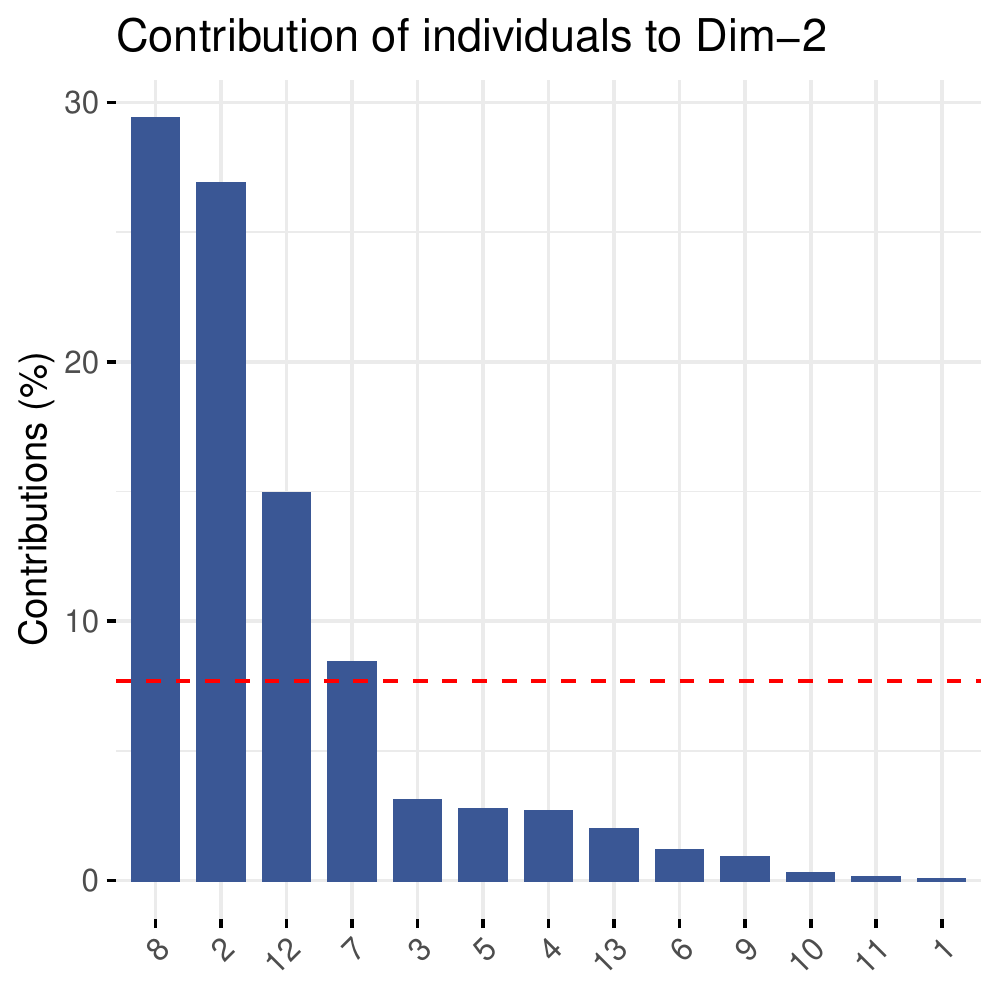}}} 
      {\scalebox{0.8}{\includegraphics[trim= 0cm 0cm 0cm 0cm, clip=true]{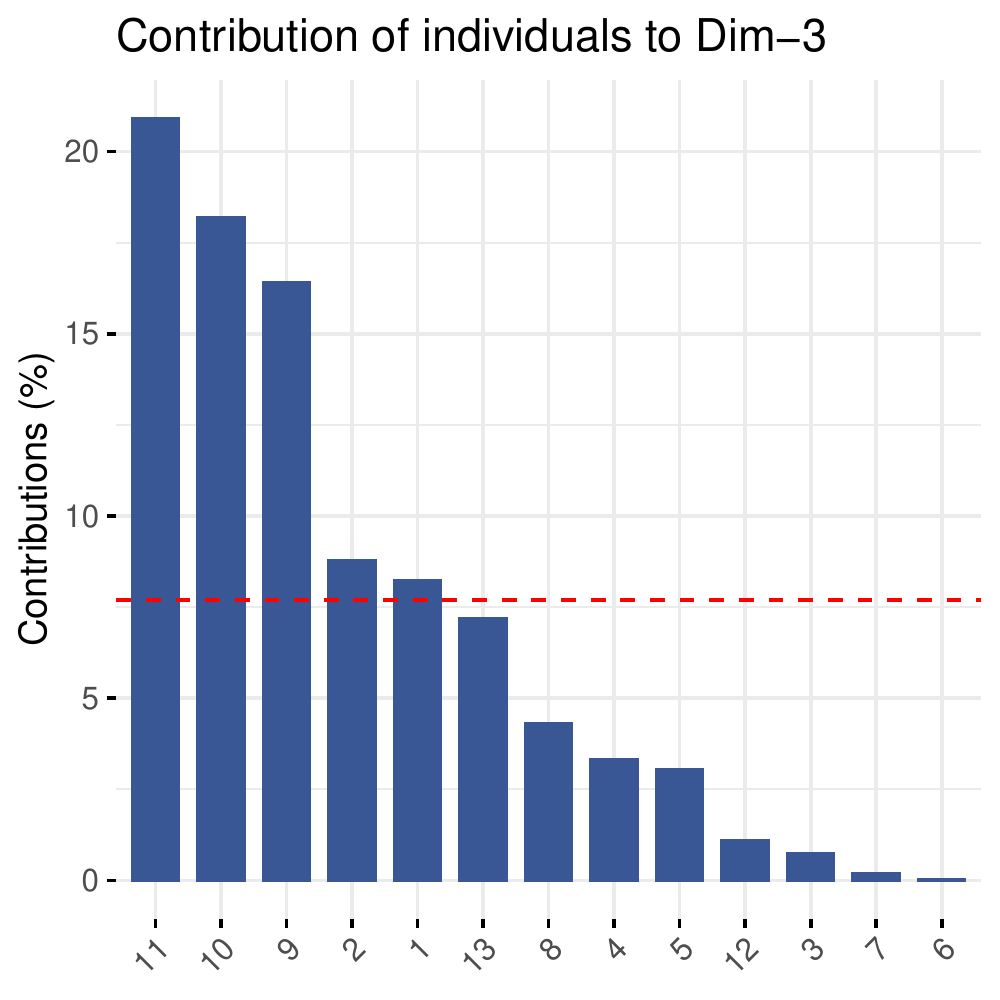}}}  
\hfill
   \caption{Upper Left: shows the contribution of individual years to the first Principal Component. The red dashed line is the reference line corresponding to the expected value if the contributions were uniform. The upper right and lower panels shows the specific years significantly contributing to the second and third principal components, respectively.}
\label{fig:indexpca}
\end{figure*}

\indent In order to further explore the role of years or epochs that possibly drive these principal components, we made the contribution plots as shown in the Figure~\ref{fig:indexpca}. The first plot (upper right) reveals the individual years that highly contributed towards the variability exhibited by the PC1. Year 13, Year 12, Year 2 and Year 1 show the significant contribution to PC1. The second principal component has significant contribution from Year 2 and Year 8 ($\sim$ 30$\%$). The third principal component reveals the epochs marked as Year 11, 10 and 9 in the order of their contributions. It is interesting to speculate that the most significant X-ray changes occurred in early years (e.g. 1 and 2), but the significant UV changes occur in later year (e.g. 9, 10, and 11). This could be suggesting of a driving mechanism if we believe the X-ray changes could have resulted in later UV changes.

\indent These results can be interpreted such that the most significant variations in the properties of the source seem to have taken place in Year 2, 8, 9, 10, 11, 12 and 13. Some of these epochs have deep observations to help understand the physical driver of variability. For instance, Year 12 and 13 are peculiar as recent observations with \xmm/\hst\ campaign have suggested the scenario of strong UV absorption \citep{Parker+2019}. Year 2 stands out as discussed in Section~\ref{xray} with respect to its unique X-ray flaring behaviour compared to other epochs (Figure~\ref{fig:corr1}) and also is marked by weaker UV absorption \citep{Longinotti+2013}. Year 8 is also the X-ray flaring year attributed to collimated, relativistic outflow that could also produce UV emission \citep{Wilkins++2015}.  It is interesting that Year 9 and Year 10 stood out in PCA results despite excluding F$_{varUV}$ from this analysis. Year 9 have shown maxima similar to Year 4 in Figure~\ref{fig:corr3} (top, left).

\indent We considered the effect on the PCA results of omitting year 12 and 13 since the X-ray data were of low counts during these epochs. There was a big change in year 12, 13 corresponding to another drop in flux and the source became X-ray weaker compared to previous years.  Excluding these years, the PCA then emphasises the X-ray flaring years (Year 2 and 8) and the UV flaring Year 4. Even though these later years have lower quality X-ray data, their contribution to the 13-year PCA is still important to include.  

\section{Discussion} 
\label{sect:disc}
Our study of the spectral energy distribution of \mrk335\ using multi-wavelength data has led to the annual estimation of bolometric luminosity and luminosity Eddington ratio. A detailed correlation analysis was performed between the measured X-ray and UV parameters spanning the monitoring period. An interesting finding from our comprehensive correlation study is that nearly all these correlations have been shown in previous studies on different samples of AGN. This is remarkable as it suggests the similar physics at work for an individual source like, \mrk335\ and in other AGNs as well.  

\indent An interesting behaviour can be seen for the soft excess emission of \mrk335\ over the monitoring period, which is found to be highly correlated with many parameters. For spectral analysis, we have used a thermal blackbody to model the thermal emission from an optically-thick accretion disc to a first order approximation. Although constraining the temperature of the soft excess using a thermal blackbody is not physically meaningful, it still provides a helpful measurement that can be utilized for comparison with other data. The blackbody kT has weak dependance on the black hole mass, thus, ruling out a thermal origin from the accretion disk. However, invoking the atomic physics could still explain the values taken by kT, for instance, in the context of strong, relativistically blurred absorption from a disc wind \citep{Gierli+2004}, kT will still reflect whether the ionized gas is lowly or highly ionized. Similarly, in the case of blurred ionized reflection which is sensitive to the ionization state of the reflecting material \citep{Crummy+2006}, kT could be a useful indicator.\newline
\indent For the 13-year monitoring data, kT shows variability, more prominently for the last two epochs. This would then imply that there are other additional processes at work. So, it is still a meaningful parameter, but for different reasons other than measuring a disc temperature.

\indent At the same time, it is to be noted that accurate constraints of parameters e.g. $\Gamma$ and kT become difficult in the low count statistics regime; at least, such is the case for the past two years with measurements of higher blackbody temperatures ($\sim$140 eV) and flatter photon indices ($\sim$0.6). It is interesting to note that the power-law shape, pertaining to the flatter photon indices, might be indicative of the onset of other processes. Nevertheless, several correlations of soft excess found in our study might give some clues to compare the physical processes weaved into realistic scenarios (e.g. blurred reflection, soft Comptonisation).

\indent Blurred reflection model can explain the observed correlation of soft excess with photon index and F$_{var}$. This scenario occasionally implies a strong suppression of the intrinsic power-law continuum to account for a rather pronounced soft excess, as justified in the context of strong light-bending effects \citep{Miniutti+2004}. An increase in the strength of soft excess emission could arise as a result of an increase in ionisation parameter and/or reflection fraction. The overall effect would be a suppressed primary power-law continuum (implying reduced variability) and an enhanced soft excess (i.e. reflected emission) \citep{Crummy+2006}. This can naturally predict the observed correlation between soft excess and photon index. 

\indent Next, the observed X-ray variability could be driven by the changes in the primary power-law continuum possibly manifested through intrinsic variations in the corona, or possible changes in its size or location throughout the monitoring period. From our analysis, we find F$_{var}$ to be inversely correlated to blackbody temperature and positively correlated to blackbody normalisation. This would suggest a suppressed power-law thereby, resulting into reduced variability. Such a scenario could possibly mimic the observed correlation between F$_{var}$ and kT. 

\indent Alternatively, in the framework of soft Comptonisation, we find a tight relationship between the soft excess and the photon index and no inverse correlation with the X-ray luminosity. A part of this enhanced soft excess emission would undergo upscattering by the hot electrons in the corona to produce the primary X-ray power-law continuum. This eventually cools off the corona, and could possibly explain the correlation between the photon index and the soft excess strength \citep{Done+2012}. Both these scenarios can explain some of the observed correlations found in this study.

\indent If we take a closer look at the SED analysis, the differences in the SED shape over these thirteen years show a systematic downward trend. Such a systematic downward trend has been investigated on the optical photometric data of \mrk335\ during 1995-2004 by \citet{Doroshenko+2005}. They found that \mrk335\ exhibit large amplitude variability reaching about 1.1, 0.9, 0.7, 0.3 and 0.3 mag in the U, B, V, R and I bands, respectively. The brightness was systematically decreasing since 1995 by about a factor of 2.5, and the variations of different amplitudes and duration were suggested. In addition, the shape of the spectral energy distribution remained constant in spite of these flux-variations.

\indent This is quite interesting as the source continued to dim in the optical over a timescale of 10 years. The abrupt change in the X-ray brightness occurred after two years of this study i.e. in 2007. It is relevant to understand the scenarios that can explain the timescales of the observed variations in this source over the long term. For the optical emission region, the timescale for the change in the accretion rate is the viscous timescale, which by considering the black hole mass of our source and assuming conservative value of disc aspect ratio h/r $\sim$ 0.01, should be tens of thousands of years, which is much longer than our current observations. However, timescales due to local temperature fluctuations i.e. thermal timescale can give rise to optical/UV variations on much shorter timescales \citep{Kelly+2009}. Despite the different timescales involved, it might be possible that the abrupt change in the X-ray flux and the more gradual change in the optical emission regions of the source are possibly connected through some process. For instance, \citet{Sun+2020} suggest a model which show that the magnetic coupling between the compact corona and the outer cold accretion disc might exist. However, based on the pre-2007 X-ray lightcurve (e.g. \citealt{Grupe+2007}), that suggests that despite the changes in the optical flux, the source probably stayed relatively normal when compared to $\alpha_{ox}$. We see that in 2006, after the changes in the optical presumably occurred, \mrk335\ was just a `normal' AGN.  

\indent The present monitoring data however, suggest relatively significant changes ascribed to the UV and X-ray spectral behaviour which is intrinsic to the central engine. The changes in the optical emission are relatively low. During the prolonged low X-ray flux period, the changes in the SED shape are found to be more dramatic for the past two years. Both obscuration and the intrinsic changes linked to either the inner disc or corona could be playing vital roles in \mrk335, while it exhibits intermittent phases of dimming and flaring over the 13-year timescale as indicated by recent studies \citep{Gallo+2019, Longinotti+2019, Parker+2019}. Regarding the changes in the inner disc; this could possibly be triggered by disc instabilities, or due to the presence of local perturbers or more distant changes in the accretion flow \citep{Stern+2018}.

\indent To investigate the observed long-term X-ray variability, one must consider the changes in the inner accretion disc, where the relevant timescales are the orbital, thermal, and the viscous timescales. \citet{Gallo+2018} have investigated these timescales previously in the context of multi-band structure functions of \mrk335. Based on a conservative value of viscosity parameter $\alpha_{v}$ $\sim$ 0.1 and disc aspect ratio h/r $\sim$ 0.01 in the standard accretion disc scenario, the characteristic timescales in the UV and optical wavebands could be attributed to the thermal and orbital timescales respectively.

\indent Another important timescale i.e. cooling and heating front timescale due to viscous effects in the accretion disc was not discussed in the previous study, possibly, because this scenario inherently proposes the characteristic timescales that are too long. In this scenario, a cooling front propagates outward in the disc and reflects back inward as a heating front and has successfully explained the dimming character of few changing look AGN (e.g. \citealt{Stern+2018, Ross+2018}). As such, these fronts propagate in the disc on timescales of $t_{front}$ $\sim$ $(h/r)^{-1}~t_{th}$ where $t_{th}$ is the thermal timescale. It remains a possibility that a revision of some of these parameters e.g. $\alpha_{v}$ might make the cooling/heating front timescale \citep{Hameury+2009} relevant in the context of \mrk335. Considering $\alpha_{v}$ $\sim$ 0.4 and h/r $\sim$ 0.2, we find that the $t_{front}$ $\sim$ 120 days is consistent with the longer break timescale which was observed in the 11-year averaged UVW2 structure function whereas the shorter break timescale could be explained by the thermal timescale $t_{th}$ $\sim$ 24 days. Also due to the quadratic dependance on h/r, $t_{vis}$ would also drop significantly.\\
\indent Assuming an albeit higher value of $\alpha_{v}$, as found by some studies (e.g. \citealt{King+2007, Tetarenko+2018}), implies more turbulence acting over the neighbouring disc annuli, thereby, leading to an increase in the propagation of fronts and probably resulting in the increased disc emission. Future work with detailed spectral modeling on these data might shed some light on the value of $\alpha_{v}$ to understand the more plausible timescale or mechanism in the inner accretion disc. 

\section{Conclusions} \label{sect:conc}
\mrk335 was once one of the brightest AGN in X-rays, allowing detailed observations even when it went into its deep low-state. We have presented the simultaneous optical-to-X-ray SEDs of \mrk335\ from the 13-year \swift\ monitoring data. For comparison, the SED for representative bright state of the source is also presented. The SEDs span the infrared, optical, ultra-violet and X-ray regimes ($\approx$ 1$\upmu$m to 2 keV) using a simple empirical model. The X-ray part of the SED is determined separately by incorporating the integrated flux between 0.3-2 keV, duly estimated from the simple X-ray spectral model. By employing the 13 epochs of multi-wavelength XRT, UVOT photometric and archival optical spectroscopic data, and reverberation mapped mass estimate of \mrk335, this work presents the following main outcomes:
\begin{enumerate}
\item{We see a gradual decreasing trend in the yearly SED models. The most prominent change in the SED is evident during the bright state transition of the source to the extended low-flux period and then during the past two years where a clear dip in the SED model can be noticed.}
\item{Bolometric Eddington luminosity ratio changes significantly over the 13-year period (full observed range is 0.1-0.07). For the bright state, it is estimated to be $\sim$ 0.2.}
\item{Relatively small changes in the optical/UV over 13-year data are found as compared to more significant changes in the X-ray flux, spectral shape and variability. The total amplitudes of variability observed in UV and X-rays are ($\sim$ 6-25$\%$) and ($\sim$ 70-110$\%$) respectively. In fact, when compared to other AGN, \mrk335 is still remarkably variable in optical/UV. But relatively rapid and higher amplitude of variability observed in X-rays rather reinforces the preference for structural or intrinsic changes in the corona and/or absorption to explain the observations.}
\item{We tested the utility of these yearly empirical SEDs by measuring the useful parameters in the optical/UV and X-rays e.g. L$_{bol}$/L$_{Edd}$, $\alpha_{uv}$. We, then performed correlation and statistical analyses on these estimated SED parameters alongwith the X-ray spectral parameters.}
\item{Our comprehensive correlation study yields several significant correlations amongst parameters, most notably, for soft excess emission and L$_{bol}$/L$_{Edd}$. These relationships have been established in previous studies of other AGN and therefore, suggest that similar physical processes apply, in general, to a majority of AGNs. We found a significant correlation between the soft excess emission and the photon index, but  no inverse correlation with X-ray luminosity. This might suggest the possibility of the role of soft Comptonization scenario to explain the long-term X-ray data. }
\item{Principal component analysis on this multi-wavelength dataset reveals higher statistical significance of the X-ray driven variability as compared to the optical/UV variability. The variability shown by PCA are mostly contributed by the X-ray flaring epochs and the recent X-ray dimmest years.}
\item{We have demonstrated that the simultaneous optical-UV and X-ray monitoring data of \mrk335\ continue to provide opportunities for understanding the physical characteristics of this source, particularly, in this work through the determination of SEDs and the statistical analyses. The detailed spectral modeling of these empirical SEDs with regards to the physical scenarios remains an interesting subject for the future work.}
\end{enumerate}

\section*{Acknowledgments}
We thank the \swift\ team for approving our various ToO requests to monitor \mrk335\ over the years. Many thanks to the anonymous referee for providing a thorough report that improved the paper. L.C.G. acknowledges financial support from the Natural Sciences and Engineering Research Council of Canada and the Canadian Space Agency. ALL acknowledges support from CONACyT grant CB-2016-01-286316. This work is partially based on observations made with the Galileo 1.22m telescope of the Asiago Astrophysical Observatory operated by the Department of Physics and Astronomy "G. Galilei" of the University of Padova. This research makes use of facilities at the Mets\"ahovi Radio Observatory, operated by the Aalto University, Finland. We thank B. M. Peterson for providing the optical data from MDM Observatory, Kitt Peak, Arizona.

\section*{Data availability}
The raw X-ray data used in this work are publicly available in the \xmm\ \url{https://www.cosmos.esa.int/web/xmm-newton/xsa} and \swift\ \url{https://www.swift.ac.uk/swift_portal/}. Optical data used in this work are available upon request by the authors.
\bibliographystyle{mnras}
\bibliography{myref}

\begin{appendix}
\section{Spectral Energy Distributions of individual epochs}
\label{sect:app2}
\begin{figure*}
\includegraphics[width=8.5cm, trim= 0cm 0cm 0cm 0cm, clip=true]{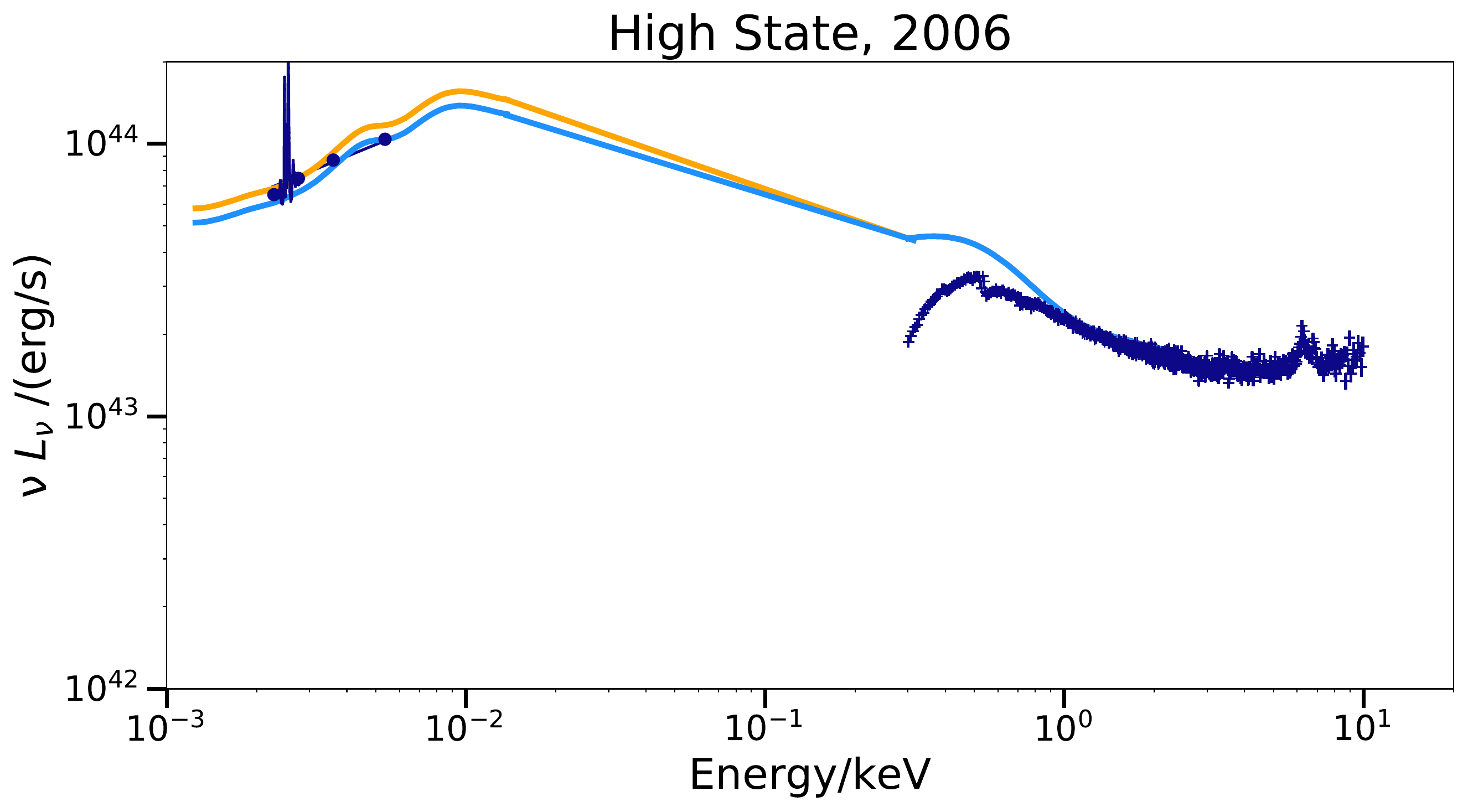}
\hfill
\includegraphics[width=8.5cm, trim= 0cm 0cm 0cm 0cm, clip=true]{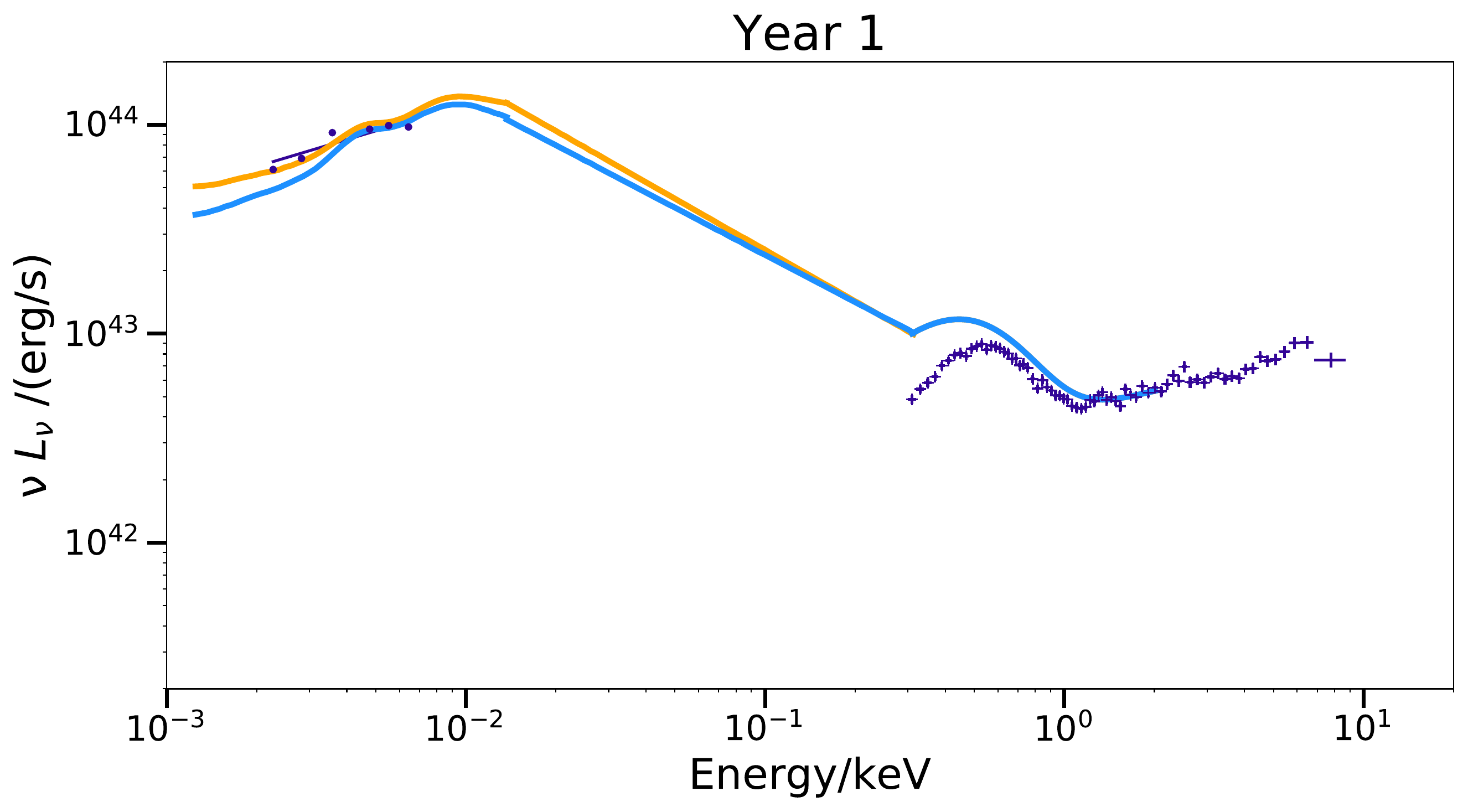}

\vspace{2.00mm}

\includegraphics[width=8.5cm, trim= 0cm 0cm 0cm 0cm, clip=true]{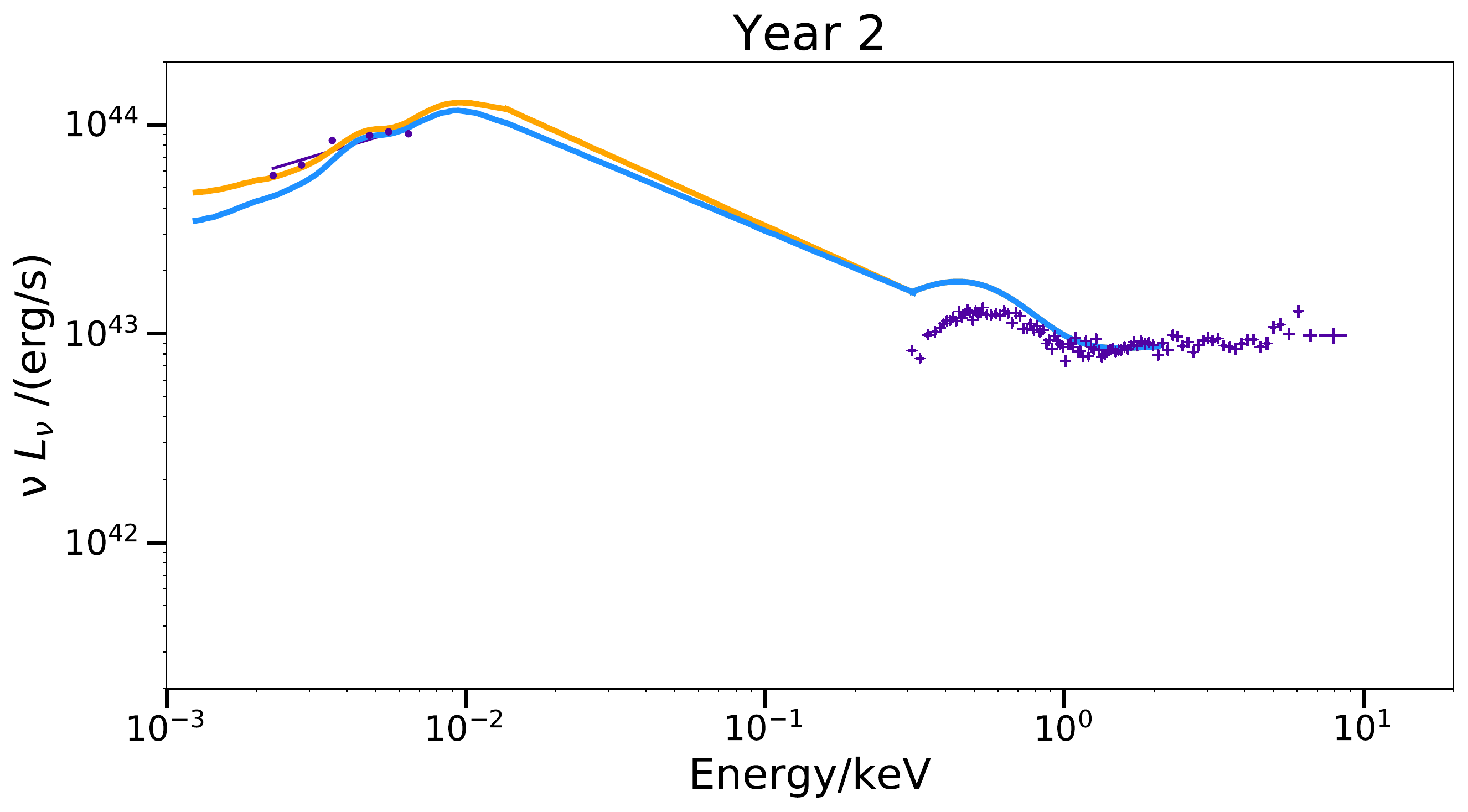}
\hfill
\includegraphics[width=8.5cm, trim= 0cm 0cm 0cm 0cm, clip=true]{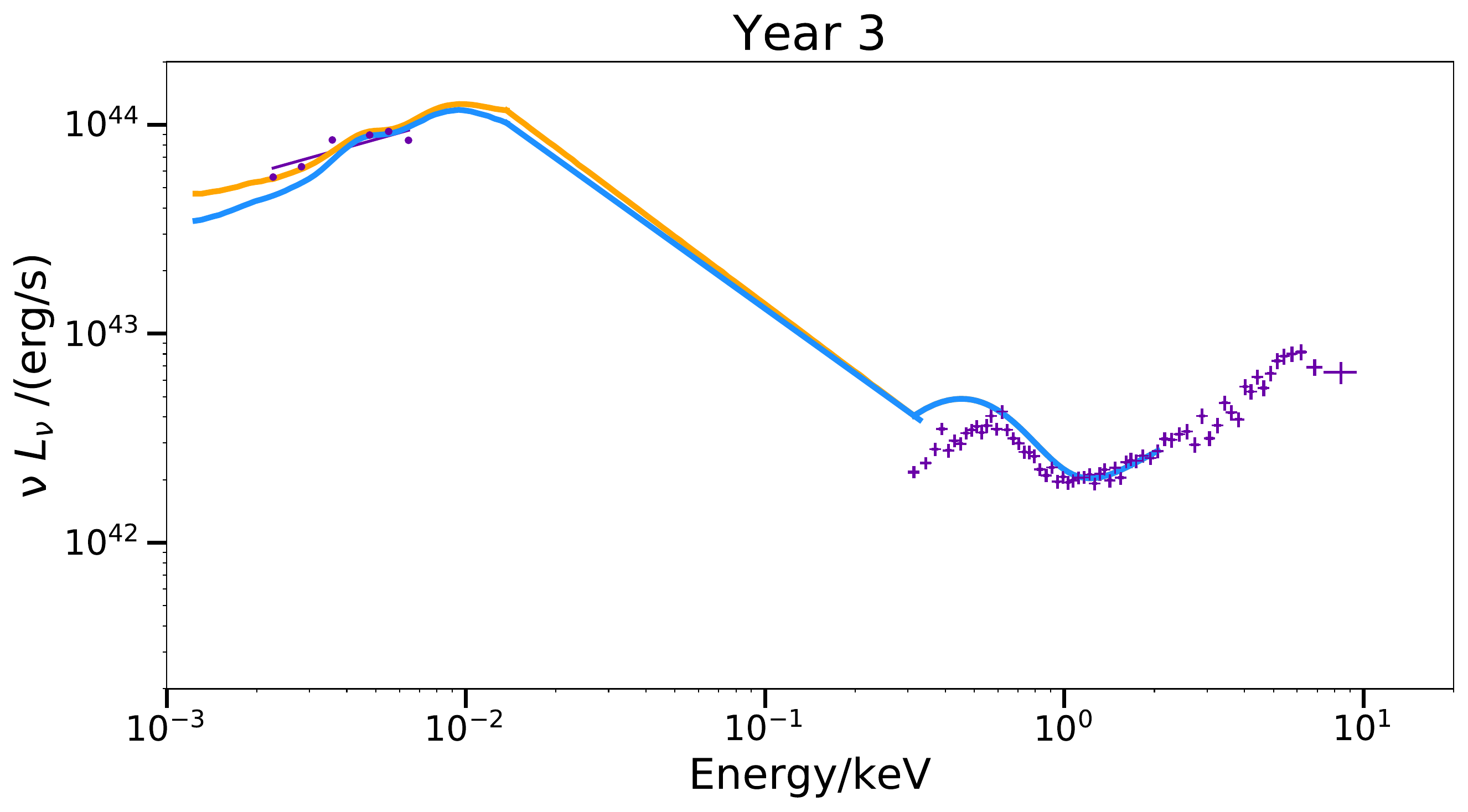}

\vspace{2.00mm}

\includegraphics[width=8.5cm, trim= 0cm 0cm 0cm 0cm, clip=true]{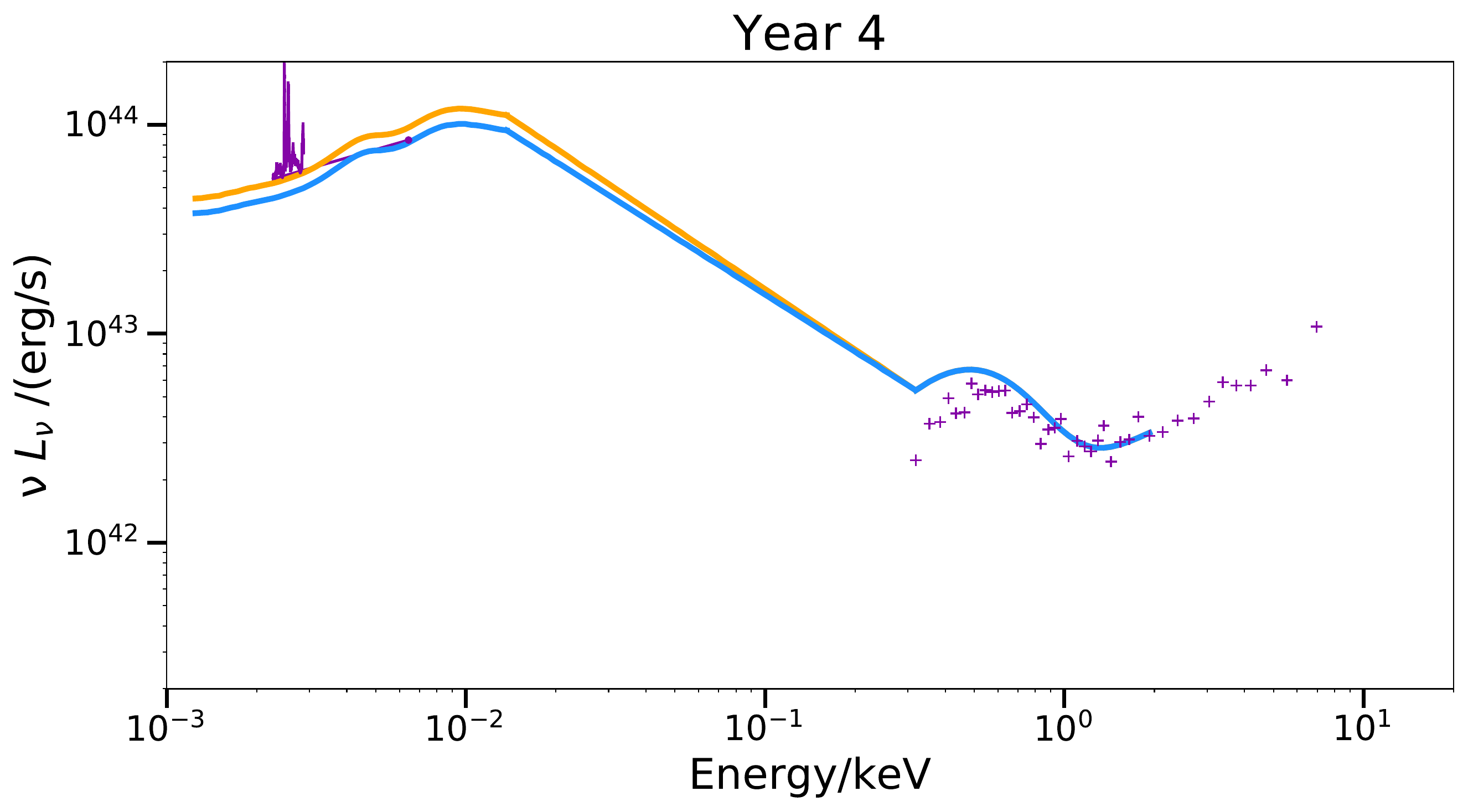}
\hfill
\includegraphics[width=8.5cm, trim= 0cm 0cm 0cm 0cm, clip=true]{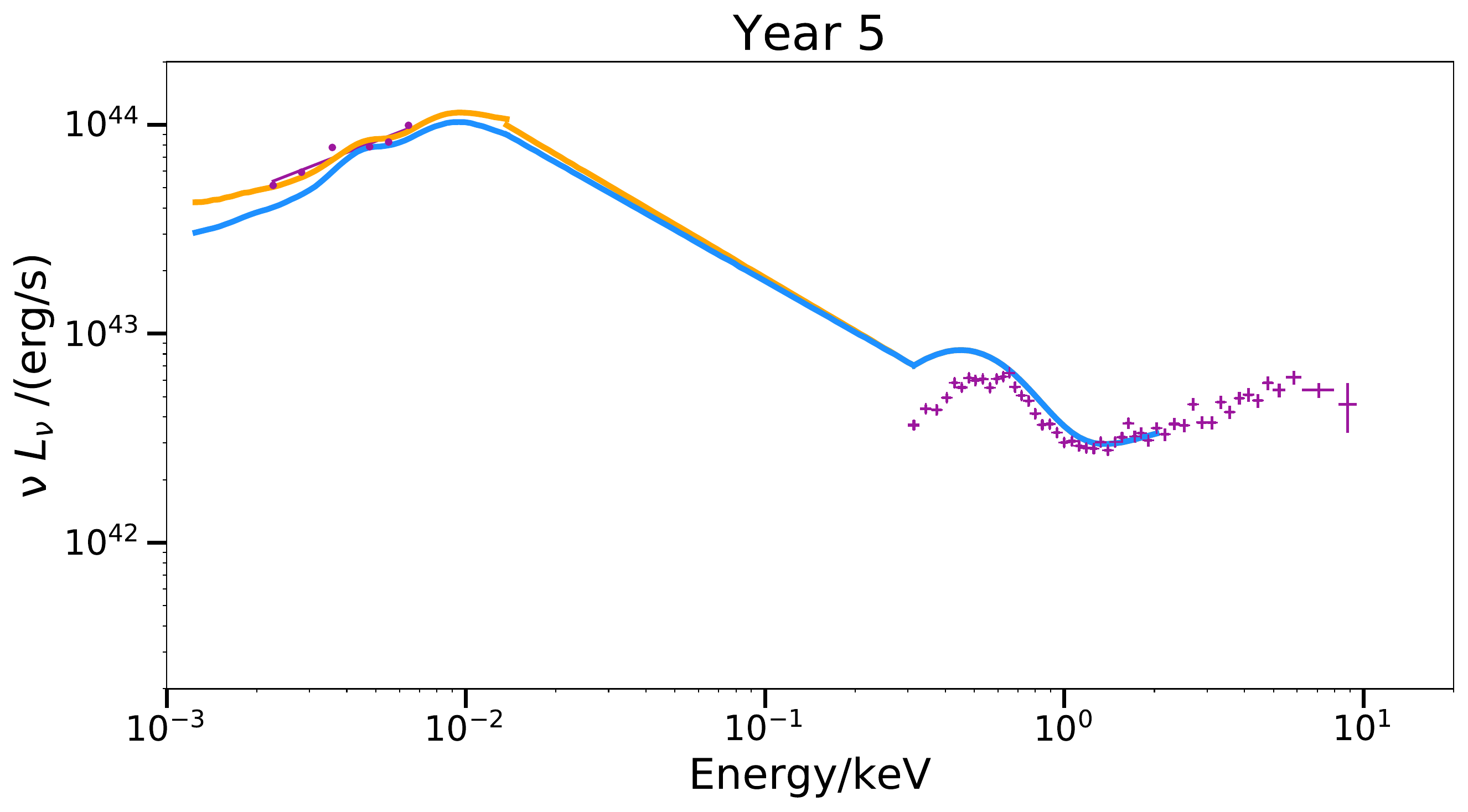}

\vspace{2.00mm}

\includegraphics[width=8.5cm, trim= 0cm 0cm 0cm 0cm, clip=true]{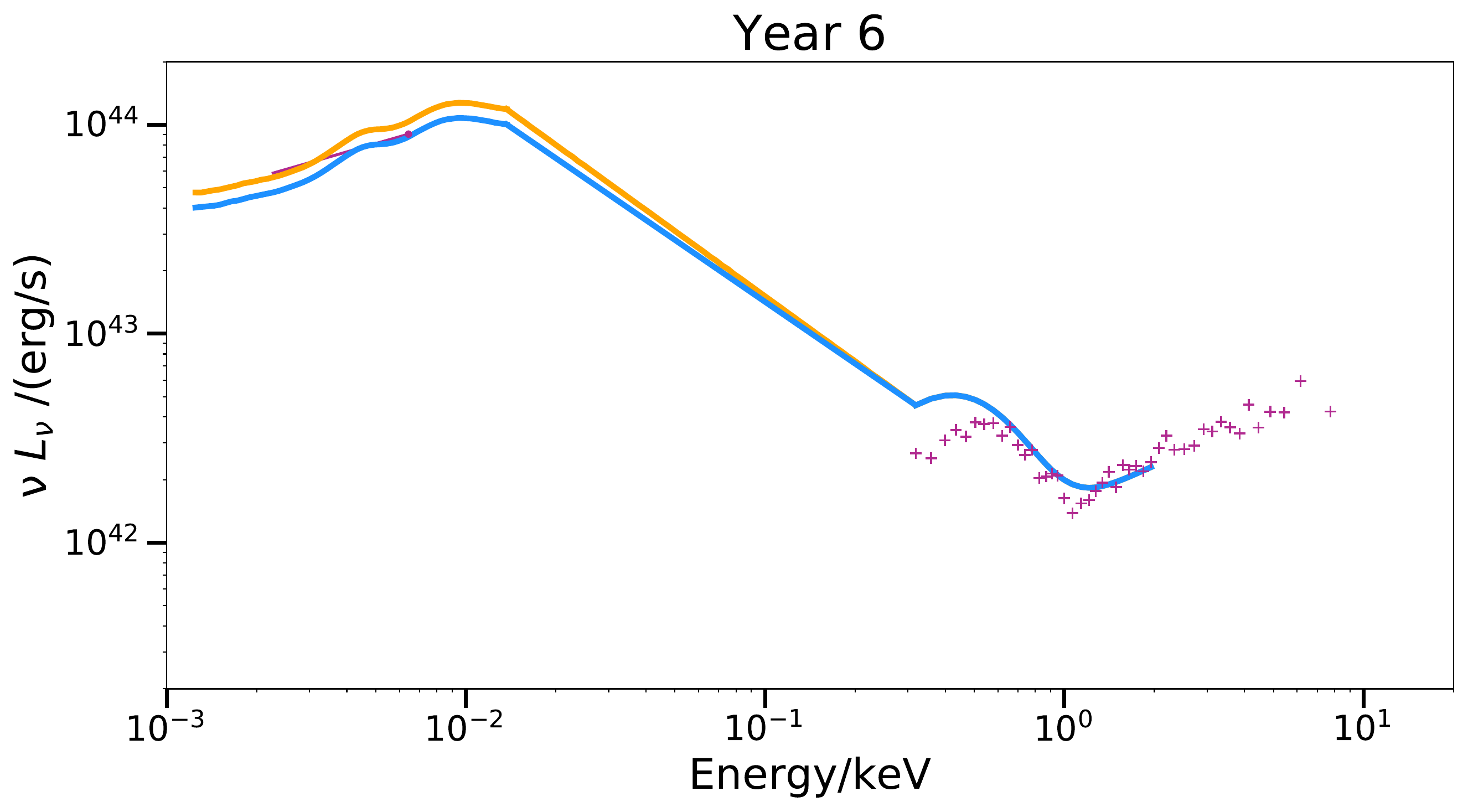}
\hfill
\includegraphics[width=8.5cm, trim= 0cm 0cm 0cm 0cm, clip=true]{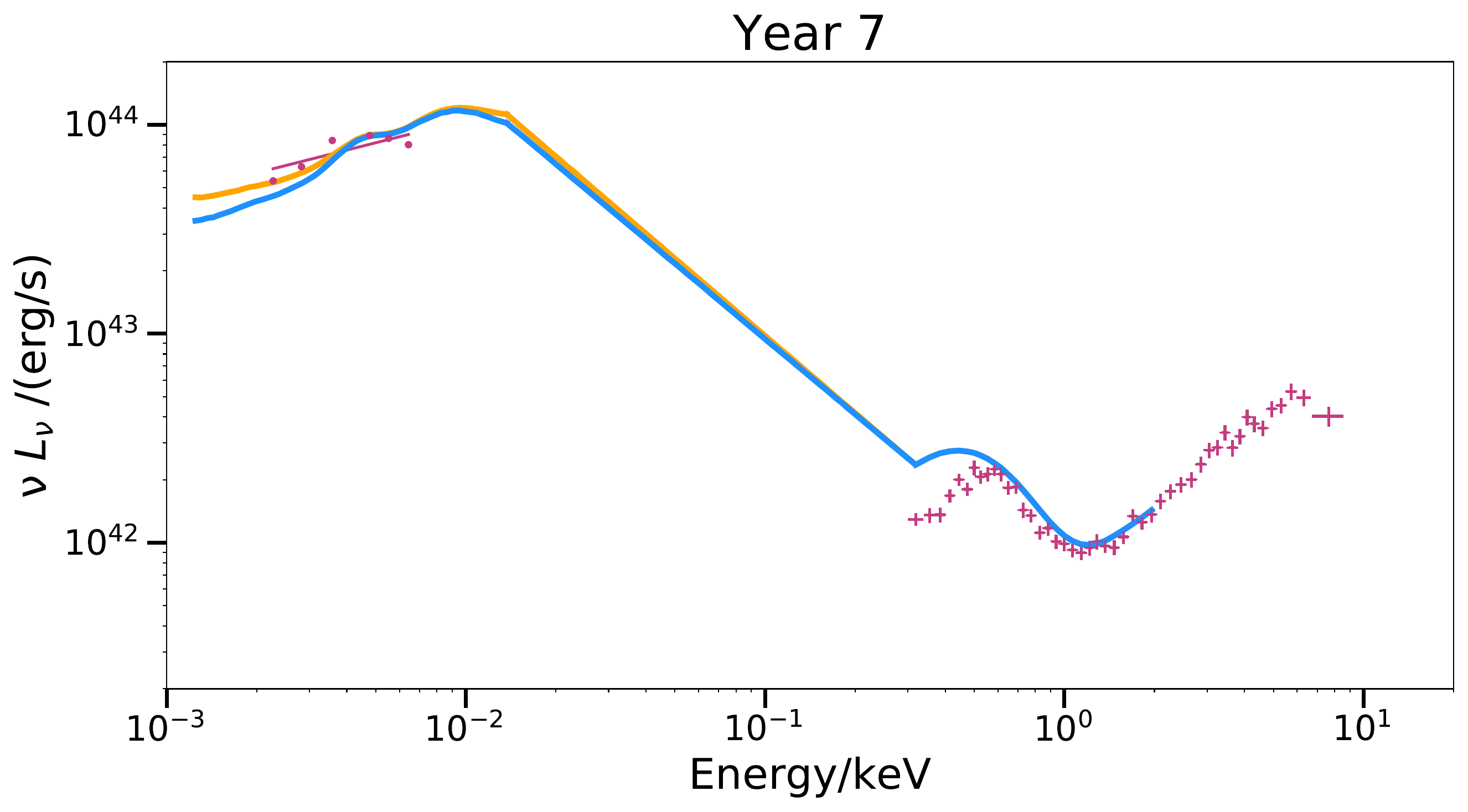}
\caption{Spectral energy distribution measurements performed on individual epoch of 13-year monitoring period of \mrk335\ are displayed. The modified Krawczyk SED, as described in Section~\ref{sect:sed}, is applied to the data in each year of the 13-year low state. The solid lines in yellow (with respect to 5100 \AA\ normalization) and blue (with respect to 2500 \AA\ normalization) show the integrated model used to evaluate bolometric luminosity for each epoch. Colours used for data in each SED diagram are consistent with Figure~\ref{fig:sed0}.}
\end{figure*}

\setcounter{figure}{0}

\begin{figure*}
\includegraphics[width=8.5cm, trim= 0cm 0cm 0cm 0cm, clip=true]{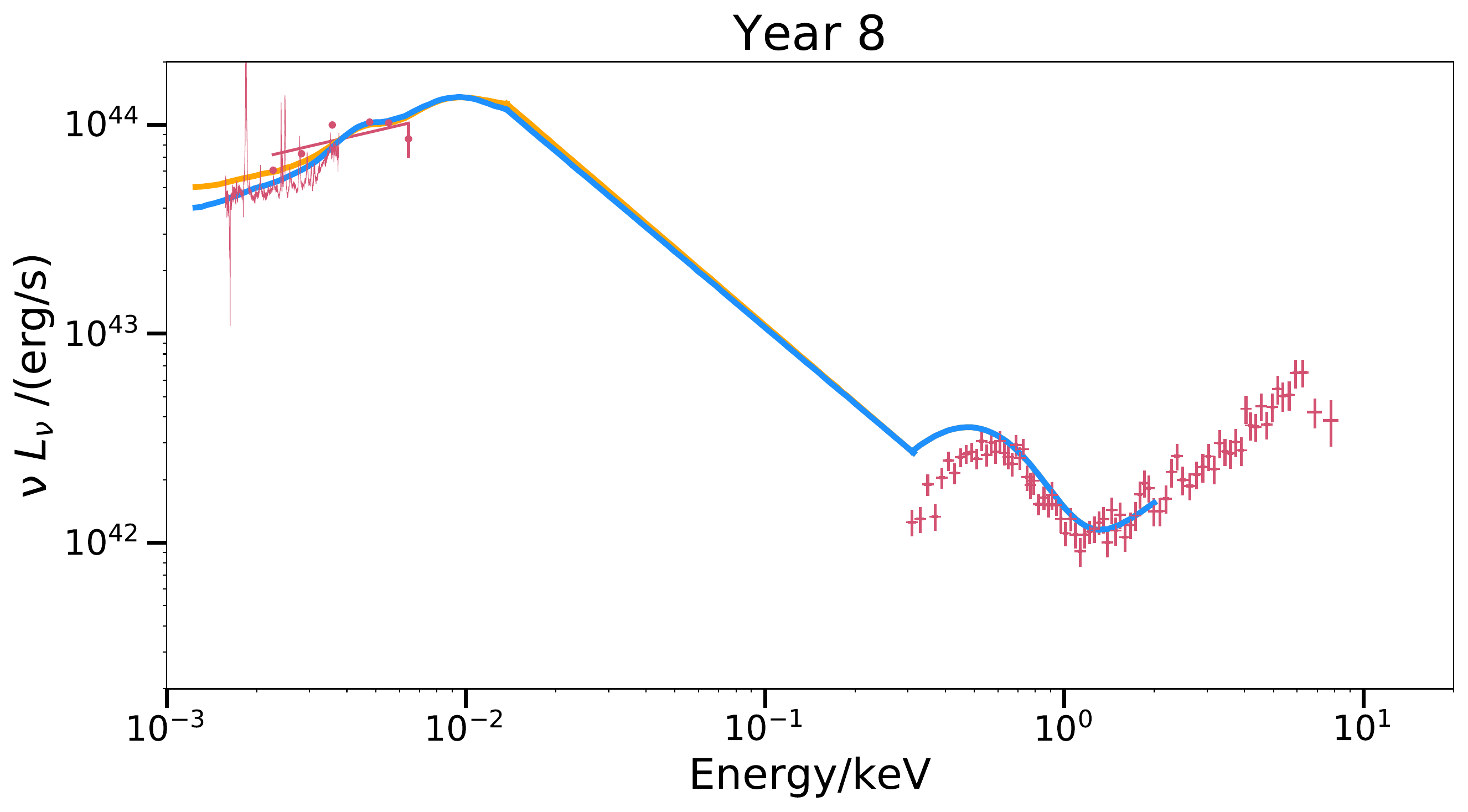}
\hfill
\includegraphics[width=8.5cm, trim= 0cm 0cm 0cm 0cm, clip=true]{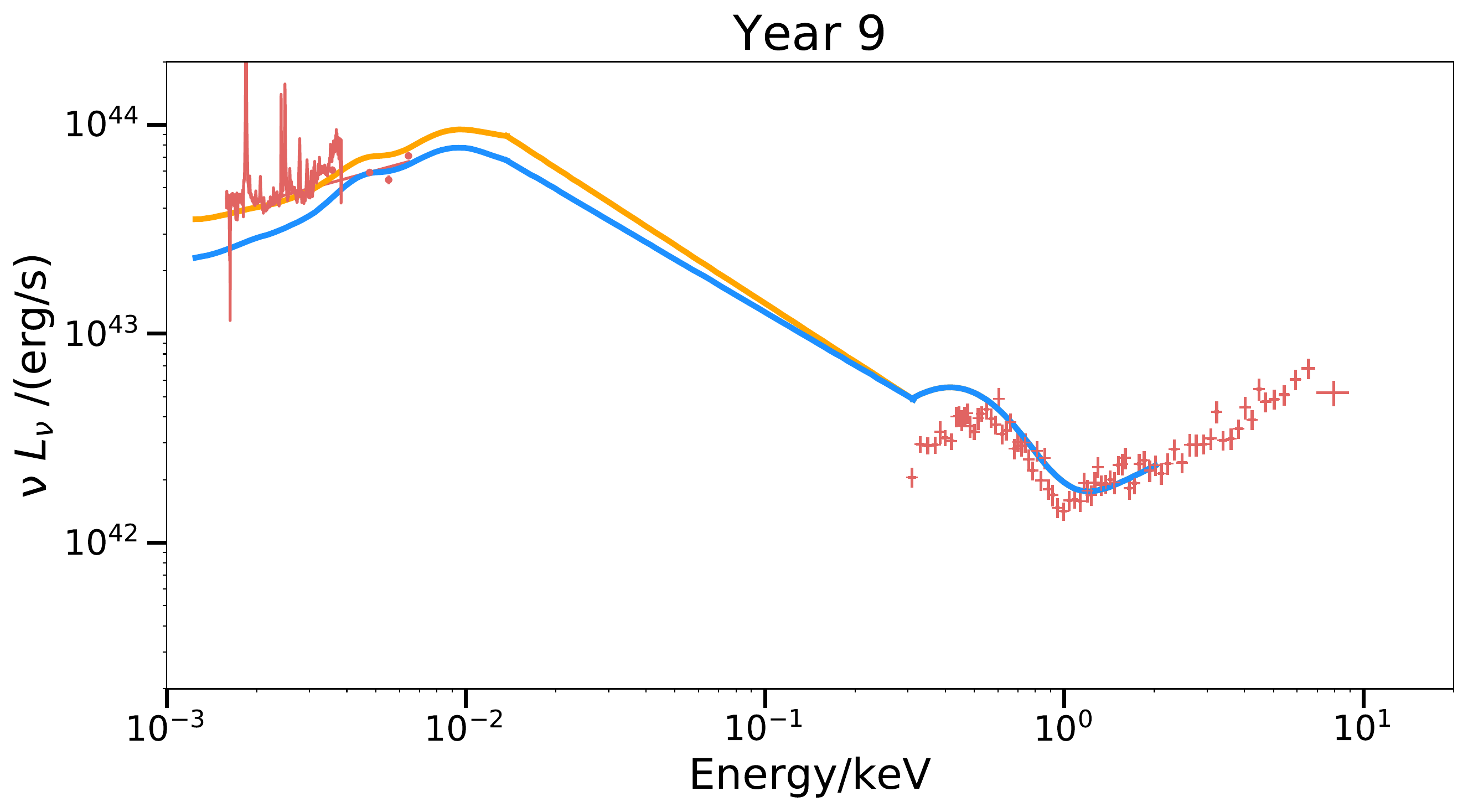}

\vspace{2.00mm}

\includegraphics[width=8.5cm, trim= 0cm 0cm 0cm 0cm, clip=true]{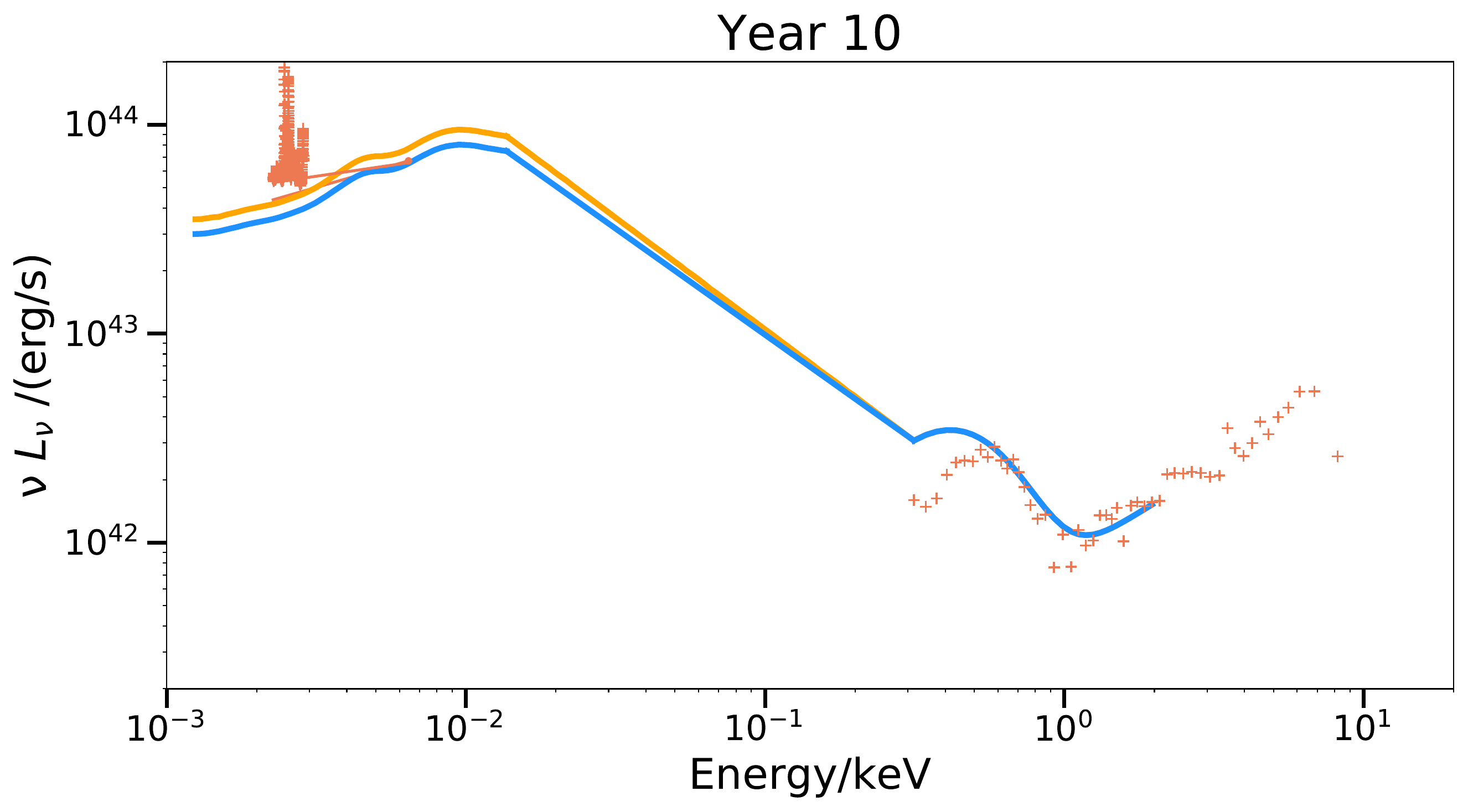}
\hfill
\includegraphics[width=8.5cm, trim= 0cm 0cm 0cm 0cm, clip=true]{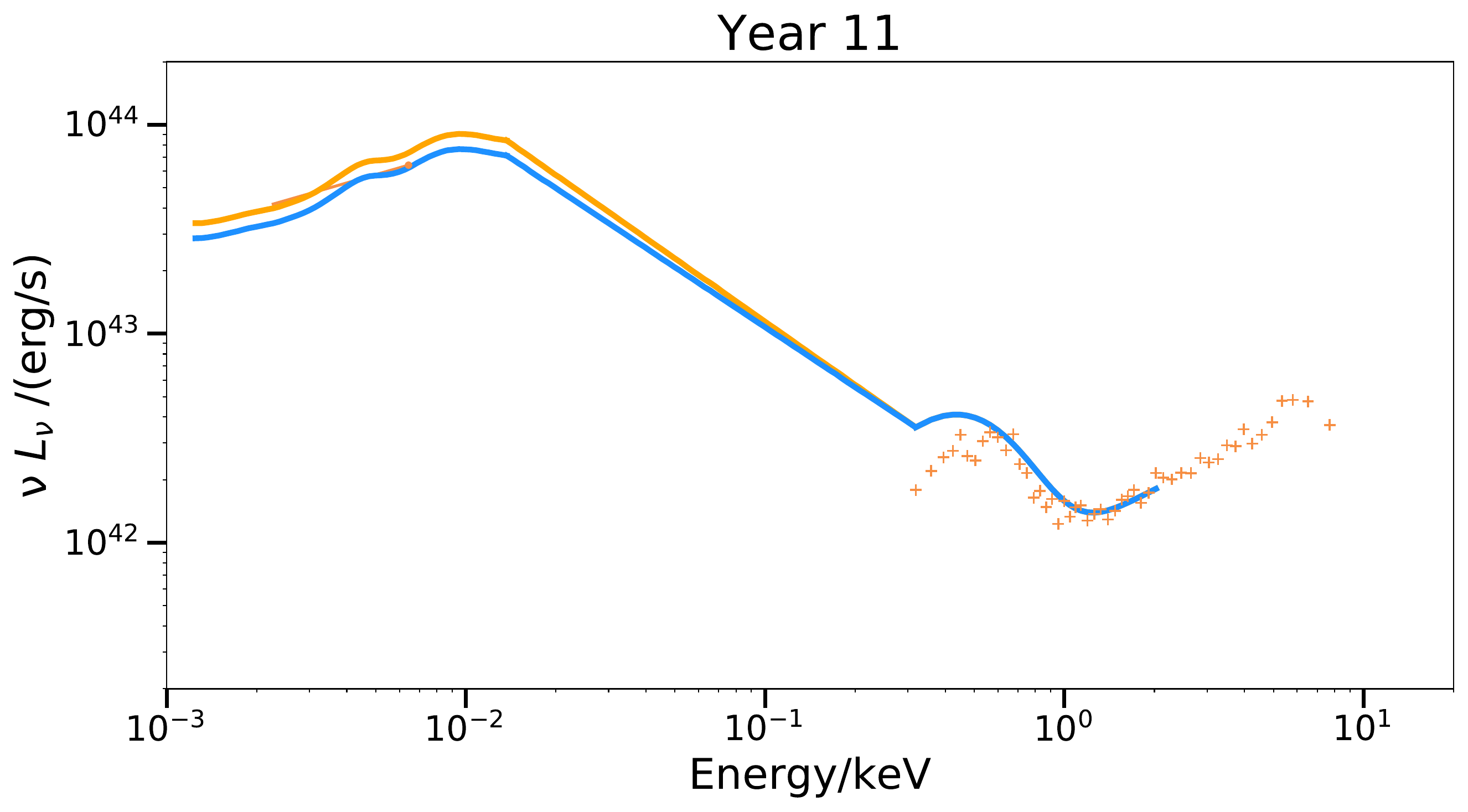}

\vspace{2.00mm}

\includegraphics[width=8.5cm, trim= 0cm 0cm 0cm 0cm, clip=true]{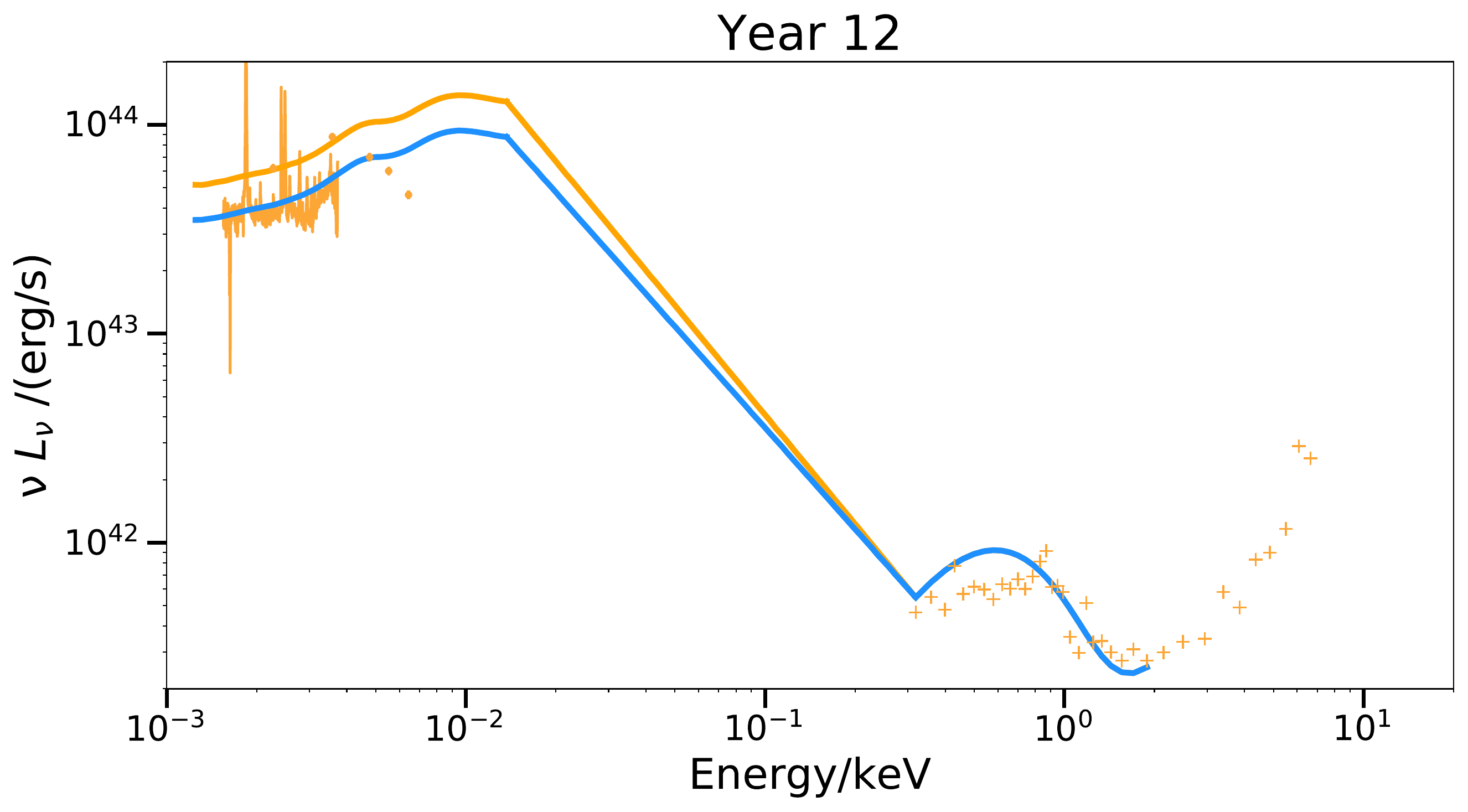}
\hfill
\includegraphics[width=8.5cm, trim= 0cm 0cm 0cm 0cm, clip=true]{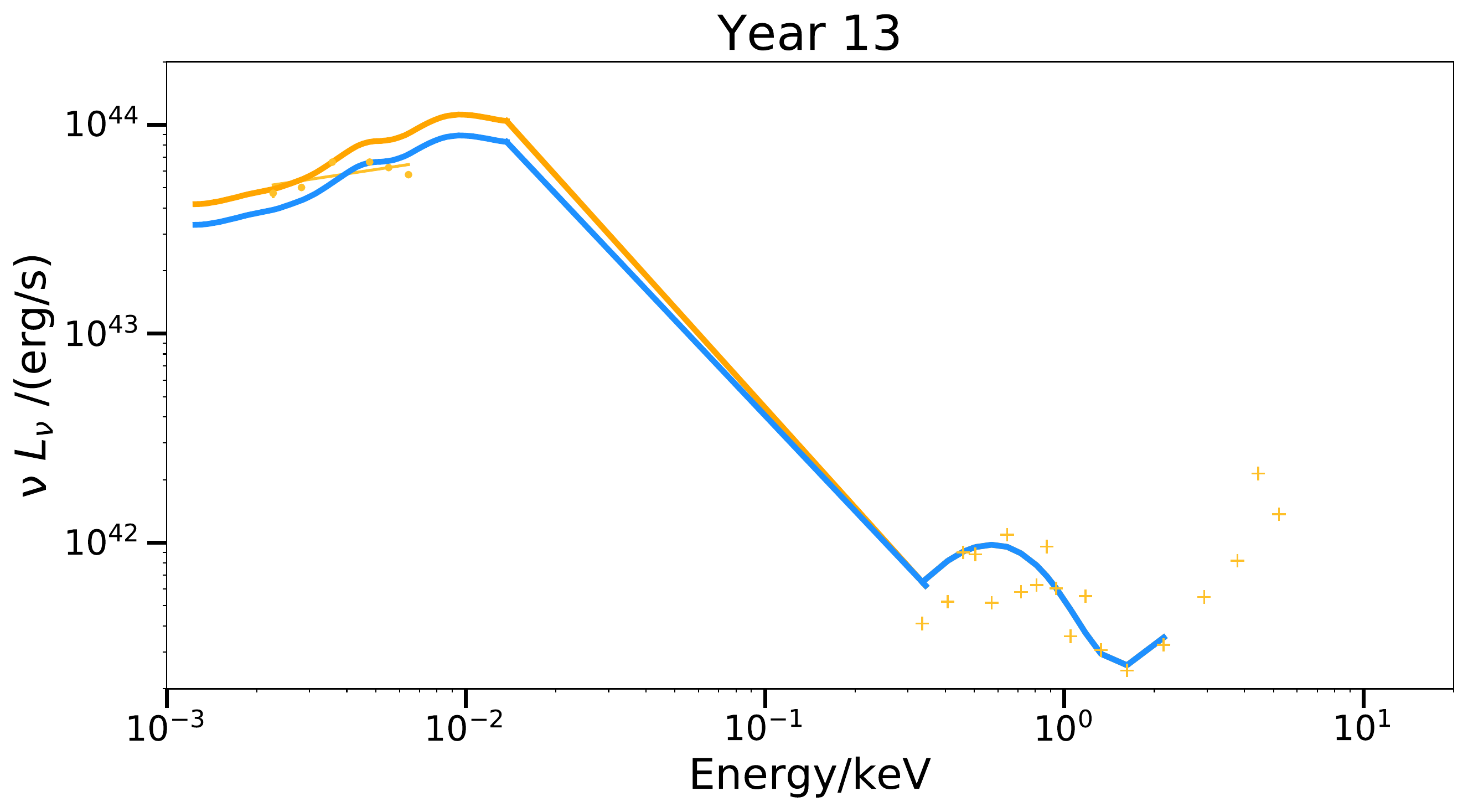}
\caption
{\label{fig:po}
continued}
\end{figure*}

\end{appendix}

\bsp
\label{lastpage}

\end{document}